\documentclass[mnsc,nonblindrev]{informs4}

\OneAndAHalfSpacedXI 

\usepackage{endnotes}
\let\footnote=\endnote
\usepackage{algorithm,wasysym}
\usepackage{algorithmicx,algpseudocode}
\usepackage{dsfont}
\usepackage{bm}
\usepackage{booktabs,multirow}
\usepackage{amsmath,amssymb,amsfonts}
\usepackage{natbib}
 \bibpunct[, ]{(}{)}{,}{a}{}{,}%
 \def\bibfont{\small}%
\usepackage{enumitem}
\usepackage{lmodern}
\usepackage{rotating}
\usepackage{fancyvrb}
\usepackage{float}

\DeclareMathOperator{\diag}{diag}
\DeclareMathOperator{\shrink}{shrink}

\def\theARTICLETOP{} %
\def\theLRHFirstLine{}
\def\theLRHSecondLine{}
\def\theRRHFirstLine{}
\def\theRRHSecondLine{}

\TheoremsNumberedThrough     
\ECRepeatTheorems
\JOURNAL{Management Science}

\EquationsNumberedThrough    

\MANUSCRIPTNO{}

\begin{document}

\RUNAUTHOR{Kong et al.}

\RUNTITLE{Data Synchronization at High Frequencies}

\TITLE{Data Synchronization at High Frequencies}

\ARTICLEAUTHORS{%
\AUTHOR{Xinbing Kong,\textsuperscript{a} Cheng Liu,\textsuperscript{b} Bin Wu,\textsuperscript{c}\textsuperscript{*}} 

\AFF{$^{a}$Southeast University; $^{b}$Wuhan University; $^{c}$University of Science and Technology of China}

\AFF{{$^{*}$Corresponding author}}
\AFF{{\bf Contact:} xinbingkong@126.com (XK), chengliu\_eco@whu.edu.cn (CL), bin.w@ustc.edu.cn (BW)}
}

\ABSTRACT{%
Asynchronous trading in high-frequency financial markets introduces significant biases into econometric analysis, distorting risk estimates and leading to suboptimal portfolio decisions. Existing synchronization methods, such as the previous-tick approach, suffer from information loss and create artificial price staleness. We introduce a novel framework that recasts the data synchronization challenge as a constrained matrix completion problem. Our approach recovers the potential matrix of high-frequency price increments by minimizing its nuclear norm---capturing the underlying low-rank factor structure---subject to a large-scale linear system derived from observed, asynchronous price changes. Theoretically, we prove the existence and uniqueness of our estimator and establish its convergence rate. A key theoretical insight is that our method accurately and robustly leverages information from both frequently and infrequently traded assets, overcoming a critical difficulty of efficiency loss in traditional methods. Empirically, using extensive simulations and a large panel of S\&P 500 stocks, we demonstrate that our method substantially outperforms established benchmarks. It not only achieves significantly lower synchronization errors, but also corrects the bias in systematic risk estimates (i.e., eigenvalues) and the estimate of betas caused by stale prices. Crucially, portfolios constructed using our synchronized data yield consistently and economically significant higher out-of-sample Sharpe ratios. Our framework provides a powerful tool for uncovering the true dynamics of asset prices, with direct implications for high-frequency risk management, algorithmic trading, and econometric inference.

}

\maketitle

\section{Introduction}\label{sec:intro}

Asynchronicity is a stylized feature of high-frequency data, rooted in the natural mechanisms of data generation. In financial markets, for instance, assets trade at distinct, irregular time instances due to factors such as price staleness, market friction, varying liquidity, and the differential speed of information flow. Price staleness, a near-universal phenomenon, means that efficient prices update at heterogeneous rates across assets, leading to non-homogeneously spaced observations. When a price does not update, the observed price is merely a repetition of the last recorded tick, creating a challenge for time-series analysis. Beyond finance, asynchronicity is also pervasive in fields like ecology, clinical medicine, and environmental science, often arising from mixed-frequency sampling schemes.

The presence of asynchronicity poses significant challenges to econometric and statistical analysis, distorting estimates and leading to flawed conclusions in applications such as covariance matrix estimation, multivariate analysis, portfolio allocation, and beta estimation. The consequences can be profound. For example, \cite{hollstein2020conditional} demonstrate that the empirical validity of the Conditional Capital Asset Pricing Model (CAPM) hinges critically on the accurate estimation of betas from high-frequency data. Their work suggests that measurement errors, exacerbated by asynchronous trading, can lead to the premature rejection of foundational economic theories. The core statistical challenges are twofold. First, simple imputation methods like the nearest-tick or previous-tick approach introduce biases in multivariate estimation (\citealt{hayashi2005covariance}). Second, in high-dimensional settings, conventional data alignment via subsampling discards a vast amount of non-synchronized data, resulting in a significant loss of efficiency. While numerous approaches have been developed to mitigate these issues (c.f., \citealt{chen2020five}; \cite{fan2016incorporating}; \citealt{kong2018testing}; \citealt{pelger2019large}; \cite{shin2023adaptive}; \citealt{kong2023discrepancy}; \citealt{cui2024regularized}), they often struggle to balance bias reduction with the full utilization of available information.

There are two streams of works that delve into synchronizing the high-frequency data to facilitate the subsequent applications. The first stream synchronizes the data by dropping some original data. Three main methods in this stream are the Previous Tick method (\citealt{zhang2011estimating}), the Refresh Time Scheme (\citealt{barndorff2011multivariate}), and the Generalized Synchronization procedure (\citealt{ait2010high}). All three methods first select the synchronized sampling time points and then, for each asset and each selected time point, choose one observation that is most close to each synchronized sampling time point from all the original observations of that asset. The drawback of those methods is the drop of a large proportion of data, especially when the number of assets is large and some assets are sparse. The second stream is considering the original data set as a set with missing values and imputing data relying on a parametric state space model and EM algorithm, see \cite{liu2014quasi} and \cite{shephard2017econometric}. Howerver, in high-dimensional settings, the EM algorithm is computationally time consuming.


In this paper, we introduce a novel method for data synchronization. We formulate the problem by minimizing the nuclear norm of the potential increment matrix that are ideally synchronized and well structured, under a large system of linear constraints of increments over non-synchronous durations, see (\ref{nuclear norm}) below. The usage of the nuclear norm is inspired by the well-known low-rank plus noise construction of the data matrix in many applications. The common factor component plus the idiosyncratic error term (discrete or continuous) is a concrete example in finance. The linear constraints over the durations serve as a natural extraction of the data generation mechanism that realizes the data asynchronicity. 

Solving this constrained low-rank optimization problem effectively removes idiosyncratic noise and recovers the ``signal'' component of the potential increment matrix from complex, disorganized data. Our method differs from the tick-sampling approaches by utilizing all data points, thereby avoiding the efficiency loss and potential biases from using overlapping increments in subsequent inference, such as covolatility estimation. It also differs from EM-based methods by being computationally stable and theoretically grounded in large-dimensional diffusion systems. At its core, our approach is a relaxed rank minimization problem, straightforward to implement and efficiently solved using the Alternating Direction Method of Multipliers (ADMM). 

Our method is also different from the price matrix completion via sampling projection operator though our data synchronization approach can be deemed as a potential increment matrix completion via solving large random linear systems. Indeed, synchronizing the data discretely sampled from a large dimensional diffusion process by price matrix completion via the projection operator (projecting a parameter matrix onto the family of matrices that is supported only on sampled entries leaving missing entries set as zeros) meets difficulty since the idiosyncratic error process is intrinsically non-stationary as a general semi-martingale with stochastic volatility. This non-stationarity makes the noisy part as strong as the signal part leading to the so-called spurious factors, see \cite{zhang2018clt} and \cite{onatski2024spurious}. Taking difference of the semi-martingale to relieve the non-stationarity and simultaneously do the sampling projection is however not applicable because of the data asynchronicity. But all durations and the increments over the durations are observable, and they are simply sum of potential (ideally synchronized but latent) increments, which forms a linear system. This is how come our procedure.


Rank minimization under linear constraints has been studied in optimization and operations research. For instance, \cite{recht2010guaranteed} proves that the nuclear norm minimization has a unique solution when the linear operator satisfies the nearly isometry condition. However, their framework is largely deterministic, assuming a noise-free data matrix and a linear operator with randomness properties that are independent of the data. In our setting, both the linear operator (related to the random trading durations) and the data-generating process are stochastic. We further allow for a stochastic idiosyncratic noise process to contaminate the low-rank signal. This introduces significant complexity, as the potential correlation between the diffusion process components and between the linear operator and the potential increment matrix makes the derivation of concentration inequalities highly non-trivial. Consequently, no statistical theory has previously existed for the solution's existence and convergence properties in the context of asynchronous, noise-contaminated, large-panel high-frequency data. This paper aims to fill this theoretical gap.


Our work makes several contributions. To the best of our knowledge, we are the first to prove, under a high-dimensional high-frequency asymptotic regime, that nuclear norm minimization under these stochastic linear constraints yields a unique solution equal to the true low-rank matrix with high probability. This is achieved by establishing the restricted isometry property via a novel concentration inequality for the self-normalized Frobenius norm of a large realized covariance matrix. We are also the first to provide a statistical convergence rate for the completed potential increment matrix. An interesting finding is that the statistical accuracy depends on the sample sizes of both sparse and dense series, in contrast to many tick-based approaches that suffer from efficiency loss.

Empirically, our contributions are equally significant. Through extensive simulations and analysis of a large panel of S\&P 500 stocks, we have demonstrated the clear superiority of our method. It not only yields substantially more accurate estimates of the underlying return and covariance matrices but also provides a more realistic depiction of systematic risk by correcting the bias in eigenvalues caused by stale prices. Most importantly, we have shown that this statistical superiority translates directly into economic value: portfolios constructed using our synchronized data generate consistently higher out-of-sample Sharpe ratios. Our analysis of spot beta dynamics during market turmoil further underscores the reliability of our approach, producing stable and economically intuitive risk profiles where traditional methods fail.



The present paper is organized as follows. In Section \ref{sec:Methodology and Model Specification}, we introduce our methodology and model set up in detail. Our main theoretical results are provided in Section \ref{sec:Main Theoretical Results}. The Monte Carlo simulations are conducted in Section \ref{sec:Monte Carlo Simulation}. Empirical studies and findings are given in Section \ref{sec:Empirical Analysis}. All technical proofs, robustness check and additional results are included in the Supplementary Appendix.


\section{Methodology and Model Specification}\label{sec:Methodology and Model Specification}

\subsection{Asynchronicity and Large-Scale Linear System}\label{subsec:Asynchronicity and Large-Scale Linear System}

Motivated by the high-frequency data analysis, we assume that the asynchronous data is discretely sampled from a large-dimensional diffusion process defined on some filtered probability space $(\Omega, \mathcal{F}, \mathcal{F}_t; 0\leq t\leq T)$ as follows.
\begin{equation}\label{model}
	(dX_t)_{N\times 1}=(\mu_t)_{N\times 1}dt+(\sigma_t)_{N\times r}(dW_t)_{r\times 1}+(\sigma_t^*)_{N\times N}(dW^*_t)_{N\times 1},
\end{equation}
where $W_t$ is a standard multivariate Brownian motion, $\sigma_t$ is the spot volatility process for the ``signal'' component, $\sigma_t^*$ is a diagonal matrix of spot volatility process for the idiosyncratic diffusion process while the $W^*_t$ is the driving Brownian motion, and $T$ is a fixed time horizon. The first term in the right hand side of (\ref{model}) represents the drift term, the second term is a low-rank common component to be reconstructed and the third one is an idiosyncratic ``noise'' term. 

Data asynchronicity means that the coordinate processes $(X_{it}, i=1,..., N)$ of $X_t$ are separately generated in completely different time instances. Assume that $X_{it}$ is sampled at time instances $\{\tau_{i0}, \tau_{i1}, \tau_{i2},..., \tau_{ii_n}\}$ which are different across $i=1,..., N$. Let $\mathcal{T}=\{t_1,...,t_n\}=\cup_{1\leq i\leq N}\{\tau_{i1},...,\tau_{ii_n}\}$ and $\tau_{i0}=0$ for all $i=1,...,N$. Then $X_{it_j}$ can be thought of as missing value if $t_j\not\in \{\tau_{i1},...,\tau_{ii_n}\}$. Let the potential increment matrix be $\Delta=[(\Delta_1,...,\Delta_N)^{\prime}]_{N\times n}$ with $\Delta_{ij}=X_{it_j}-X_{it_{j-1}}$, and the potential matrix be $X=(X_{it_j})_{N\times n}$. Though $X_{\cdot t_j}$, the $j$-th column of $X$, has at least one observed data, $\Delta_{j\cdot}$, the $j$-th row of $\Delta$, may not. This makes the projection operator approach for matrix completion based on $\Delta$ not applicable, see for example \cite{candes2012exact} and \cite{chen2019inference}. The same method applied to $X$ is of difficult also because the semi-martingale is generally nonstationary making the idiosyncratic ``noise'' part as strong as the ``signal'' part leading to spurious factors, see \cite{zhang2018clt} and \cite{onatski2024spurious}.

\begin{figure}
	\FIGURE
	{\includegraphics[scale=0.55]{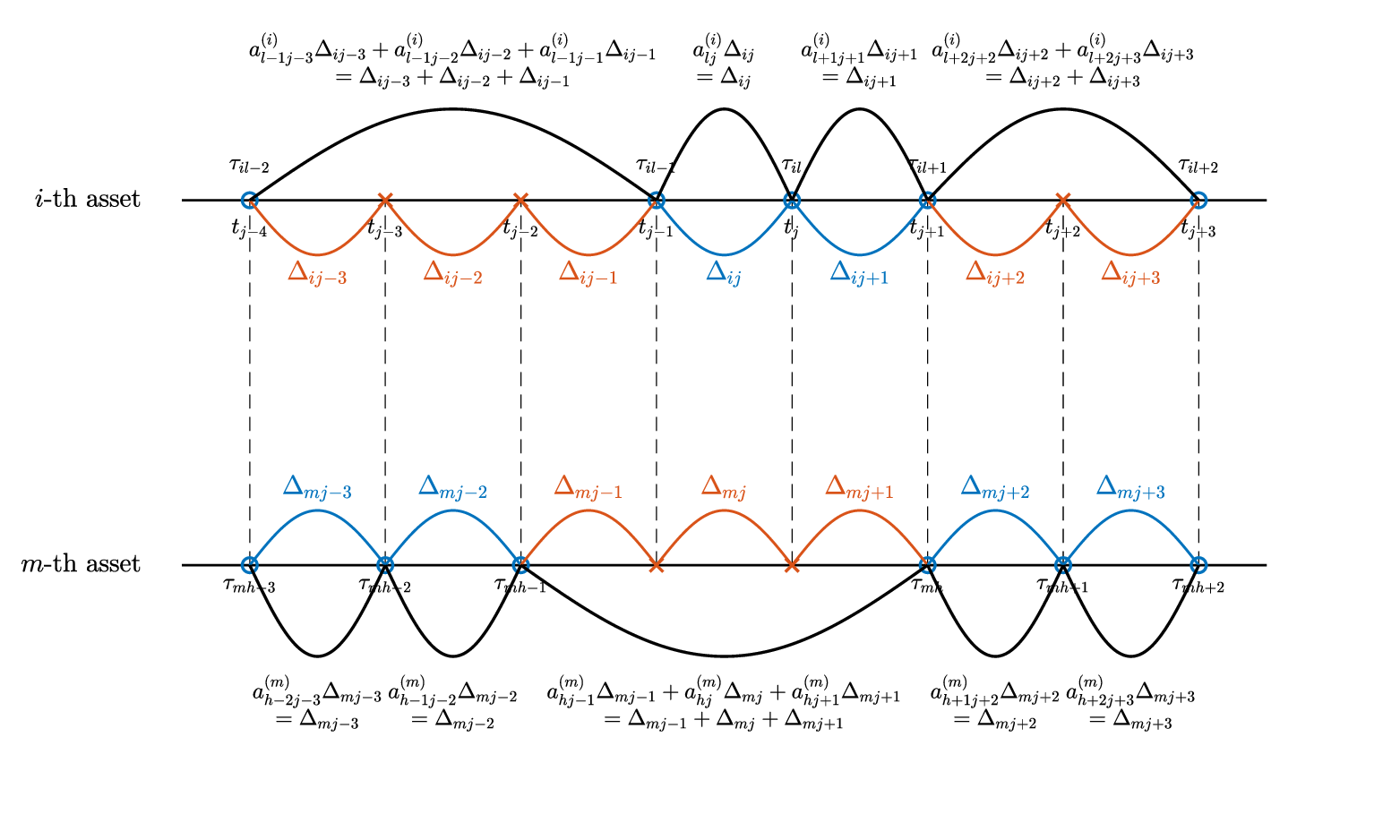}}
	{Asynchronous observations of two assets. \label{fig:asynchronous price}}
	{This figure illustrates the problem of asynchronous trading for two assets ($i$-th and $m$-th) relative to a synchronous time grid, $\mathcal{T}$. The red line in the figure frames the range of returns that are in $\mathcal{T}$ and unobservable (potential), the blue line frames the range of returns that are in $\mathcal{T}$ and observable, and the black line frames the range of all observable returns. The equations represent the linear constraints.}
\end{figure}

Though the potential increment matrix $\Delta$ is far from fully observed, the increments over durations, a linear transform of $\Delta$, can be fully observed. Define a linear operator $\mathcal{A}$ as follows.
$$
\mathcal{A}(\Delta)=\mbox{diag}\{A_1,...,A_N\}\mbox{vec}(\Delta^{\prime})=: A \mbox{vec}(\Delta^{\prime}),
$$
where $A_i=(a_{jk}^{(i)})_{n_i\times n}$ is a matrix of zeros and ones so that $A_i\Delta_i=b_i$ where $b_i=(X_{i\tau_{i1}}-X_{i\tau_{i0}},..., X_{i\tau_{in_i}}-X_{i\tau_{i(n_i-1)}})^{\prime}$ which is the vector of observed increments of $X_{it}$ over durations, and $\text{vec}(\cdot)$  is the standard vectorization operator. A simple example of $A_i$ is as follows.
$$
A_i=\left(\begin{array}{cccc}
	a^{(i)}_{11} & a^{(i)}_{12} & \cdots & a^{(i)}_{1n} \\
	a^{(i)}_{21} & a^{(i)}_{22} & \cdots & a^{(i)}_{2n} \\
	&    \cdots     & \cdots &              \\
	a^{(i)}_{n_i1} & a^{(i)}_{n_i2} & \cdots & a^{(i)}_{n_in}
\end{array}
\right)=\left(\begin{array}{ccccccccccc}
	1 & 1 & 0 & 0 & , & \cdots & , & 0 & 0 & 0 & 0 \\
	0 & 0 & 1 & 0 & , & \cdots & , & 0 & 0 & 0 & 0 \\
	&   &   &   & , & \cdots & , &   &   &   &   \\
	0 & 0 & 0 & 1 & , & \cdots & , & 0 & 0 & 0 & 0 \\
	0 & 0 & 0 & 0 & , & \cdots & , & 0 & 1 & 1 & 1
\end{array}
\right).
$$
The first row of $A_i$ amounts to saying that we can only observe the sum of the first two potential increments of $X_{it}$ but not any of them. In finance, this is caused by the trading mechanism so that the observation times are typically random and unequally spaced.
Let $b=(b_1^{\prime},...,b_N^{\prime})^{\prime}$. The constraint for $\Delta$ is $\mathcal{A}(\Delta)=b$. An illustration of the asynchronicity for two assets is in Figure \ref{fig:asynchronous price}

Both in theory and applications, we assume that the signal term $\{\sigma_tdW_t| 0\leq t\leq T\}$ is in some manifold of low-dimension. Suppose that for each sample path $\omega$, $\sigma_t(\omega)=(\sigma_0)_{N\times r}(\Sigma_t(\omega))_{r\times r}$ where $\sigma_0$ is a constant volatility level and the time variation of the volatility process $\sigma_t\sigma_t^{\prime}$ is due to $\Sigma_t$. Typically, we assume that $\Sigma_t$ is always of full rank for any time $t$. Without loss of generality, we assume, as in \cite{kong2017number,kong2018systematic}, that $\sigma_t^*$ is a diagonal matrix with locally bounded entries of stochastic processes, and $W^*_t$ is a $N$-dimensional Brownian motion with correlation matrix $(\rho^*)_{N\times N}$. So the parameter space is
$$
\Theta=\{(\sigma^0, \Sigma_t(\omega)dW_t(\omega)); (\sigma^{0})^{\prime}\sigma^0=N^{\alpha}, \omega\in \Omega\},
$$
where $\alpha$ is a constant in $(0, 1]$ that controls the strength of the low-rank signal component, $\Omega$ stands for the sample space. While the $\omega$ will be involved in the probability calculation, the deterministic parameter is $\sigma^0\in \Sigma^0$.

When the idiosyncratic error term is vanishing, the problem to solve is to recover the potential increment matrix $\Delta=\Pi$. One proposal is to find a low-rank matrix so that the following optimization has a solution.
\begin{equation}\label{nuclear norm}
	\widehat{\Pi}=\mbox{arg}\min_{\Pi} \|\Pi\|_{*} \ \ \mbox{subject to} \ \ \mathcal{A}(\Pi)=b,
\end{equation}
where $\|\cdot\|_*$ stands for the nuclear norm of some matrix. 
When the idiosyncratic error term is present, (\ref{nuclear norm}) only identifies the low-rank signal part of the potential increment matrix ignoring completing the increments inside $b$ that includes idiosyncratic errors. To simultaneously recover the low-rank component $\Pi$ and impute the missing increments in $\Delta$, we consider the following optimization problem.
\begin{equation}\label{nuclear norm2}
	(\widehat{\Pi},\widehat{\Delta})=\mbox{arg}\min_{\Pi,\Delta}\|\Pi\|_* \ \ \mbox{subject to} \ \ \mathcal{A}(\Delta)=b \ \& \ \Delta=\Pi+\Pi^*.
\end{equation}
In the context of the model (\ref{model}), $\Pi$ and $\Pi^*$ in (\ref{nuclear norm2}) correspond to the increment matrices contributed by $\sigma_tdW_t$ and $\mu_tdt+\sigma_t^*dW_t^*$, respectively.

\subsection{Computational Issues}\label{subsec:Computational Issues}

To solve the optimization problem (\ref{nuclear norm2}), we propose to apply the ADMM algorithm, which is simple but powerful. The ADMM decomposes a large global optimization problem into small local subproblems, such that it can be efficient when both the dimension and the sample size of the data are large, see \cite{scheinberg2010sparse} and \cite{boyd2011distributed} for details.

\subsubsection{Scaled ADMM Algorithm}\label{subsubsec:Scaled ADMM Algorithm}
Since $\Delta=\Pi+\Pi^*$ and $\Pi^*$ can be considered as the residuals, the equivalent Lagrangian formula is 
\begin{equation*}
	{\min}_{(\Delta,\Pi)}
	\frac{1}{2}{\Vert \mathcal{A}}(\Delta) -b\Vert^2_F  +  \frac{\mu}{2}\Vert \Delta-\Pi\Vert^2_{F} +\lambda\Vert \Pi\Vert_*,
\end{equation*}
which is further equivalent to the following relaxed form 
\begin{equation}\label{New_variable2}
	\begin{split}
		&{\min}_{(\Delta,\Pi,Z_{\Delta},Z_{\Pi})}~
		\frac{1}{2}\Vert \mathcal{A}(\Delta) -b\Vert^2_F  + \frac{\mu}{2}\Vert Z_{\Delta}-\Pi\Vert^2_{F} +\lambda\Vert Z_{\Pi}\Vert _*\\
		&~~~~~~~~~~~~~~~~~~\mbox{subject to} \quad  \Delta=Z_{\Delta},~ \Pi=Z_{\Pi},
	\end{split}
\end{equation}
where $Z_{\Delta}$ and $Z_{\Pi}$  are auxiliary variables.

To solve (\ref{New_variable2}), we borrow the idea from \cite{scheinberg2010sparse} and propose a {scaled version} of the ADMM algorithm, which relies on the following augmented Lagrangian:
\begin{equation}\label{Augmented Lagrangian}
	\begin{split}
		\widetilde{\mathcal{L}}(\Delta,\Pi,Z_{\Delta},Z_{\Pi},U_{\Delta},U_{\Pi})
		= &\frac{1}{2}\Vert \mathcal{A}(\Delta) -b\Vert^2_F  +  \frac{\mu}{2}\Vert Z_{\Delta}-\Pi\Vert^2_{F} +\lambda\Vert Z_{\Pi}\Vert _* \\
		+ &\frac{\eta}{2} \Vert  Z_{\Delta}-\Delta+U_{\Delta} \Vert ^2_F + \frac{\eta}{2} \Vert  Z_{\Pi}-\Pi+U_{\Pi} \Vert ^2_F,
	\end{split}
\end{equation}
where $\eta>0$ is a penalty parameter, $U_{\Delta}$ and $U_{\Pi}$ are scaled dual variables corresponding to the constraints in (\ref{New_variable2}). We then solve problem (\ref{Augmented Lagrangian}) by updating the unknown terms one by one. Let $\left(\Delta^{k},\Pi^{k},Z_{\Delta}^{k},Z_{\Pi}^{k},U_{\Delta}^{k},U_{\Pi}^{k}\right)$ be the solution at step $k$, for $k=0,1,2, \cdots$. We first update $\Delta$ according to 
$$
\begin{aligned}
	\Delta^{k+1}&=\underset{\Delta}{\arg \min}~
	\widetilde{\mathcal{L}}(\Delta,\Pi^{k},Z_{\Delta}^{k},Z_{\Pi}^{k},U_{\Delta}^{k},U_{\Pi}^{k})\\
	& =\underset{\Delta}{\arg \min}~\{
	\frac{1}{2}\Vert \mathcal{A}(\Delta) -b\Vert^2_F
	+ \frac{\eta}{2} \Vert  Z_{\Delta}^{k}-\Delta+U_{\Delta}^{k} \Vert ^2_F \}\\
	& =\underset{\Delta}{\arg \min}~ \{\frac{1}{2}\Vert \mathcal{A}(\Delta) -b\Vert^2_F
	+ \frac{\eta}{2} {\mbox{vec}}((Z_{\Delta}^{k}-\Delta+U_{\Delta}^{k})')' {\mbox{vec}}((Z_{\Delta}^{k}-\Delta +U_{\Delta}^{k})')\}.
\end{aligned}
$$
Denote $\widetilde{\Delta}={\mbox{vec}}(\Delta')$, we rewrite above result as 
$$
\begin{aligned}
	\widetilde{\Delta}^{k+1}&=\underset{\widetilde{\Delta}}{\arg \min}~\left\{
	\frac{1}{2}\Vert A\widetilde{{\Delta}} -b\Vert^2_F + \frac{\eta}{2}
	\left[{\mbox{vec}}((Z_{\Delta}^{k}+U_{\Delta}^{k})')'-\widetilde{\Delta}' \right]
	\left[{\mbox{vec}}((Z_{\Delta}^{k}+U_{\Delta}^{k})')-\widetilde{\Delta} \right]\right\}\\
	&=(A'A+\eta I)^{-1}\left[A'b+\eta \mbox{vec}((Z_{\Delta}^{k}+U_{\Delta}^{k})')\right],
\end{aligned}
$$
where the last equation can be obtained by taking the derivative of $\widetilde{\Delta}$ for the above objective function and solving the equation
$$A'A\widetilde{\Delta}-Ab'+\eta \widetilde{\Delta}-\eta\mbox{vec}(Z_{\Delta}^{k}+U_{\Delta}^{k})=0 .$$
We then have $$\Delta^{k+1} = \mbox{vec}^{-1}\left((A'A+\eta I)^{-1}\left[A'b+\eta \mbox{vec}((Z_{\Delta}^{k}+U_{\Delta}^{k})')\right]\right)^{\prime},$$ where $\mbox{vec}^{-1}$ is the inverse vectorization (or matricization) operator. The closed form of $\Pi^{k+1}$ is 
\begin{align*}
\Pi^{k+1}=&\underset{\Pi}{\arg \min}~
\widetilde{\mathcal{L}}(\Delta^{k+1},\Pi,Z_{\Delta}^{k},Z_{\Pi}^{k},U_{\Delta}^{k},U_{\Pi}^{k})=\frac{1}{\mu+\eta}(\mu Z_{\Delta}^{k}+\eta Z_{\Pi}^{k}+\eta U_{\Pi}^{k}),
\end{align*}
and similarly, for $Z_{\Delta}^{k+1}$ and $Z_{\Pi}^{k+1}$,
\begin{align*}
    Z_{\Delta}^{k+1} =\frac{1}{\mu+\eta}(\mu \Pi^{k+1}+\eta \Delta^{k+1}-\eta U_{\Delta}^{k}),\quad Z_{\Pi}^{k+1}= \shrink(\Pi^{k+1}-U_{\Pi}^{k},\frac{\lambda}{\eta}).
\end{align*}
To get $Z_{\Pi}^{k+1}$, we have used equation (8) in \cite{gandy2011tensor}, where for a scalar $\psi>0$, $\shrink(\Psi, \psi)$ is an operator that gives a soft-threshold to each singular value of the matrix $\Psi$ such that $\shrink(\Psi, \psi)=U\diag(\max(\rho_1-\psi, 0), \cdots, \max(\rho_N-\psi, 0))V^*$ with $U{\diag}(\rho_1, \cdots, \rho_N)V^*$ to be the singular value decomposition of $\Psi$. For convenience, we set 
$$U_{\Delta}^{k+1}=U_{\Delta}^{k}+Z_{\Delta}^{k+1}-\Delta^{k+1},\quad
U_{\Pi}^{k+1}=U_{\Pi}^{k}+Z_{\Pi}^{k+1}-{\Pi}^{k+1}.$$ 
The computational steps are summarized in Algorithm 1. 

\begin{algorithm}[t!]
    \caption{Estimation of $\Delta$ and $\Pi$ via scaled ADMM}\label{alg1}
    \label{alg:admm_estimation}
    \begin{algorithmic}[1]
        \Require 
        Observations of log-prices: $X=\{X_{i,\tau_{i0}}, X_{i,\tau_{i1}}, \dots, X_{i,\tau_{ii_n}}; i=1, \dots, N\}$.
        Initial estimates: $\Pi^0, Z_{\Delta}^0, Z_{\Pi}^0, U_{\Delta}^0, U_{\Pi}^0$ (typically initialized to be zero).
        Tuning parameters: $\mu, \lambda, \eta$.

        \Ensure Estimated matrices $\widehat{\Delta}$ and $\widehat{\Pi}$.

        \State Construct matrices $A_1, \dots, A_N$ and vector $b$ from the observations of $X$.
        \State Set iteration counter $k \leftarrow 0$.

        \While{not converged}
    \Statex \textit{$\RHD$ Update primal variables} 
    \State $\Delta^{k+1} \leftarrow \mbox{vec}^{-1}\left((A'A+\eta I)^{-1}\left[A'b+\eta \mbox{vec}((Z_{\Delta}^{k}+U_{\Delta}^{k})')\right]\right)^{\prime}$
    \State $\Pi^{k+1} \leftarrow \frac{1}{\mu+\eta} \left( \mu Z_{\Delta}^{k} + \eta Z_{\Pi}^{k} +\eta U_{\Pi}^{k} \right)$
    
    \Statex \textit{$\RHD$ Update auxiliary variables}
    \State $Z_{\Delta}^{k+1} \leftarrow \frac{1}{\mu+\eta} \left( \mu \Pi^{k+1} + \eta \Delta^{k+1} - \eta U_{\Delta}^{k} \right)$
    \State $Z_{\Pi}^{k+1} \leftarrow \shrink\left(\Pi^{k+1} - U_{\Pi}^{k}, \frac{\lambda}{\eta}\right)$
    
    \Statex \textit{$\RHD$ Update dual variables}
    \State $U_{\Delta}^{k+1} \leftarrow U_{\Delta}^{k}+Z_{\Delta}^{k+1}-\Delta^{k+1}$
    \State $U_{\Pi}^{k+1} \leftarrow U_{\Pi}^{k}+Z_{\Pi}^{k+1}-{\Pi}^{k+1}$
    
    \State $k \leftarrow k+1$

   \EndWhile

        \State Set $\widehat{\Delta} \leftarrow \Delta^k$ and $\widehat{\Pi} \leftarrow \Pi^k$.
    \end{algorithmic}
\end{algorithm}

\begin{remark}[Computational Efficiency and Implementation]
	The main computational bottleneck in Algorithm \ref{alg1} is the inversion of the matrix $(A'A+\eta I)^{-1}$, as $A'A$ can be a huge, non-diagonal matrix. To maintain computational efficiency, we leverage the Woodbury matrix identity:
	$$(A'A+\eta I)^{-1}=\eta^{-1}I -\eta^{-1}IA' \left[I+ A \eta^{-1}I A'\right]^{-1} A \eta^{-1}I =  \eta^{-1}I -\eta^{-2}A' \left[I^{-1}+\eta^{-1} A  A'\right]^{-1} A,$$
as calculating the inverse of the diagonal matrix $I+AA'/\eta$ is fast. In addition, since we have all the closed forms of $\Delta^{k+1},\Pi^{k+1},Z_{\Delta}^{k+1},Z_{\Pi}^{k+1},U_{\Delta}^{k+1}$, and $U_{\Pi}^{k+1}$, the Algorithm 1 is not time-consuming even if the sample size and dimension of $X$ are high. In practice, the algorithm is considered to have converged when the following condition is satisfied for a tolerance level of $\epsilon=10^{-5}$:
\footnotesize\begin{equation*}
\begin{aligned}
\max\left\{\frac{\Vert\Delta^{k+1}-\Delta^{k}\Vert_{F}}{\max \left(1, \Vert \Delta^{k}\Vert_{F}, \Vert \Delta^{k+1}\Vert_{F} \right)}, \frac{\Vert\Pi^{k+1}-\Pi^{k}\Vert_{F}}{\max\left(1, \Vert \Pi^{k}\Vert_{F},\Vert \Pi^{k+1}\Vert_{F}  \right)}, \frac{\Vert \Delta^{k}-Z_{\Delta}^{k}\Vert_{F}}{\max
		\left(1, \Vert \Delta^{k} \Vert_{F}, \Vert Z_{\Delta}^{k} \Vert_{F}\right)}, \frac{\Vert \Pi^{k}-Z_{\Pi}^{k}\Vert_{F}}{\max
		\left(1, \Vert \Pi^{k} \Vert_{F}, \Vert Z_{\Pi}^{k}\Vert_{F}\right)}\right\} < \epsilon.
\end{aligned}
\end{equation*}
\end{remark}

\subsubsection{Choices of Initial Estimates and Tuning Parameters}\label{subsubsec:Choices of Initial Estimates and Tuning Parameters}
The above algorithm is not sensitive to the initial estimates of $\Pi$, $Z_{\Delta}$, $Z_{\Pi}$, $U_{\Delta}$, $U_{\Pi}$. We can set them to the zero matrices. First of all, to improve the convergence speed of the algorithm, we can use the previous tick method to get a full observed $X$, which means we let $X_{it_1}$ to be the first observation of the $i$th asset and $X_{it_j}=X_{it_{j-1}}$ if the $i$th asset has no observation at time $t_j$. We then set the initial values of $Z_{\Delta}$ to be the first-order difference of the initial estimate of $X$ and run the PCA to get the initial estimate of $\Pi$. The initial estimate of $U_{\Delta}$ and $U_{\Pi}$ can be set to zero matrices.

To select the optimal tuning parameter ($\mu$, $\lambda$, $\eta$), we propose a data-driven validation scheme based on artificial masking. The procedure aims to find the parameter that minimizes imputation error on a held-out portion of the data. Our scheme operates as follows. Given the input log-price matrix $\mathcal{P}$, a set of candidate masking probabilities $\{p_1,p_2,..,p_m\}$, and a number of repetitions $Q$, we sequentially do the following. (i) Mask generation: For each probability $p_i$ and each repetition $j\in\{1,...,Q\}$, we generate a binary mask matrix $\mathcal{M}_{ij}$. This is done by first identifying all observable entries in $\mathcal{P}$ and then randomly selecting a fraction $p_i$ of them to be masked. In $\mathcal{M}_{ij}$, these masked positions are marked with 1, and all others with 0. (ii) Imputation: The Algorithm 1 is then applied onto the partially observed data (the entries of $\mathcal{P}\circ\mathcal{M}_{ij}$) to produce a completed matrix $\hat{\mathcal{P}}_{ij}(\mu,\lambda, \eta)$. (iii) Error calculation: We then calculate the imputation error between the imputed matrix $\hat{\mathcal{P}}_{ij}(\mu,\lambda, \eta)$ and the original matrix $\mathcal{P}$, but only on the set of entries that were artificially masked (corresponding position of $\mathcal{M}_{ij}$ is 1).

The optimal parameter is chosen as the one that yields the lowest average imputation error across all masks and repetitions. We evaluate the imputation accuracy using the following two error metrics:
\begin{align}\label{eq:compute error0}
	\begin{aligned}
		\mathcal{R}^{\text{absolute}}(\mu,\lambda, \eta):=&\frac{1}{mQ}\sum_{i=1}^m\sum_{j=1}^Q\frac{\|(\hat{\mathcal{P}}_{ij}(\mu,\lambda, \eta)-\mathcal{P})\circ\mathcal{M}_{ij}\|_F}{\|\mathcal{M}_{ij}\|_F},\\
		\mathcal{R}^{\text{relative}}(\mu,\lambda, \eta):=&\frac{1}{mQ}\sum_{i=1}^m\sum_{j=1}^Q\frac{\|(\hat{\mathcal{P}}_{ij}(\mu,\lambda, \eta)-\mathcal{P})\circ\mathcal{M}_{ij}\|_F}{\|\mathcal{P}\circ\mathcal{M}_{ij}\|_F},
	\end{aligned}
\end{align}
where $\circ$ denotes the Hadamard product. The Frobenius norm $\|\cdot\|_F$ in the numerator is calculated only over the set of masked entries.

\begin{figure}
	\FIGURE
	{\includegraphics[scale=0.42]{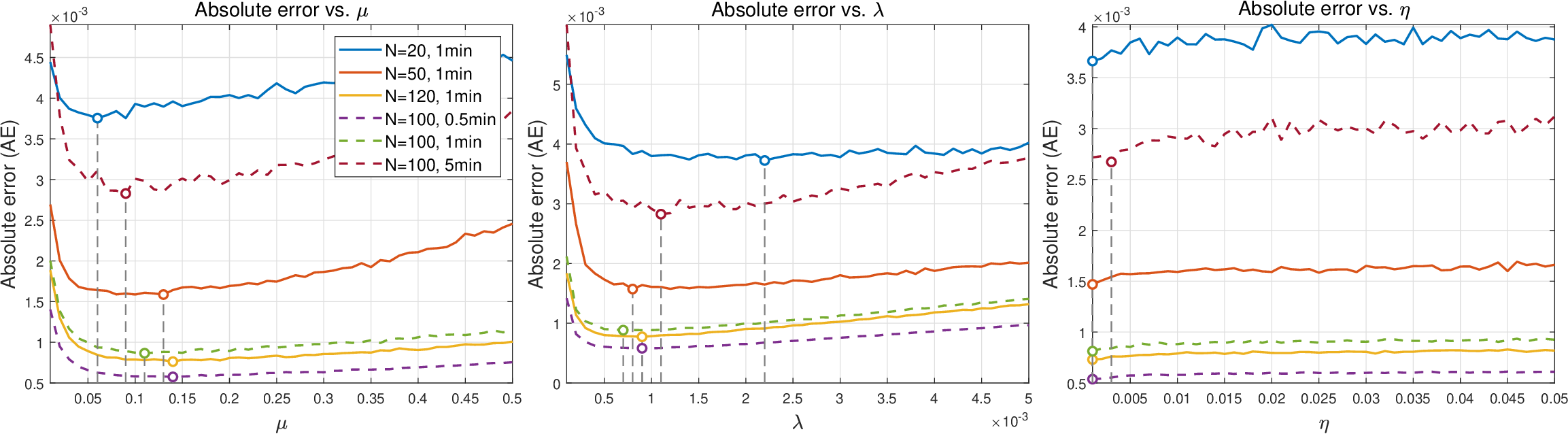}}
	{Absolute error as a function of tuning parameters $\mu$, $\lambda$ and $\eta$. \label{fig:TuningPara_AE}}
	{The absolute error is calculated according to \eqref{eq:compute error0}. The vertical dotted line in each panel indicates the parameter value that minimizes the error.}
\end{figure}

To demonstrate the effect of these tuning parameters, we conduct the validation procedure using the following candidate sets: $\mu\in\{0.01,0.02,...,0.5\}$, $\lambda\in\{0.0001,0.0002,...,0.005\}$, $\eta\in\{0.001,0.002,...,0.05\}$. The set of masking probabilities is $\{0.1,0.2,...,0.7\}$, and the number of repetitions is set to $Q=1$. Figure \ref{fig:TuningPara_AE} illustrates the impact of each tuning parameter on the absolute error. The corresponding results for the relative error are provided in the Supplementary Appendix.

Figure \ref{fig:TuningPara_AE} illustrates the sensitivity of the absolute error to the tuning parameters $\mu$, $\lambda$, and $\eta$. Several key observations guide our parameter selection strategy. First, the estimation error appears insensitive to the specific value of $\eta$. Based on this observation and to enhance computational efficiency, we simplify the three-dimensional search by fixing the ratio between the two parameters, setting $\lambda/\eta=0.1$. Second, the figure shows that the optimal values for $\mu$ and $\lambda$ are quite stable, consistently falling within a relatively narrow range. This suggests that the optimal parameter configuration is robust to the specific characteristics of the dataset (such as the number of assets or observation frequency, as implicitly varied across the panels). This stability allows us to streamline the tuning process in practice. Instead of searching over a wide grid, we can focus on more refined candidate sets, $\mathbb{M}$ and $\mathbb{L}$, for $\mu$ and $\lambda$, respectively. Formally, the optimal parameters $(\hat{\mu},\hat{\lambda})$ can be determined by solving the following optimization problem:
\begin{align*}
	(\hat{\mu},\hat{\lambda})=\argmin_{\mu\in\mathbb{M},\lambda\in\mathbb{L}}\mathcal{R}^{\text{absolute}}(\mu,\lambda, 10\lambda).
\end{align*}
	
\subsection{Technical Assumptions}\label{subsec:Technical Assumptions}
To connect the matrix parameter $\Pi$ to the generative semi-martingale, one can relate $\Pi$ to $\sigma^0$ and $\Sigma_t$ so that the varying of $\Pi$ is caused by the smooth change of $\sigma^0$ and $\Sigma_t$ for every sample path $\omega\in\Omega$. That is the parameter $\Pi$ is simply a potential increment matrix of some regular semi-martingale for some volatility parameter given $\omega$. Thus we impose the technical conditions on the dynamics of the generative model (\ref{model}) and the durations that realizes the operator $\mathcal{A}$.

The first assumption is a regular condition for the durations of the coordinate processes. Before stating it, we introduce one more notation. For a generic process $Y_t\in R^{d}$, let $$
d_{ik}:=\sharp\{j; \tau_{i(l-1)}\leq t_j\leq t_{k-1}\leq \tau_{il} \ \mbox{for some} \ 1\leq l\leq n_i\}
$$
to be the number of potential increments before $\Delta^n_{k}Y:=Y_{t_j}-Y_{t_{j-1}}$ in the observed interval $(\tau_{i(l-1)}, \tau_{il}]$ which contains the time point $t_{k-1}$. We make a convention that $d_{ik}:=0$ if the set is empty and the resulting sum $\sum^{0}_{l=1}\Delta^n_{k-l}Y=0$.

\begin{assumption}\label{ass1}
	\begin{enumerate}
		\item
		There exists positive random variables $R_{j}$'s independent of $\mathcal{F}$, such that $t_j-t_{j-1}\leq R_{j}$, $\sum^n_{j=1}R_j\leq C$, $\max_jR_j=o(1)$, and $(\sum^n_{j=1}R_j)^{3/4}/\max_jR_j\geq c$ for some $c>0$.
		\item The maximum diagonal element of the diagonal matrix $AA^{\prime}$ satisfies
		$$
		P\left(\|AA^{\prime}\|_\infty>2\left(1+\sqrt{Nn/n^*}\right)\right)\leq \exp\{-\gamma Nn\},
		$$
		where $n^*=\sum^N_{i=1}n_i$ and $\gamma$ is a positive constant.
		
		\item There exists a sequence of numbers $\{D_{ik}\}$ that are independent of $\mathcal{F}$ and satisfy $d_{ik}\leq D_{ik}$.
	\end{enumerate}
\end{assumption}

Assumption \ref{ass1}-1 allows for random and unequally spaced sampling times. Though the upper bound $R_j$'s are assumed to be independent of the process $X_t$, the coordinate durations can be stopping times that are dependent on $X_t$. Assumption \ref{ass1}-2 regularizes the maximum number of potential increments missed in an observed duration. A simple example is that $\|AA^{\prime}\|_\infty$ is bounded. In this case, the probability upper bound is zero for large enough $N$ and $n$ once $\sqrt{Nn/n^*}\rightarrow \infty$, and $t_j-t_{j-1}$'s are of order $1/n$ and all conditions in Assumption \ref{ass1} are satisfied.

The next assumption assumes that the systematic and idiosyncratic spot volatility processes are locally bounded, and the correlation matrix $\rho^*$ is sparse in some sense.

\begin{assumption}\label{ass2}
	\begin{enumerate}
		\item
		There exists a sequence of stopping times $s_m\rightarrow\infty$ and a sequence of numbers $c_m$, such that $\max_{i\leq N}(|\sigma_{i(t\wedge s_m)}|+|\sigma_{i(t\wedge s_m)}^*|)\leq c_m$.
		\item $\max_{i\leq N}\sum_{j=1}^N|\rho^*_{ij}|\leq C$.
		\item $\|IV_1\|_F^2:=\|\sum^T_0\Sigma_t\Sigma_t^{\prime}dt\|_F^2\geq cr$ for some $c>0$.
		\item $|\sigma_{i(t+h)}^*-\sigma_{it}|\leq Ch^{1/2-\epsilon}$ for some constant $C$ and arbitrary constant $\epsilon$.
	\end{enumerate}
\end{assumption}

\section{Main Theoretical Results}\label{sec:Main Theoretical Results}
The first theoretical result gives the nearly-isometry property for $\mathcal{A}$ operating on matrix generated from large-dimensional semi-martingales.  Let $P_{R, D}$ be a probability measure conditional on $\{R_j, D_{ik}; i\leq N, j\leq n, k\leq n\}$.
\begin{theorem}\label{th1}
	Let the quantities $\overline{L}_1=\sum^n_{k=1}R_k\sum^N_{i=1}\|\sigma^0(i,\cdot)\|^2(\sum^{D_{ik}}_{l=1}R_{k-l})\log{(\sum^{D_{ik}}_{l=1}R_{k-l})}$ and $\overline{L}_2=\sum^n_{k=1}R_k\sum^N_{i=1}(\sum^{D_{ik}}_{l=1}R_{k-l})\log{(\sum^{D_{ik}}_{l=1}R_{k-l})}$, where $\sigma^0(i,\cdot)$ is the $i$-th row of $\sigma^0$. For some small $c>0$,
	\begin{eqnarray}\label{bound-th1}
		P_{R, D}\left(|\frac{\|\mathcal{A}(\Pi)\|_F^2-\|\Pi\|_F^2}{\|\Pi\|_F^2}|>\delta\right)
		\leq  C\exp\left\{-c^2\delta^2N^{\alpha}/\overline{L}_1\right\}+2\exp\left\{-c^2r^2/\left(\sum^n_{j=1}R_j^2\right)^{1/2}\right\},
	\end{eqnarray}
	and
	\begin{align}\label{bound-th1-1}
	\notag	& P_{R, D}\left(|\frac{\|\mathcal{A}(\Delta)\|_F^2-\|\Delta\|_F^2}{\|\Delta\|_F^2}|>\delta\right)\\
 \leq & 2\exp\left\{-\frac{N^{2-\alpha}c^2\delta^2}{144Cr^2\overline{L}_1}\right\} + 2 \exp\left\{-\frac{N^2c^2\delta^2}{144C\overline{L}_2}\right\}+2N\exp\left\{-\frac{c}{24C\max_{j}R_j^{2-\epsilon}}\right\}\\
\notag 	&+ 2\exp\left\{-c^2N^{2-2\alpha}/\left(128C^2\sum^n_{j=1}R_j^2\right)\right\} + 2\exp\left\{-\frac{Nc\epsilon^*}{6\sqrt{2C}\max_jR_j}\right\}.
	\end{align}
\end{theorem}

\begin{remark}[Interpretation of $\overline{L}_1$ and $\overline{L}_2$]
The number $\overline{L}_2/N$ is an approximate measure of the average length of durations across all assets within the time window. As an example when the diagonal entries of $AA^{\prime}$ are bounded so that $t_j-t_{j-1}=O(1/n)$, $\overline{L}_2$ is upper bounded by $\max_k\sum^N_{i=1}D_{ik}\log{(n)}/n\leq CN\log{(n)}/n$. In particular, if the vector $(D_{1k}, \cdots, D_{Nk})'$ is sparse when the asynchronicity is rare, the upper bound could be of smaller order than $O(N\log{(n)}/n)$. Then the reciprocal $N/\overline{L}_2$ is approximately equal to the average number of observed durations across all assets, which is larger than the smallest sample size of the most sparsely sampled asset. This implies that our approach makes use of all the data points through the large-scale linear system, in constrast with the tick subsampling method listed in previous sections which depends much on the sample size of most illiquid assets, thus loss of efficiency in estimation and subsequent applications. The number $\overline{L}_1$ is a cross-sectionally weighted version of $\overline{L}_2$ and hence has a similar interpretation as $\overline{L}_2$. 
\end{remark}

Before we prove a uniform result on $|\|\mathcal{A}(\Pi)\|_F^2-\|\Pi\|_F^2|/\|\Pi\|_F^2$ and $|\|\mathcal{A}(\Delta)\|_F^2-\|\Delta\|_F^2|/\|\Delta\|_F^2$ over the set of matrices $\Pi$ and $\Delta$, we introduce some facts of the Grassmannian manifold. The set of all $d$-dimensional subspaces of $R^D$ is known as the Grassmannian manifold which is denoted by $\mathcal{B}(D, d)$. First of all, let $U$ be an arbitrary subspace of $N\times n$ matrices with dimension $r\leq N\wedge n$, then there exists a finite set $\Omega$ of at most $(12/\delta)^r$ points such that for every $\Pi\in U$ with $\|\Pi\|_F^2/N^{\alpha}\leq 1$, there exists a $Q\in \Omega$ such that $\|\Pi-Q\|_F/N^{\alpha/2}\leq \delta/4$. Define the natural distance between two subspaces by $\rho(T_1, T_2):=\|P_{T_1}-P_{T_2}\|$, where $T_1$ and $T_2$ are subspaces and $P_{T_i}$ is the orthogonal projection associated with each subspace. This equals to the sine of the largest principal angle between $T_1$ and $T_2$. As demonstrated by the work of Szarek on $\epsilon$-nets of the Grassmannian, c.f., \cite{szarek1983finite}, the covering number of $\mathcal{B}(N, r)$ at resolution $\epsilon$ (i.e. the smallest number of subspaces $U_i$ such that for any subspace $U$, there is an $i$ with $\rho(U, U_i)\leq \epsilon$) is at most $(\frac{2C_0}{\epsilon})^{r(N-r)}$ where $C_0$ is a constant independent of $\epsilon$, $N$ and $r$.

By the union bound and Theorem \ref{th1}, we have the following theorem.
\begin{theorem}\label{th2}
	Suppose that Assumptions \ref{ass1}-\ref{ass2} hold. If
	$$
	r(N-r)\log{\left(\sqrt{\frac{Nn}{n^*}}+1\right)}=o\left\{\frac{N^{\alpha}}{\overline{L}_1}\right\} \ \mbox{and} \ N^{\alpha}/\overline{L}_1\rightarrow \infty,
	$$
	then for any $0<\delta<1$,
 \begin{align}\label{equation-th2}
 \begin{aligned}
     &P_{R, D}\left(|\frac{\|\mathcal{A}(\Pi)\|_F^2-\|\Pi\|_F^2}{\|\Pi\|_F^2}|>\delta, \ \mbox{for some} \ \sigma^0\in\mathcal{B}(N, r) \right)\\
     &\leq C\exp\{-c^2\delta^2N^{\alpha}/\overline{L}_1\}+2\exp\left\{-c^2r^2/\left(\sum^n_{j=1}R_j^2\right)^{1/2}\right\},
     \end{aligned}
 \end{align}
	for $c$ small enough and $C$ large enough which might depend on $\delta$.
	
	If
	$$
	r(N-r)\log{\left(\sqrt{\frac{Nn}{n^*}}+1\right)}=o\left\{\frac{N^{2-\alpha}}{\overline{L}_1}\right\} \ \mbox{and} \ N^{2-\alpha}/\overline{L}_1 \rightarrow \infty,
	$$
	then for any $0<\delta<1$,
	\begin{eqnarray}\label{equation-th2-1}
		&&P_{R, D}\left(|\frac{\|\mathcal{A}(\Delta)\|_F^2-\|\Delta\|_F^2}{\|\Delta\|_F^2}|>\delta, \ \mbox{for some} \ \sigma^0\in\mathcal{B}(N, r) \right)\nonumber\\
		&\leq & 2\exp\left\{-\frac{N^{2-\alpha}c^2\delta^2}{144Cr^2\overline{L}_1}\right\} + 2 \exp\left\{-\frac{N^2c^2\delta^2}{144C\overline{L}_2}\right\}+ 2\exp\left\{-\frac{Nc\epsilon^*}{6\sqrt{2C}\max_jR_j}\right\}\\
		&& + 2\exp\left\{-c^2N^{2-2\alpha}/\left(128C^2\sum^n_{j=1}R_j^2\right)\right\}+2N\exp\left\{-\frac{c}{24C\max_{j}R_j^{2-\epsilon}}\right\}\nonumber.
	\end{eqnarray}
	
	Let $\Pi_0$ be a matrix of rank less than or equal to $r$ and satisfy the constraints in (\ref{nuclear norm2}). The problem (\ref{nuclear norm2}) has a solution for $\Pi$ that equals to $\Pi_0$ with probability ($P_{R, D}$) at least
	$$
	1-C\exp\left\{-c^2\delta^2N^{\alpha}/\overline{L}_1\right\}-2\exp\left\{-c^2r^2/\left(\sum^n_{j=1}R_j^2\right)^{1/2}\right\}.
	$$
	
	Moreover, Let $\Delta_0$ be a matrix satisfying the constraints in (\ref{nuclear norm2}). The problem (\ref{nuclear norm2}) has a solution for $\Delta$ that equals to $\Delta_0$ with probability ($P_{R, D}$) at least
	\begin{eqnarray}
		&& 1-2\exp\left\{-\frac{N^{2-\alpha}c^2\delta^2}{144Cr^2\overline{L}_1}\right\}- 2 \exp\left\{-\frac{N^2c^2\delta^2}{144C\overline{L}_2}\right\}- 2\exp\left\{-\frac{Nc\epsilon^*}{6\sqrt{2C}\max_jR_j}\right\}\nonumber\\
		&& - 2\exp\left\{-c^2N^{2-2\alpha}/\left(128C^2\sum^n_{j=1}R_j^2\right)\right\}-2N\exp\left\{-\frac{c}{24C\max_{j}R_j^{2-\epsilon}}\right\}.
	\end{eqnarray}
\end{theorem}

\begin{remark}[Dependence on Data Frequency]
	Similar to Theorem \ref{th1}, an interesting finding is that the uniform rate depends also on the average length of the observed durations which is different from the previous-tick approach in large-panel high-frequency data analysis literature. For the previous-tick method, the statistically efficiency rests on the length of the time lags of the most illiquid asset where data comes with the lowest frequency.
\end{remark}

Theorem \ref{th2} implies that
\begin{equation}\label{closeness}
	\sup_{\sigma^0\in\mathcal{B}(N, r)}|\|\mathcal{A}(\Delta)\|_F^2-\|\Delta\|_F^2|=O_p\left(N\sqrt{\overline{L}_1/N^{2-\alpha}}\right)=:O_p(Na_{Nn}).
\end{equation}
Let
$$
G_{n}(\Pi, \Delta):=\frac{1}{2}(\|\mathcal{A}(\Delta)-b\|_F^2)+\frac{1}{2}\mu\|\Delta-\Pi\|_F^2+\lambda \|\Pi\|_*,
$$
and
$$
G_{n0}(\Pi, \Delta):=\frac{1}{2}\|\Delta-\Delta_0\|^2_F+\frac{1}{2}\mu\|\Delta-\Pi\|_F^2+\lambda \|\Pi\|_*.
$$
(\ref{closeness}) shows that $G_{n}(\Pi, \Delta)$ and $G_{n0}(\Pi, \Delta)$ are uniformly close enough. The minimizers of $G_{n}(\Pi, \Delta)$ and $G_{n0}(\Pi, \Delta)$ are denoted by $(\widehat{\Pi}, \widehat{\Delta})$ and $(\widehat{\Pi}_0, \widehat{\Delta}_0)$, respectively. The next theorem shows how close are the two minimizers.

\begin{theorem}\label{th3}
	Under Assumptions \ref{ass1}-\ref{ass2},
	$\|\widehat{\Pi}-\widehat{\Pi}_0\|_F/\sqrt{N}=O_p\left(\sqrt{\frac{a_{Nn}}{1+\lambda}}\right)$ and $\|\widehat{\Delta}-\widehat{\Delta}_0\|_F/\sqrt{N}=O_p\left(\sqrt{\frac{a_{Nn}}{1+\lambda}}\right)$.
\end{theorem}

The minimizer of $G_{n0}(\Pi,\Delta)$ satisfies the following first order condition.
\begin{equation}\label{first-order}
	(1+\mu)\widehat{\Delta}_0=\Delta_0+\mu\widehat{\Pi}_0, \ \widehat{\Pi}_0=U_0D_{\lambda/\mu+}V_0^{\prime},
\end{equation}
where $D_{v+}$ always denotes a thresholded singular value matrix whose $j$-th singular value is $\max\{\lambda_j-v, 0\}$ with $\lambda_j$ being the $j$th largest singular value of $\widehat{\Delta}_0$ (the $v$ in the above equation is $\lambda/\mu$), and $U_0$ and $V_0$ are the left and right singular vectors of $\widehat{\Delta}_0$, respectively. Let $D$ be the the matrix of singular values of $\widehat{\Delta}_0$. Combining the two conditions in (\ref{first-order}) leads to the following equation
$$
U_0\left(D-\frac{\mu}{1+\mu}D_{\lambda/\mu+}\right)V_0^{\prime}=\frac{1}{1+\mu}\Delta_0.
$$
A solution to this equation is $D-\frac{\mu}{1+\mu}D_{\lambda/\mu+}=\frac{1}{1+\mu}D_*$ where $D_*$ is the matrix of singular values of $\Delta_0$. That being said, 
$$
\widehat{\Delta}_0=U_{*}\{(D_{*})_{\lambda+}+D_{**}\}V_*^{\prime} \ \mbox{and} \ \widehat{\Pi}_0=U_{*}\{(D_{*})_{\lambda+}+D_{**}\}_{\lambda/\mu+}V_*^{\prime},
$$ 
where $U_*$ and $V_*$ are the left and right singular vector matrices of $\Delta_0$, respectively, and
$$
D_{**}=\frac{1}{1+\mu}\left\{D_*-\left[(D_*)_{\lambda+}+\lambda\left(\begin{array}{ll}
	I_k & 0\\
	0   & 0 \\
\end{array}\right)\right]\right\},
$$
with $k$ being the number of singular values of $D_*$ greater than or equal to $\lambda$. Let $J$ be the number of singular values of $D_{*}$ greater than or equal to $\lambda(1+1/\mu)$. $\widehat{\Delta}_0$ shrinks the first $k$ singular values by subtracting $\lambda$ from them and scaling down the remaining singular values with a factor of $1/(1+\mu)$. $\widehat{\Pi}_0$ shrinks the first $J$ singular values of $D_*$ by subtracting $\lambda(1+1/\mu)$ from them and truncate the remaining singular values to zero. The shrinkage of the largest singular values for $\widehat{\Pi}_0$ is more than that for $\widehat{\Delta}_0$ because the former is of low rank but the latter is not, and thus $J\geq k$. When $\mu=0$, $\widehat{\Pi}_0=U_{*}(D_{*})_{\lambda+}V_*^{\prime}$ and $\widehat{\Delta}_0=\Delta_0$, which boils down to the solution to (\ref{nuclear norm}). Notice here that $\Delta_0$ as a true potential increment matrix is unknown and has to be numerically solved and approximated by $\widehat{\Delta}$, but this is not achievable when setting $\mu=0$ in the optimization. Also notice the difference between $(\widehat{\Pi}_0, \widehat{\Delta}_0)$ and $(\widehat{\Pi}, \widehat{\Delta})$. The latter is the computationally feasible version while the former is a theoretical approximation that is easy to analyze mathematically. Theorem \ref{th3} demonstrates that they are close in terms of the averaged Frobenius norm.

Let $\|\cdot\|$ be the operator norm of a matrix. Next, we analyze the closeness of $(\widehat{\Pi}, \widehat{\Delta})$ to $(\Pi_0, \Delta_0)$. 
\begin{theorem}\label{th4}
	Assume that $\Pi_0=U_{*r}\mbox{diag}\{\lambda_1^*,...,\lambda_r^*\}V_{*r}^{\prime}$ where $U_{*r}$ and $V_{*r}$ are matrices consisting of $r$ left and right singular vectors of $\Delta_0$ corresponding to $\lambda_1^*\geq\lambda_2^*\geq\cdots\geq\lambda_r^*$, respectively.
	Under Assumptions \ref{ass1}-\ref{ass2},
	\begin{eqnarray*}
		\|\widehat{\Delta}-\Delta_0\|&\leq & \lambda\vee \frac{\mu}{1+\mu}\lambda^*_{k+1}+\sqrt{Na_{Nn}/(1+\lambda)}, \\ \|\widehat{\Delta}-\Delta_0\|_F&\leq & \lambda k\vee \frac{\mu}{1+\mu}\sum^{N\wedge n}_{j=k+1}\lambda^*_{j}+\sqrt{Na_{Nn}/(1+\lambda)},
	\end{eqnarray*}
	and
	\begin{eqnarray*}
		\|\widehat{\Pi}-\Pi_0\|&\leq & \lambda(1+\frac{1}{\mu})\vee \left[\left(\lambda_{r+1}^*-\lambda\left(1+\frac{1}{\mu}\right)\right)I(J>r)+\lambda_{J+1}^*I(J\leq r)\right]+\sqrt{N\frac{a_{Nn}}{1+\lambda}},\\
		\|\widehat{\Pi}-\Pi_0\|_F&\leq & \left[r\lambda\left(1+\frac{1}{\mu}\right)+(J-r)\left(\lambda^*_{r+l}-\lambda\left(1+\frac{1}{\mu}\right)\right)\right]I(J>r)\\
		&&+[J\lambda(1+1/\mu)+(r-J)\lambda^*_{J+1}]I(J\leq r)+\sqrt{N\frac{a_{Nn}}{1+\lambda}},
	\end{eqnarray*}
	with probability approaching one.
\end{theorem}

Theorem \ref{th4} gives the convergence rates of the estimators $\widehat{\Delta}$ and $\widehat{\Pi}$. In each upper bound, the first term is a bias due to the shrinkage brought by the low-rank penalty, while the last term comes from the data synchronization or approximation error of $\|\mathcal{A}(\Delta)\|$ by $\|\Delta\|_F$. In estimating $\Delta_0$, setting $\lambda=\mu=0$ is optimal to reduce the bias, but large value of $\lambda$ decreases the approximation error. In estimating $\Pi_0$, small value of $\mu$ and large value of $\lambda$ enlarge the bias term $\lambda(1+1/\mu)$ but decrease the bias term $\lambda^*_{r+1}-\lambda(1+1/\mu)$ when $J>r$. So there is a tradeoff between the bias terms, between the bias and approximation error, and between estimating $\Pi_0$ and $\Delta_0$. 

To relieve the complexity in choosing the tuning parameter, we introduce a bias-corrected version of the estimators. Given $\widehat{\Delta}$ at hand, we do the SVD, and correct the singular value matrix by adding $\lambda$ onto its first $k$ (assume a priori known first) largest singular value and multipling $1+\mu$ onto the remaining singular values. We denote the resulting estimator by $\widetilde{\Delta}$. Given the $\widehat{\Pi}$, we do the SVD also, and correct the singular value matrix by adding $\lambda(1+1/\mu)$ onto its first $J$ (again known a priori first) singular values. We denote the resulting estimator by $\widetilde{\Pi}$. Now, we have the following asymptotic results for the de-biased estimators.

\begin{theorem}\label{th5}
	Under Assumptions \ref{ass1}-\ref{ass2},
	\begin{align*}
		\|\widetilde{\Delta}-\Delta_0\|/\sqrt{N} = & O_p\left(\left(\lambda\vee \frac{\mu}{1+\mu}\lambda_{k+1}^*+1\right)\sqrt{a_{Nn}/(1+\lambda)}\right), \\ \|\widetilde{\Delta}-\Delta_0\|_F/\sqrt{N} = & O_p\left(\left(\lambda\vee \frac{\mu}{1+\mu}\lambda_{k+1}^*+1\right)\sqrt{a_{Nn}/(1+\lambda)}\right),
	\end{align*}
	and
\begin{align*}
	\frac{\|\widetilde{\Pi}-\Pi_0\|}{\sqrt{N}}=&O_p\left\{\left[\lambda\left(1+\frac{1}{\mu}\right)+1\right]\sqrt{\frac{a_{Nn}}{1+\lambda}}+\frac{\lambda^*_{J+1}I(J<r)+\lambda^*_{r+1}I(J>r)+0I(J=r)}{\sqrt{N}}\right\},\\
	\frac{\|\widetilde{\Pi}-\Pi_0\|_F}{\sqrt{N}}=&O_p\left\{\left[\lambda\left(1+\frac{1}{\mu}\right)+1\right]\sqrt{\frac{a_{Nn}}{1+\lambda}}+\frac{\sum\limits^r_{l=J+1}\lambda^*_lI(J<r)+\sum\limits^J_{l=r+1}\lambda^*_{r+1}I(J>r)+0I(J=r)}{\sqrt{N}}\right\}.
\end{align*} 
\end{theorem}

Theorem \ref{th5} shows that as in the typical case where $\sqrt{\frac{a_{Nn}}{1+\lambda}}=o(1)$, the de-biased estimator decreases the bias with a downward scaling of $\sqrt{\frac{a_{Nn}}{1+\lambda}}$ for $\Delta_0$. As for $\Pi$, a good choice is setting $J=r$ (or equivalently tuning $\lambda(1+1/\mu)$). In this case, the convergence rate for $\widetilde{\Pi}$ simplifies to $\lambda(1+1/\mu)\sqrt{\frac{a_{Nn}}{1+\lambda}}$ in both averaged operator norm or Frobenius norm, and the bias is also scaled down by a factor of $\sqrt{\frac{a_{Nn}}{1+\lambda}}$. To further understand the choice of $J=r$, we separate $\Delta$ into the sum of $\Delta_{1}+\Delta_{2}$ where $\Delta_{1}$ is the potential increment matrix attributed by low-rank diffusion $\sigma_tdW_t$ and $\Delta_{2}$ is that attributed by $\mu_tdt+\sigma_t^*dW^*_t$. In traditional high-dimensional factor analysis, $\|E(\Delta_2\Delta_2^{\prime})\|$ is bounded by $C_{\mbox{noise}}$ while $\|E(\Delta_1\Delta_1^{\prime})\|\geq crN^{\alpha}>C_{\mbox{noise}}$ for large $N$ and $n$ (recall the parameter space $\Theta$). Then choosing $\lambda(1+1/\mu)$ so that $J=r$ is selecting a threshold in between $C_{\mbox{noise}}$ and $crN^{\alpha}$ to differentiate the low-rank component and the idiosyncratic component. As an end of this section, it is worthy of notice that our results hold for weak factor cases where $\alpha<1$, that being said the strong factor condition is not necessary. 

\section{Monte Carlo Simulation}\label{sec:Monte Carlo Simulation}
In this section, we use Monte Carlo simulations to demonstrate the effectiveness of our methodology. 
\subsection{Simulation Settings}\label{subsec:Simulation Settings}
We adopt a stochastic volatility model without jumps, similar to the framework in \cite{kong2023discrepancy}. The latent log-price process, $X_t$, and the latent factor process, $V_t$, are specified as follows:
\begin{align*}
	dX_t =\beta_tdV_t+dZ_t \quad \mbox{and} \quad dV_t=\mu_v dt+\sigma_{v,t}\rho^{1/2}dW_{t}^v,
\end{align*}
where $Z_t$ is an idiosyncratic error process and $W_t^v$ is a multivariate ($r$-dimensional) Brownian motion. The matrix $\rho=(\rho_{ij})_{r\times r}=\{{\diag}(H)\}^{-1/2}HH'\{{\diag}(H)\}^{-1/2}$, where $H=(h_{ij})_{r\times r}$ is a lower triangular matrix with elements $h_{ij}=0.6^{|i-j|}$ for $i\geq j$. The diagonal volatility matrix,
$\sigma_{v,t}={\diag}(\sigma_{v,t1}, ..., \sigma_{v,tr})$, contains individual factor volatilities, each following a Heston-type square-root process:
$$
d\sigma_{v,tl}^2  =\kappa (\bar{\sigma}^2-\sigma_{v,tl}^2)dt+s\sqrt{\sigma_{v,tl}^2}d\bar{W}_{v,tl},
$$
where $\bar{W}_{v,t}=(\bar{W}_{v,t1}, ..., \bar{W}_{v,tr})'$ is a multivariate Brownian motion independent of $W_t^v$, with correlation matrix $I_r$. The initial value for each variance process, $\sigma_{v,0l}^2$, is drawn from a uniform distribution $U[0.8\bar{\sigma}^2, 1.2\bar{\sigma}^2]$. We set $r=3$, $(\kappa, \bar{\sigma}^2, s)=(3, 0.3^2, 0.3)$ and $\mu_v=(0.05, 0.03, 0.02)'$.

The process $Z_t=\sqrt{\mu}\sigma_t^*d{W}^*_t$, where $\sigma_{tj}^{*2}$ is generated by the same procedure as $\sigma_{tl}^2$ with the same parameters, and ${W}_t^*$ is a $N$-dimensional Brownian motion with the correlation matrix $\rho^*$ being a block diagonal matrix with each block being $\left(\tilde{\rho}^*_{ij}=0.6^{|i-j|}\right)_{10\times 10}$.  This setting is similar to that in \cite{ait2017using}.
We let $\mu=0.1$ such that the averaged standard errors of all elements of $Z_t$ is around $18\%$ of that of $Y_t$.

The factor loading matrix $\beta_t=\widetilde{\beta}^0M$ where $M=(\widetilde{\beta}^{0\prime}\widetilde{\beta}^0/p^{\alpha})^{-1/2}$
such that  $\beta'_t\beta_t/p^{\alpha}=I_p$,
and $\widetilde{\beta}^0=\left(\tilde{\beta}^0_{j}(l)\right)_{j=1, ..., N}^{l=1, ..., r}$ with $\tilde{\beta}^0_{j}(1)\sim U[0.25, 1.75]$ associated with the market factor, and $\tilde{\beta}^0_{j}(l)\sim N(0, 0.5^2)$ for $l=2, ..., r$. In the base case, the factor strength $\alpha$ is set to 1.

Our simulation is configured with $N=100$ assets over a time horizon of $T=5$ trading days. Each day is discretized into a grid of $n=390$ equally-spaced intervals, corresponding to a 1-minute frequency with a time step of $t_j-t_{j-1}=\delta_n=1/390$. Asynchronous trading times for each asset are then generated from a standard Poisson process with an arrival intensity of $\lambda_{\mathrm{asy}}=1$. For our proposed algorithm, the tuning parameters are initialized to $\lambda=0.001$, $\mu=0.1$, and $\eta=0.01$.

To assess the robustness of our method, we systematically vary key parameters of the simulation, keeping all other settings fixed according to the baseline configuration described above. We consider four scenarios:
\begin{enumerate}
	\item[(i)] \textbf{Number of Assets}: We set the number of assets $N$ to 20, 50, 120.
	\item[(ii)] \textbf{Observation Frequency}: We vary the length of the observation interval $\delta_n$, setting it to $1/780$ (30-second), $1/390$ (1-minute), and $1/78$ (5-minute).
	\item[(iii)] \textbf{Asynchronous Intensity}: We divide the $N=100$ assets into two equally-sized groups and assign different Poisson arrival intensities, ($\lambda_{\mathrm{asy},1}$, $\lambda_{\mathrm{asy},2}$), to each. We test three configurations: uniform low intensity (0.5, 0.5), mixed intensity (0.5, 3), and uniform high intensity (3, 3).
	\item[(iv)] \textbf{Factor Strength}: We examine the impact of different factor strengths by setting $\alpha$ to 0.8 (strong), 0.5 (moderate), and 0.2 (weak).
\end{enumerate}

We conduct 200 Monte Carlo replications for each Data Generating Process (DGP) to evaluate the finite-sample performance of our proposed Nuclear Norm (NN) method. We compare it against several widely-used methods for handling asynchronous data, which can be broadly classified into two categories: 
\begin{enumerate}
	\item [(i)]Imputation Methods: These methods aim to fill in missing observations. In addition to our NN method, we include:
	\begin{itemize}
		\item Previous-Tick interpolation (PI), as used in \cite{zhang2011estimating}.
		\item Linear Interpolation (LI).
	\end{itemize}
	\item [(ii)]Subsampling Methods: These methods create a synchronized dataset by selecting a sparse subset of the original data. We consider:
	\begin{itemize}
		\item Refresh Time (RT), proposed by \cite{barndorff2008designing}.
		\item Pre-Averaging (PA), as used in \cite{mykland2019algebra}.
	\end{itemize}
\end{enumerate}
For all methods, after obtaining a synchronized return matrix, we estimate the covariance matrix using the Principal Component Analysis (PCA)-based approach developed by \cite{ait2017using}.

We evaluate the estimation accuracy of increment matrix $\widehat{\Pi}$ by reporting scaled $L_1$ norm of relative error $(\widehat{\Pi}-\Pi)\oslash\Pi$, that is, $\frac{1}{nN}\|(\widehat{\Pi}-\Pi)\oslash\Pi\|_{L_1}$, where $(a_{ij})_{n\times m}\oslash(b_{ij})_{n\times m}=(a_{ij}/b_{ij})_{n\times m}$. We evaluate the estimation accuracy of covariance matrix $\widehat{\Sigma}$ by reporting various relative estimation error of norms, including the Frobenius $\|\widehat{\Sigma}-\Sigma\|_F/\|\Sigma\|_F$, the matrix $L_2$ $\|\widehat{\Sigma}-\Sigma\|/\|\Sigma\|$, and the maximum $\|\widehat{\Sigma}-\Sigma\|_{max}/\|\Sigma\|_{max}$ norms, where $\Sigma$ represents the true integrated covariance.

\subsection{Simulation Results}\label{subsec:Simulation Results}
\begin{figure}
	\FIGURE
	{\includegraphics[scale=0.5]{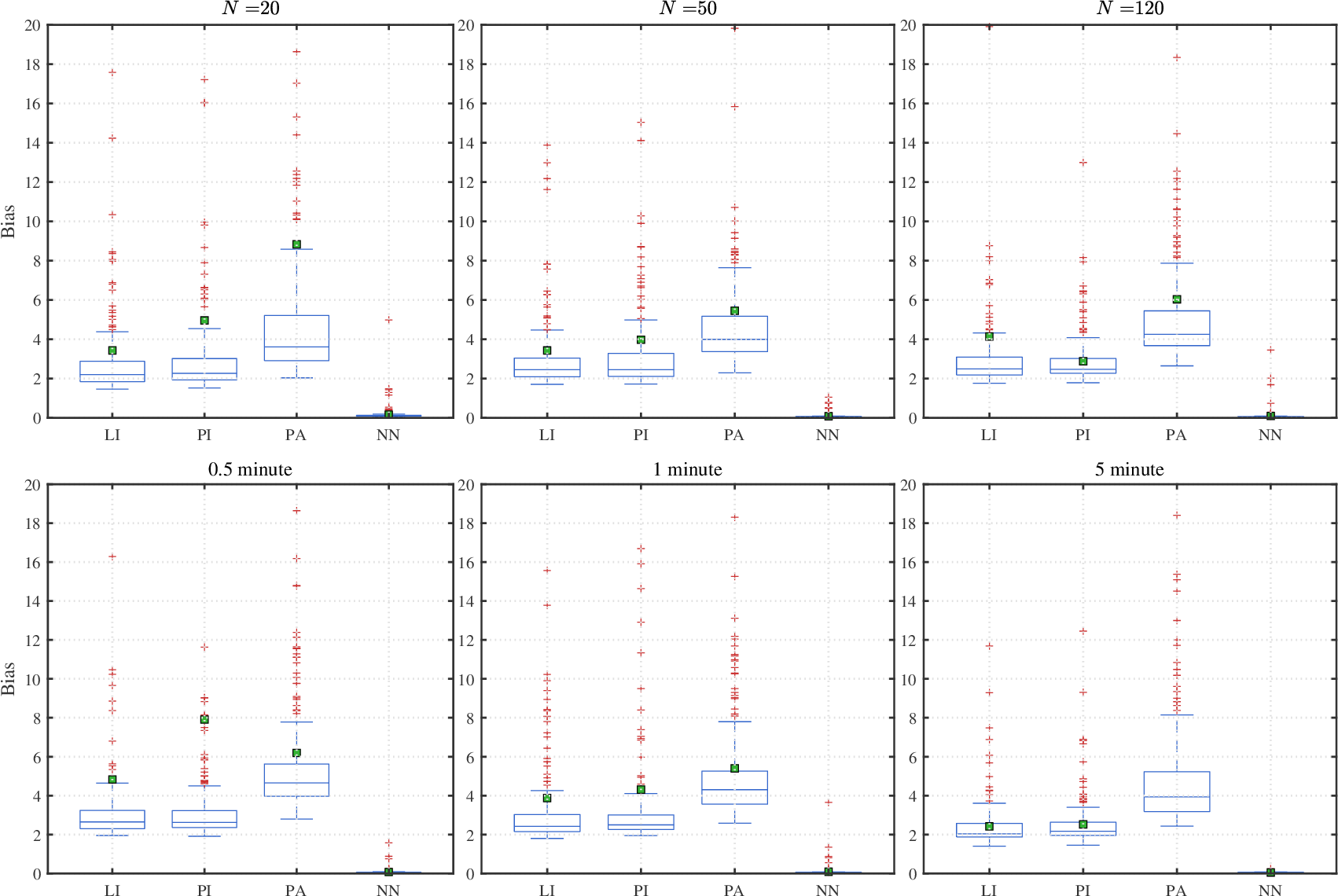}}
	{Box plots of the $L_1$ norm of the incremental relative error matrix. \label{fig:ErrorRet_SampleSize}}
	{Upper panel: varying number of assets; lower panel: different observation frequencies. Here, ``LI'' stands for linear interpolate; ``PI'' stands for previous-tick interpolate; ``PA'' for pre-averaging; ``NN'' stands for our nuclear norm. Small green squares denote the mean values.}
\end{figure}
Figure \ref{fig:ErrorRet_SampleSize} presents the recovery bias of the returns matrix across various sample sizes and number of assets. Notably, our nuclear norm approach consistently outperforms alternative methods,\footnote{The RT method is excluded from this comparison. This is because RT is a subsampling technique, not an imputation method; it only retains a sparse subset of true observed returns and does not generate any estimated values to be compared against the ground truth.} with low estimation errors even with few assets ($N=20$). Furthermore, our method demonstrates superior stability, as evidenced by its minimal performance variation, whereas other techniques exhibit substantially larger fluctuations.

\begin{figure}
	\FIGURE
	{\includegraphics[scale=0.5]{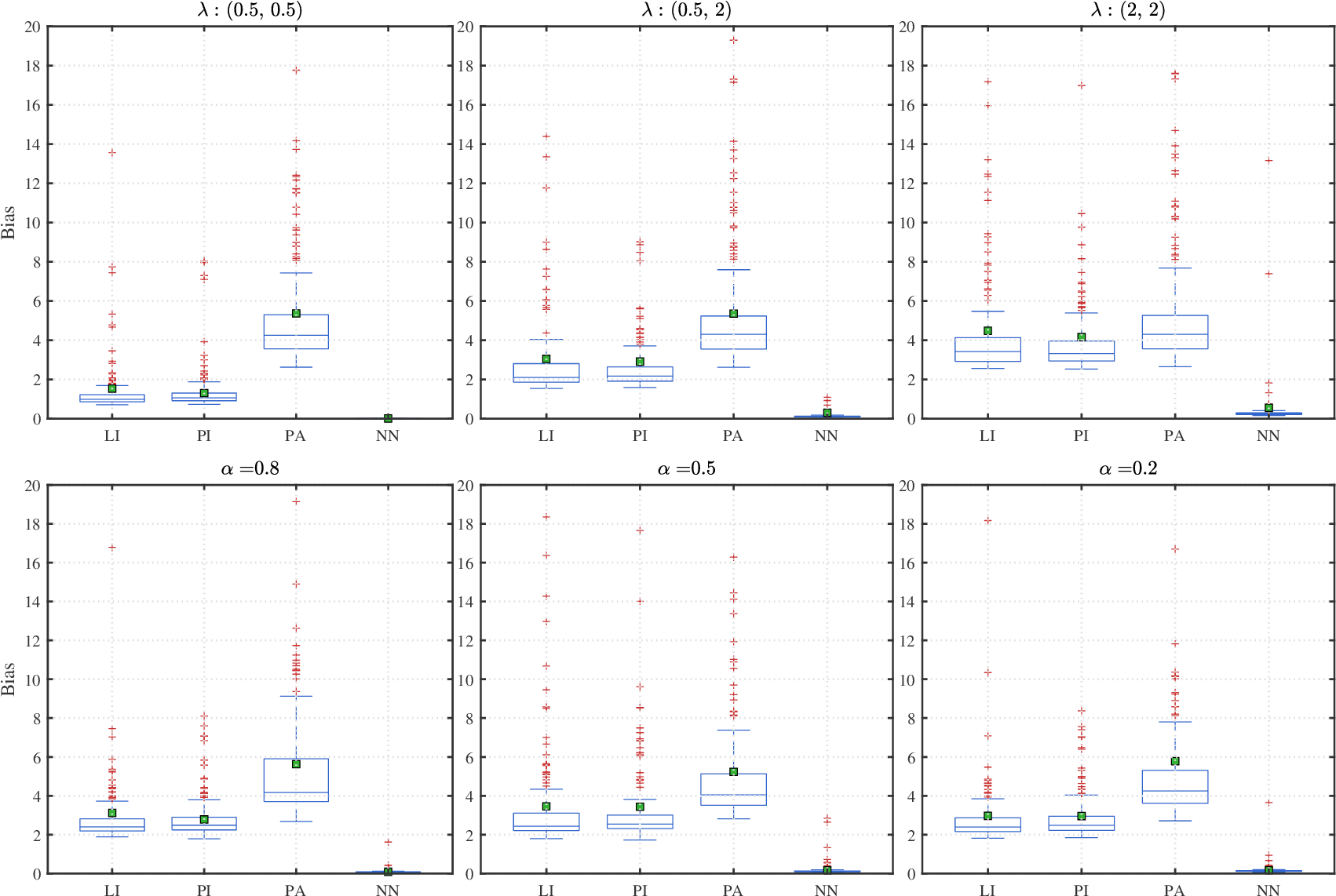}}
	{Box plots of the $L_1$ norm of the incremental relative error matrix. \label{fig:ErrorRet_Scenario}}
	{Upper panel: different asynchronous intensities; lower panel: different factor strengths. Here, ``LI'' stands for linear interpolate; ``PI'' stands for previous-tick interpolate; ``PA'' for pre-averaging; ``NN'' stands for our nuclear norm. Small green squares denote the mean values.}
\end{figure}
Figure \ref{fig:ErrorRet_Scenario} illustrates the recovery bias of the returns matrix under varying levels of asynchrony and different factor strengths. Across all scenarios, our nuclear norm method leads the field, maintaining the lowest estimation errors. As expected, every method's performance deteriorates as asynchrony intensifies---unobservable prices become excessive at extreme levels of asynchrony. Nevertheless, our approach consistently yields small, stable errors. Moreover, variations in factor strength have only a minor impact on its performance.

We further evaluate the accuracy of the covariance matrix estimation across different asynchronous recovery methods, examining how it is influenced by four key variables: the number of assets and the sampling frequency (Table {\ref{tab:ErrorCov_N_fre}}), the magnitude of asynchronous intensity and the strength of underlying factors (Table \ref{tab:ErrorCov_lambda_alpha}).

\begin{table}
	\TABLE
	{Covariance estimation accuracy: Varying assets and frequencies. \label{tab:ErrorCov_N_fre}}
	{
	\setlength\tabcolsep{2pt}\begin{tabular*}{\textwidth}{@{\extracolsep{\fill}}cccccccccccccccccc}
		\toprule
		\multicolumn{6}{c}{Dimension}                 &       & \multicolumn{5}{c}{Frequency} \\
		\cmidrule{2-6}\cmidrule{8-12}    $N$   & \multicolumn{1}{c}{RT} & \multicolumn{1}{c}{LI} & \multicolumn{1}{c}{PI} & \multicolumn{1}{c}{PA} & \multicolumn{1}{c}{NN} & Minute & \multicolumn{1}{c}{RT} & \multicolumn{1}{c}{LI} & \multicolumn{1}{c}{PI} & \multicolumn{1}{c}{PA} & \multicolumn{1}{c}{NN} \\
		\midrule
		\multicolumn{12}{c}{$\|\hat{\Sigma}-\Sigma\|_F/\|\Sigma\|_F$} \\
		\midrule
		\multirow{2}[1]{*}{$20$} & \multicolumn{1}{c}{0.131} & \multicolumn{1}{c}{0.134} & \multicolumn{1}{c}{0.132} & \multicolumn{1}{c}{0.296} & \multicolumn{1}{c}{\textbf{0.073}} & \multirow{2}[1]{*}{0.5 min} & \multicolumn{1}{c}{0.121} & 0.080 & 0.080 & \multicolumn{1}{c}{0.252} & \textbf{0.060} \\
		& (0.055) & (0.035) & (0.029) & (0.159) & (0.032) &       & (0.049) & (0.029) & (0.028) & (0.123) & (0.025) \\
		\multirow{2}[0]{*}{$50$} & \multicolumn{1}{c}{0.143} & \multicolumn{1}{c}{0.137} & \multicolumn{1}{c}{0.133} & \multicolumn{1}{c}{0.301} & \multicolumn{1}{c}{\textbf{0.077}} & \multirow{2}[0]{*}{1 min} & \multicolumn{1}{c}{0.151} & \multicolumn{1}{c}{0.139} & \multicolumn{1}{c}{0.135} & \multicolumn{1}{c}{0.301} & \multicolumn{1}{c}{\textbf{0.079}} \\
		& (0.060) & (0.037) & (0.033) & (0.156) & (0.031) &       & (0.067) & (0.037) & (0.035) & (0.152) & (0.032) \\
		\multirow{2}[1]{*}{$120$} & \multicolumn{1}{c}{0.149} & \multicolumn{1}{c}{0.141} & \multicolumn{1}{c}{0.137} & \multicolumn{1}{c}{0.301} & \multicolumn{1}{c}{\textbf{0.076}} & \multirow{2}[1]{*}{5 min} & \multicolumn{1}{c}{0.307} & \multicolumn{1}{c}{0.233} & \multicolumn{1}{c}{0.227} & \multicolumn{1}{c}{0.431} & \multicolumn{1}{c}{\textbf{0.145}} \\
		& (0.062) & (0.036) & (0.034) & (0.153) & (0.032) &       & (0.127) & (0.058) & (0.053) & (0.197) & (0.058) \\
		\midrule
		\multicolumn{12}{c}{$\|\hat{\Sigma}-\Sigma\|/\|\Sigma\|$} \\
		\midrule
		\multirow{2}[1]{*}{$20$} & \multicolumn{1}{c}{0.126} & \multicolumn{1}{c}{0.133} & \multicolumn{1}{c}{0.115} & \multicolumn{1}{c}{0.289} & \multicolumn{1}{c}{\textbf{0.074}} & \multirow{2}[1]{*}{0.5 min} & \multicolumn{1}{c}{0.119} & 0.080 & 0.078 & \multicolumn{1}{c}{0.246} & \multicolumn{1}{c}{\textbf{0.058}} \\
		& (0.063) & (0.041) & (0.039) & (0.178) & (0.036) &       & (0.056) & (0.033) & (0.032) & (0.137) & (0.028) \\
		\multirow{2}[0]{*}{$50$} & \multicolumn{1}{c}{0.138} & \multicolumn{1}{c}{0.137} & \multicolumn{1}{c}{0.125} & \multicolumn{1}{c}{0.296} & \multicolumn{1}{c}{\textbf{0.077}} & \multirow{2}[0]{*}{1 min} & \multicolumn{1}{c}{0.147} & \multicolumn{1}{c}{0.139} & \multicolumn{1}{c}{0.131} & \multicolumn{1}{c}{0.292} & \multicolumn{1}{c}{\textbf{0.079}} \\
		& (0.069) & (0.042) & (0.041) & (0.175) & (0.035) &       & (0.076) & (0.043) & (0.042) & (0.171) & (0.037) \\
		\multirow{2}[1]{*}{$120$} & \multicolumn{1}{c}{0.144} & \multicolumn{1}{c}{0.141} & \multicolumn{1}{c}{0.133} & \multicolumn{1}{c}{0.293} & \multicolumn{1}{c}{\textbf{0.076}} & \multirow{2}[1]{*}{5 min} & \multicolumn{1}{c}{0.294} & \multicolumn{1}{c}{0.233} & \multicolumn{1}{c}{0.216} & \multicolumn{1}{c}{0.419} & \multicolumn{1}{c}{\textbf{0.145}} \\
		& (0.072) & (0.041) & (0.041) & (0.172) & (0.036) &       & (0.145) & (0.066) & (0.065) & (0.220) & (0.064) \\
		\midrule
		\multicolumn{12}{c}{$\|\hat{\Sigma}-\Sigma\|_{max}/\|\Sigma\|_{max}$} \\
		\midrule
		\multirow{2}[1]{*}{$20$} & \multicolumn{1}{c}{0.155} & \multicolumn{1}{c}{0.119} & \multicolumn{1}{c}{0.135} & \multicolumn{1}{c}{0.277} & \textbf{0.070} & \multirow{2}[1]{*}{0.5 min} & \multicolumn{1}{c}{0.149} & \multicolumn{1}{c}{0.071} & \multicolumn{1}{c}{0.084} & 0.230 & \multicolumn{1}{c}{\textbf{0.053}} \\
		& (0.078) & (0.038) & (0.035) & (0.194) & (0.036) &       & (0.064) & (0.030) & (0.027) & (0.132) & (0.026) \\
		\multirow{2}[0]{*}{$50$} & \multicolumn{1}{c}{0.162} & \multicolumn{1}{c}{0.121} & \multicolumn{1}{c}{0.139} & 0.270 & \multicolumn{1}{c}{\textbf{0.071}} & \multirow{2}[0]{*}{1 min} & \multicolumn{1}{c}{0.169} & \multicolumn{1}{c}{0.126} & \multicolumn{1}{c}{0.147} & \multicolumn{1}{c}{0.274} & \multicolumn{1}{c}{\textbf{0.071}} \\
		& (0.080) & (0.038) & (0.035) & (0.179) & (0.034) &       & (0.083) & (0.040) & (0.037) & (0.164) & (0.036) \\
		\multirow{2}[1]{*}{$120$} & \multicolumn{1}{c}{0.164} & 0.130 & \multicolumn{1}{c}{0.152} & 0.280 & \multicolumn{1}{c}{\textbf{0.071}} & \multirow{2}[1]{*}{5 min} & \multicolumn{1}{c}{0.304} & \multicolumn{1}{c}{0.224} & \multicolumn{1}{c}{0.274} & \multicolumn{1}{c}{0.388} & \multicolumn{1}{c}{\textbf{0.138}} \\
		& (0.079) & (0.040) & (0.039) & (0.170) & (0.036) &       & (0.159) & (0.069) & (0.068) & (0.208) & (0.065) \\
		\bottomrule
	\end{tabular*}%
}
	{This table summarizes results from 200 Monte Carlo simulations. The reported values are averages, with standard deviations in parentheses. The left and right panels show results under varying numbers of assets ($N$) and observation frequencies, respectively. Covariance matrices are estimated using a PCA-based approach \citep{ait2017using}. The methods compared are: Refresh Time (RT), Linear Interpolation (LI), Previous-Tick Interpolation (PI), Pre-Averaging (PA), and our proposed Nuclear Norm (NN) method. Bold entries highlight the best-performing method in each scenario.}
\end{table}%

The left panel of Table \ref{tab:ErrorCov_N_fre} illustrates the impact of number of assets ($N$) on covariance estimation accuracy. Two key findings emerge. First, our NN method consistently outperforms all competing methods across both error norms and all sizes ($N=20$, $50$, $120$). Second, the NN method also exhibits the lowest standard deviation, highlighting its superior stability and reliability.

The right panel examines the effect of observation frequency. As expected, the estimation accuracy of all methods degrades as the frequency decreases. Despite this general trend, the NN method maintains its top-ranking performance, demonstrating its robustness even with less frequent data. In contrast, while the interpolation methods (LI and PI) perform reasonably well, particularly at higher frequencies, the PA method consistently yields the poorest results.

\begin{table}
	\TABLE
{Covariance estimation accuracy: Varying asynchrony and factor strength.\label{tab:ErrorCov_lambda_alpha}}
	{
	\setlength\tabcolsep{2pt}{\begin{tabular*}{\textwidth}{@{\extracolsep{\fill}}ccccccccccccccccc}
		\toprule
		\multicolumn{6}{c}{ Asynchronous intensity}   &       & \multicolumn{5}{c}{Factor strength} \\
		\cmidrule{2-6}\cmidrule{8-12}    $\lambda_{\mathrm{asy}}$ & \multicolumn{1}{c}{RT} & \multicolumn{1}{c}{LI} & \multicolumn{1}{c}{PI} & \multicolumn{1}{c}{PA} & \multicolumn{1}{c}{NN} & $\alpha$ & \multicolumn{1}{c}{RT} & \multicolumn{1}{c}{LI} & \multicolumn{1}{c}{PI} & \multicolumn{1}{c}{PA} & \multicolumn{1}{c}{NN} \\
		\midrule
		\multicolumn{12}{c}{$\|\hat{\Sigma}-\Sigma\|_F/\|\Sigma\|_F$} \\
		\midrule
		\multirow{2}[1]{*}{$(0.5, 0.5)$} & \multicolumn{1}{c}{0.134} & \textbf{0.060} & \multicolumn{1}{c}{0.062} & \multicolumn{1}{c}{0.301} & \multicolumn{1}{c}{0.079} & \multirow{2}[1]{*}{$0.8$} & \multicolumn{1}{c}{0.158} & \multicolumn{1}{c}{0.139} & \multicolumn{1}{c}{0.135} & \multicolumn{1}{c}{0.314} & \multicolumn{1}{c}{\textbf{0.084}} \\
		& (0.054) & (0.023) & (0.023) & (0.152) & (0.033) &       & (0.066) & (0.036) & (0.034) & (0.150) & (0.031) \\
		\multirow{2}[0]{*}{$(0.5, 2.0)$} & \multicolumn{1}{c}{0.192} & \multicolumn{1}{c}{0.243} & \multicolumn{1}{c}{0.252} & \multicolumn{1}{c}{0.301} & \multicolumn{1}{c}{\textbf{0.078}} & \multirow{2}[0]{*}{$0.5$} & \multicolumn{1}{c}{0.201} & \multicolumn{1}{c}{0.148} & \multicolumn{1}{c}{0.146} & \multicolumn{1}{c}{0.382} & \textbf{0.120} \\
		& (0.091) & (0.034) & (0.033) & (0.152) & (0.032) &       & (0.061) & (0.030) & (0.028) & (0.140) & (0.025) \\
		\multirow{2}[1]{*}{$(2.0, 2.0)$} & \multicolumn{1}{c}{0.201} & \multicolumn{1}{c}{0.361} & 0.340 & \multicolumn{1}{c}{0.301} & \multicolumn{1}{c}{\textbf{0.077}} & \multirow{2}[1]{*}{$0.2$} & \multicolumn{1}{c}{0.383} & \multicolumn{1}{c}{\textbf{0.258}} & \multicolumn{1}{c}{0.261} & \multicolumn{1}{c}{0.615} & \multicolumn{1}{c}{0.285} \\
		& (0.093) & (0.036) & (0.035) & (0.152) & (0.032) &       & (0.045) & (0.015) & (0.014) & (0.112) & (0.018) \\
		\midrule
		\multicolumn{12}{c}{$\|\hat{\Sigma}-\Sigma\|/\|\Sigma\|$} \\
		\midrule
		\multirow{2}[1]{*}{$(0.5, 0.5)$} & \multicolumn{1}{c}{0.134} & \multicolumn{1}{c}{\textbf{0.057}} & \multicolumn{1}{c}{0.058} & \multicolumn{1}{c}{0.292} & \multicolumn{1}{c}{0.079} & \multirow{2}[1]{*}{$0.8$} & \multicolumn{1}{c}{0.151} & \multicolumn{1}{c}{0.138} & 0.130 & 0.300 & \multicolumn{1}{c}{\textbf{0.081}} \\
		& (0.061) & (0.027) & (0.027) & (0.171) & (0.037) &       & (0.077) & (0.042) & (0.042) & (0.171) & (0.037) \\
		\multirow{2}[0]{*}{$(0.5, 2.0)$} & \multicolumn{1}{c}{0.185} & \multicolumn{1}{c}{0.234} & \multicolumn{1}{c}{0.234} & \multicolumn{1}{c}{0.292} & \multicolumn{1}{c}{\textbf{0.078}} & \multirow{2}[0]{*}{$0.5$} & \multicolumn{1}{c}{0.172} & \multicolumn{1}{c}{0.136} & \multicolumn{1}{c}{0.127} & \multicolumn{1}{c}{0.338} & \multicolumn{1}{c}{\textbf{0.097}} \\
		& (0.103) & (0.043) & (0.044) & (0.171) & (0.037) &       & (0.078) & (0.040) & (0.039) & (0.172) & (0.035) \\
		\multirow{2}[1]{*}{$(2.0, 2.0)$} & \multicolumn{1}{c}{0.193} & \multicolumn{1}{c}{0.361} & \multicolumn{1}{c}{0.333} & \multicolumn{1}{c}{0.292} & \multicolumn{1}{c}{\textbf{0.077}} & \multirow{2}[1]{*}{$0.2$} & \multicolumn{1}{c}{0.274} & \textbf{0.160} & \multicolumn{1}{c}{0.153} & \multicolumn{1}{c}{0.503} & \multicolumn{1}{c}{0.181} \\
		& (0.105) & (0.042) & (0.043) & (0.171) & (0.036) &       & (0.072) & (0.024) & (0.022) & (0.169) & (0.027) \\
		\midrule
		\multicolumn{12}{c}{$\|\hat{\Sigma}-\Sigma\|_{max}/\|\Sigma\|_{max}$} \\
		\midrule
		\multirow{2}[1]{*}{$(0.5, 0.5)$} & 0.150 & \multicolumn{1}{c}{\textbf{0.061}} & 0.080 & \multicolumn{1}{c}{0.274} & \multicolumn{1}{c}{0.072} & \multirow{2}[1]{*}{$0.8$} & 0.170 & 0.120 & \multicolumn{1}{c}{0.139} & \multicolumn{1}{c}{0.279} & \multicolumn{1}{c}{\textbf{0.072}} \\
		& (0.067) & (0.029) & (0.035) & (0.164) & (0.036) &       & (0.082) & (0.040) & (0.032) & (0.161) & (0.036) \\
		\multirow{2}[0]{*}{$(0.5, 2.0)$} & \multicolumn{1}{c}{0.198} & \multicolumn{1}{c}{0.259} & \multicolumn{1}{c}{0.273} & \multicolumn{1}{c}{0.274} & \multicolumn{1}{c}{\textbf{0.071}} & \multirow{2}[0]{*}{$0.5$} & \multicolumn{1}{c}{0.178} & \multicolumn{1}{c}{0.119} & \multicolumn{1}{c}{0.132} & \multicolumn{1}{c}{0.306} & \multicolumn{1}{c}{\textbf{0.081}} \\
		& (0.109) & (0.072) & (0.076) & (0.164) & (0.036) &       & (0.076) & (0.037) & (0.030) & (0.155) & (0.032) \\
		\multirow{2}[1]{*}{$(2.0, 2.0)$} & \multicolumn{1}{c}{0.212} & \multicolumn{1}{c}{0.319} & \multicolumn{1}{c}{0.342} & \multicolumn{1}{c}{0.274} & \multicolumn{1}{c}{\textbf{0.069}} & \multirow{2}[1]{*}{$0.2$} & \multicolumn{1}{c}{0.228} & \multicolumn{1}{c}{0.156} & \multicolumn{1}{c}{0.161} & 0.390 & \multicolumn{1}{c}{\textbf{0.154}} \\
		& (0.114) & (0.055) & (0.074) & (0.164) & (0.036) &       & (0.063) & (0.037) & (0.035) & (0.141) & (0.034) \\
		\bottomrule 
	\end{tabular*}}
}
	{This table summarizes results from 200 Monte Carlo simulations. The reported values are averages, with standard deviations in parentheses. The left and right panels show results under varying asynchronous intensities ($\lambda_{\mathrm{asy}}$) and factor strengths, respectively. Covariance matrices are estimated using a PCA-based approach \citep{ait2017using}. The methods compared are: Refresh Time (RT), Linear Interpolation (LI), Previous-Tick Interpolation (PI), Pre-Averaging (PA), and our proposed Nuclear Norm (NN) method. Bold entries highlight the best-performing method in each scenario.}
\end{table}%
Next, we verify the effect of asynchronous intensity and factor intensity on covariance estimation. First, we examine the impact of asynchronous intensity. As expected, the performance of most methods deteriorates as the intensity of asynchrony increases (i.e., as more prices become unobservable). However, our proposed NN method demonstrates remarkable robustness, showing only a slight degradation in performance. In contrast, the interpolation-based methods (LI and PI) are highly sensitive to this parameter. While they perform reasonably well at a low asynchronous intensity (e.g., 0.5), their accuracy collapses when the intensity is high (e.g., 2). This is because these methods rely on nearby observed prices for imputation. As asynchrony intensifies, the average time between valid price observations widens, forcing interpolation over longer gaps and inevitably introducing significant bias.

Second, we analyze the effect of factor strength ($\alpha$). The PA method performs very poorly, especially when the underlying factors are weak. The NN method achieves the best performance across nearly all scenarios. The only exception occurs at $\alpha=0.2$, where the PI and LI methods yield comparable results. This is likely because when the common factor is extremely weak, the cross-sectional dependence is minimal, which reduces the relative advantage of our global, factor-based approach. Nevertheless, even in this edge case, the NN method still provides the most accurate recovery of the underlying return matrix, as shown in Figure \ref{fig:ErrorRet_Scenario}.

\section{Empirical Analysis}\label{sec:Empirical Analysis}
To demonstrate the practical applicability and utility of our proposed methodology, we conduct several empirical applications. The remainder of this section is organized as follows. Section \ref{subsec:Data Description} describes the asynchronous characteristics of high-frequency data. Section \ref{subsec:Eigenvalues} highlights the difference between the eigenvalues derived from traditional stale prices (derived from the PI method) and those recovered by our methodology. Section \ref{subsec:Imputation Error} compares the imputation errors across different approaches. Section \ref{subsec:Applications to Portfolio Selection} evaluates the impact of these methods on portfolio selection. Finally, Section \ref{subsec:Spot Beta} analyzes the beta discrepancies during periods of market turbulence.

\subsection{Data Description}\label{subsec:Data Description}
We combine data for S\&P 500 constituents from the TAQ database with data for the SPDR S\&P 500 ETF Trust (SPY) from Pi Trading, covering the period from January 2008 to December 2022. Following \cite{bollerslev2024optimal}, we exclude holidays and days with shortened trading hours. Our analysis is based on high-frequency data at 1-minute, 5-minute, and 10-minute frequencies. For each frequency, an observation is deemed ``missing'' if no trade for the respective asset occurs within the corresponding time interval (e.g., within a given one-minute window for the 1-minute series). We construct a balanced panel, which ultimately comprises 360 stocks.  

Asynchrony is a pervasive feature of high-frequency data. In our analysis, we define the 1-minute interval as the base observational unit, corresponding to a discrete time grid $\mathcal{T}=\{t_1, ... ,t_n\}$.\footnote{We select the 1-minute frequency, rather than a higher frequency (e.g., 1-second), to mitigate the effects of microstructure noise. Nevertheless, our methodology is also applicable to such ultra-high-frequency data, a direction we reserve for future research.} The challenge of asynchronous trading can thus be framed as an imputation or data completion problem. Our primary goal is to recover complete time series of prices at 1-minute, 5-minute, and 10-minute frequencies, respectively, for every stock in the panel.

\begin{figure}
	\FIGURE
	{\includegraphics[scale=0.55]{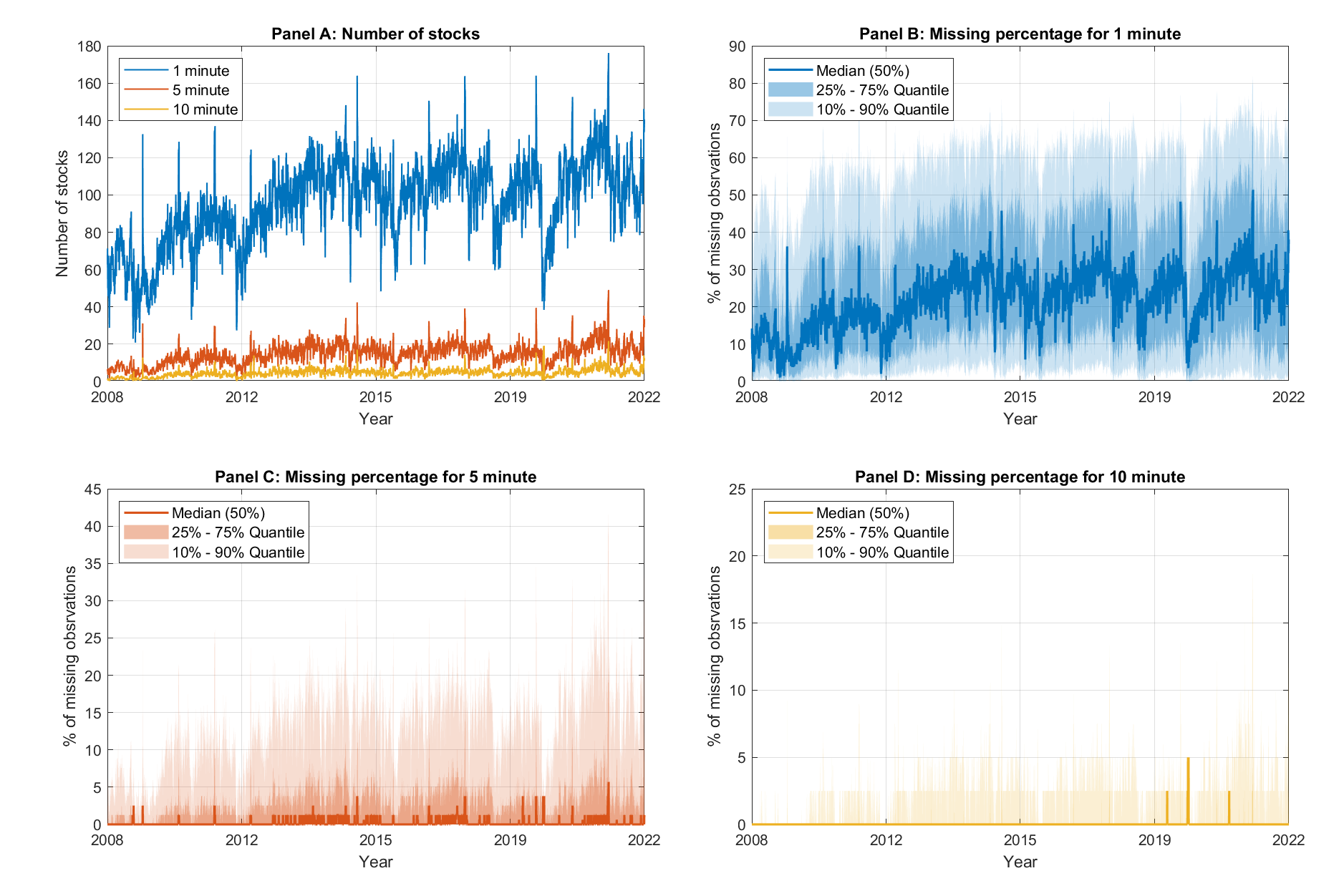}}	
	{Missing values over time. \label{fig:asynchronous price description}} 
	{This figure illustrates patterns of missing data in our sample. Panel A displays the daily average number of stocks with missing observations at the 1-minute frequency. Panels B, C, and D depict the daily evolution of the cross-sectional quantiles of missing observations for data sampled at 1-minute, 5-minute, and 10-minute frequencies, respectively.}
\end{figure}

Let $\mathcal{T}=\{t_1,...,t_n\}$ represent the complete grid of discrete observation times within a given period (e.g., a trading day with $n=390$ one-minute intervals starting from an initial time $t_0$). For any stock $i$, let $\{\tau_{i1},...,\tau_{ii_n}\}$ denote the subset of times at which its price is actually observed. Asynchronous trading occurs when a stock's observation set does not cover the entire grid, i.e., $\mathcal{T}\setminus \{\tau_{i1},...,\tau_{ii_n}\}\neq\{\emptyset\}$. Consequently, the set of time points where the price of stock $i$ is considered missing is defined by the set difference $\mathcal{T}\setminus \{\tau_{i1},...,\tau_{ii_n}\}$. Figure \ref{fig:asynchronous price description} provides a summary of such missingness across our high-frequency panel data.

Panel A of Figure \ref{fig:asynchronous price description} plots the daily average number of missing stocks at the 1-minute frequency. Despite a well-documented increase in overall trading activity over the past 15 years, the number of missing stocks at this high frequency has remained persistently high, showing no significant downward trend. While this number temporarily declined during the COVID-19 pandemic, the trend was short-lived, reverting to its previous level of approximately 100 within a year. In contrast, at lower frequencies (5- and 10-minute intervals), the issue of missing data is substantially less severe, with the count of missing stocks typically remaining below 40.

Panel B of Figure \ref{fig:asynchronous price description} illustrates the intraday evolution of the cross-sectional quantiles of the missing probability---also known as staleness probability\footnote{Staleness probability refers to the likelihood that a price update does not occur within a given interval. See, e.g., \cite{bandi2017excess}, \cite{bandi2020zeros}, and \cite{kong2024staleness}.}---at the 1-minute frequency. We observe a median missing rate of 20–30\%, a level consistent with the findings of \cite{bandi2020zeros} for NYSE-listed stocks. The prevalence of such missing data, or price stagnation (i.e., price staleness), is a critical issue in high-frequency analysis. The problem becomes even more acute at finer time scales; for instance, \cite{bandi2024systematic} reports a missing rate exceeding 50\% for highly liquid stocks at the 10-second frequency. Using these stale prices as if they were true observations of the underlying price process can introduce significant biases into key financial analyses, such as volatility and correlation estimation. Returning to our specific findings, Panel B reveals that the extent of missingness can be extreme for some assets, with the 90th percentile reaching 60\% even at the 1-minute frequency. As we decrease the sampling frequency, this issue is substantially mitigated. Panel C shows that the 90th percentile of the missing rate drops to approximately 15\% for 5-minute data. For 10-minute data (Panel D), the missing rate becomes negligible, with the 90th percentile falling below 5\%. This trend is intuitive, as non-trading spells of 10 minutes are relatively uncommon for the stocks in our sample.

\begin{figure}
     \FIGURE
	{\includegraphics[scale=0.55]{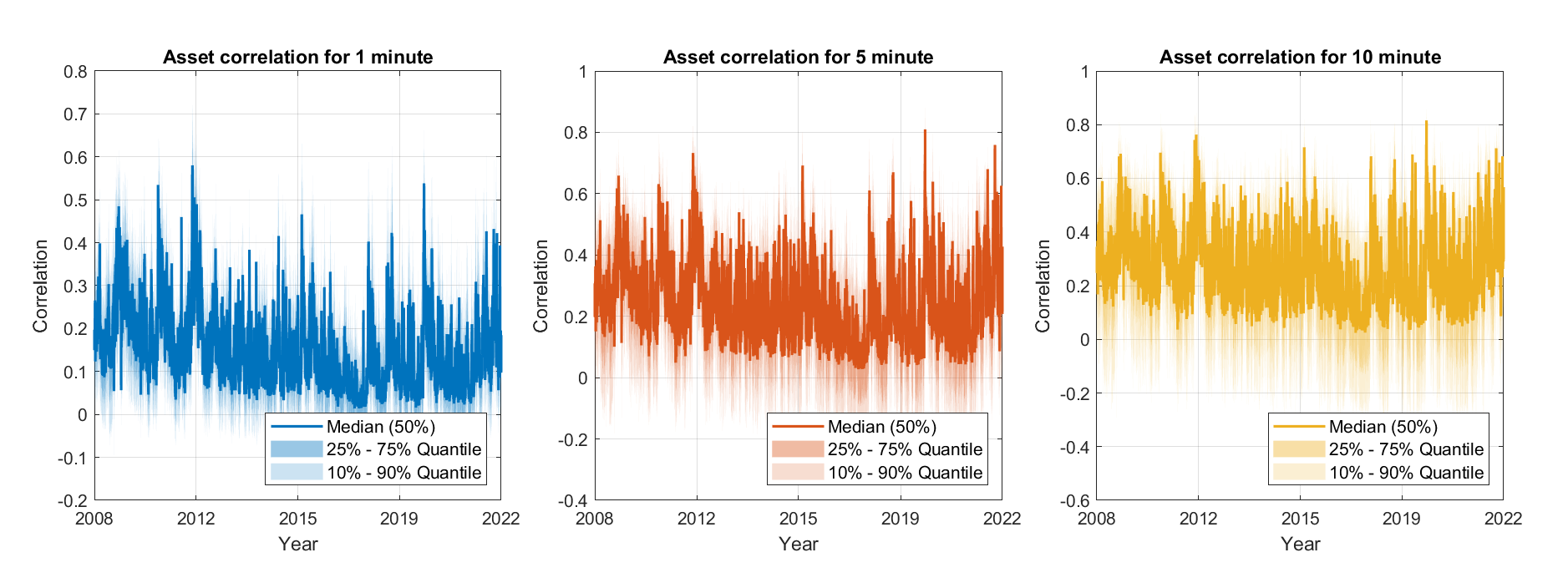}}
	{Asset realized correlation over time. \label{fig:asset correlation}}
	{This figure reports the results of the cross-sectional quantiles for realized correlation (daily).}
\end{figure}
Figure \ref{fig:asset correlation} confirms the presence of significant cross-sectional correlations among stocks, although the magnitude of this correlation tends to decrease as the sampling frequency increases. This strong inter-dependence, often driven by common factors like industry effects, represents a valuable source of information for handling asynchrony. Ignoring it when imputing prices risks discarding crucial data.

However, conventional imputation methods that rely solely on an asset's own time series fail to exploit this cross-sectional information. For instance: The PI method, which carries forward the last observed price, utilizes only past univariate information and ignores contemporaneous trades in related stocks. The LI method artificially smooths prices between an asset's trades. Consequently, these univariate approaches cannot propagate the impact of market-wide information—reflected in the trading of correlated assets—to the price of an asynchronously traded stock. Therefore, any imputation scheme based purely on an individual time series is inherently limited, as it neglects vital information from the cross-section, such as latent factors. Capturing this complex dependence structure effectively requires sophisticated modeling tools \citep[see, e.g., the discussion in][]{pelger2020understanding}.

An alternative approach, distinct from imputation, is the refresh time scheme, which synchronizes data by subsampling, retaining only the time points where a specified set of assets has traded \citep{ait2010high}. This method, however, faces a significant trade-off in high-dimensional settings. A full-cross-section refresh time discards a vast amount of temporal data, whereas a pairwise refresh time is restricted to two assets and fails to scale. Ultimately, both univariate imputation and subsampling schemes struggle to effectively balance the preservation of temporal data with the integration of high-dimensional cross-sectional information.

\subsection{Eigenvalues}\label{subsec:Eigenvalues}
\begin{figure}
\FIGURE
	{\includegraphics[scale=0.55]{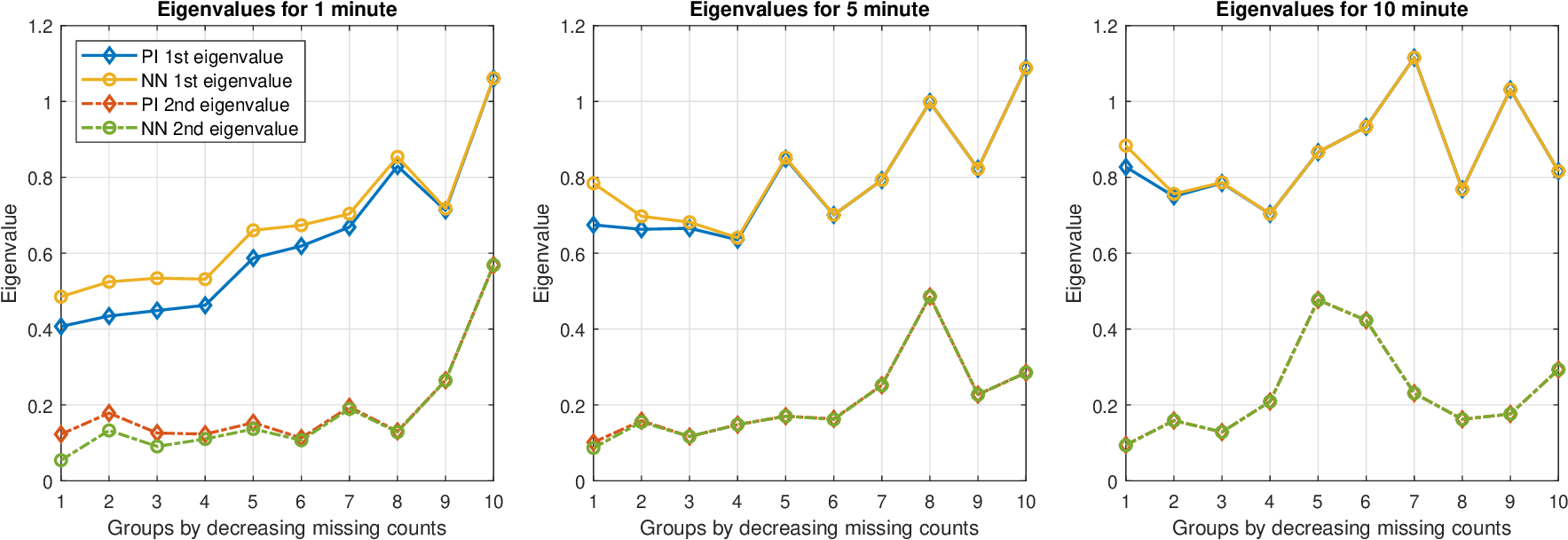}}
	{Eigenvalues for covariance. \label{fig:Eigenvalues}}
	{This figure shows the first two eigenvalues of the panel covariance matrix for different groups, which are divided sequentially according to the missingness of the original data, with group 1 indicating the highest number of missing values and group 10 indicating the lowest number of missing values.}
\end{figure}

The eigenvalues of the covariance matrix provide direct insights into the structure of systematic risk. To investigate this, we first categorize our 360 stocks into 10 equally-sized groups (36 stocks per group) based on their total number of missing observations, sorted from highest to lowest. Figure \ref{fig:Eigenvalues} then plots the magnitudes of the first two eigenvalues of the covariance matrix for each of these groups. These leading eigenvalues represent the amount of variance explained by the first two principal components, which are often interpreted as dominant systematic factors.

To evaluate our methodology, we compare it against the PI imputation, a standard benchmark in high-frequency finance. The PI method is the primary source of price staleness, as it simply carries forward the last observed price, creating an artificially static price series. Our analysis contrasts the eigenvalue structure of the covariance matrix derived from these stale, PI-imputed prices with that derived from the ``effective prices'' recovered by our proposed method.

Our findings reveal that price staleness significantly distorts the covariance matrix's eigenvalue structure, an effect that is most pronounced under two conditions: at higher frequencies (1-minute) and for stocks with more missing data.
\begin{enumerate}
	\item[(i)] At the 1-minute frequency, the discrepancy is stark. In Group 1 (stocks with the most missing data), the leading eigenvalue from the PI method is approximately 0.4, whereas our method yields a value of around 0.5. This gap diminishes as data quality improves, becoming minimal in Group 10. A similar, substantial difference is also observed for the second eigenvalue.
	\item[(ii)] At lower frequencies (5- and 10-minute), the distortion from the PI method is less severe, with significant differences in eigenvalues confined primarily to the first few groups (i.e., those with the most missing data).
\end{enumerate}

Furthermore, focusing on the more reliable eigenvalues generated by our method for the 1-minute data, we observe a clear increasing trend in their magnitude from Group 1 to Group 10. This suggests a fundamental difference in the risk composition of these stocks. Stocks that trade frequently (e.g., in Group 10) exhibit stronger co-movement and are more exposed to systematic factors. Conversely, stocks prone to infrequent trading (e.g., in Group 1) possess a larger component of idiosyncratic risk, which is not captured by the leading principal components, even after our advanced imputation.

\subsection{Imputation Error}\label{subsec:Imputation Error}
In this subsection, we conduct an extensive study to compare the imputation accuracy of our proposed method against several benchmarks, including the LI and PI methods. We evaluate performance based on the imputation errors of log-prices across various sampling frequencies.\footnote{This choice is motivated by the semi-martingale nature of high-frequency asset price dynamics. Unlike low-frequency data, the properties of high-frequency price increments depend on the length of the time interval, making a direct analysis of log-prices more appropriate across different frequencies (e.g., 1-minute, 5-minute, and 10-minute).}

To create a controlled setting for evaluating imputation errors, we follow a two-step procedure to artificially mask observed data points. For each day $t$ in our sample ($t=1,...,T$):
\begin{enumerate}[label=Step \arabic*:, leftmargin=*]
	\item Identify testable data: We start with the potential log-price matrix for that day, denoted as $\mathcal{P}_t$ (an $N\times n$ matrix, where $N$ the number of stocks and $n$ is the number of intraday observations). We first exclude any prices that were already missing in the original dataset. The remaining set of observed prices forms our ``ground truth'' data for the test.
	\item Construct mask matrix: We then create a binary mask matrix, $\mathcal{M}_t$, of the same dimensions. For each position corresponding to an observable price in $\mathcal{P}_t$, we randomly assign a value of $1$ (indicating the price will be masked) with a pre-specified probability $p$, and 0 otherwise. We vary this masking probability $p$ across the set $\{0.1,0.2,...,0.7\}$ to assess performance under different levels of data sparsity. \footnote{Four different missing observations in low-frequency data were investigated by \cite{duan2024factor}: (i) missing-at-random, (ii) simultaneous adoption, (iii) staggered adoption, (iv) switchback. Random missing in high-frequency data, or random missing but with missing probabilities that are not constant but semi-martingale processes may be more acceptable (e.g., \citealt{bandi2017excess} and \citealt{kong2024staleness}).}
\end{enumerate}

The observed data available to the imputation methods is thus the element-wise product $\mathcal{P}_t^{obs}=\mathcal{P}_t\circ\mathcal{M}_t^c$, where $\mathcal{M}_t^c$ is the complement of $\mathcal{M}_t$ and $\circ$ denotes the Hadamard product. After applying an imputation method to obtain an estimate $\hat{\mathcal{P}}_t$, we measure its accuracy only on the artificially masked data points. The main results are summarized in Table \ref{tab:Imputation error}. We report two error metrics, averaged over all trading days:
\begin{align}\label{eq:compute error}
	\begin{aligned}
		&\text{Absolute error}:=\frac{1}{T}\sum_{t=1}^T\frac{\|(\hat{\mathcal{P}}_t-\mathcal{P}_t)\circ\mathcal{M}_t\|_F}{\|\mathcal{M}_t\|_F},\\
		&\text{Relative error}:=\frac{1}{T}\sum_{t=1}^T\frac{\|(\hat{\mathcal{P}}_t-\mathcal{P}_t)\circ\mathcal{M}_t\|_F}{\|\mathcal{P}_t\circ\mathcal{M}_t\|_F},
	\end{aligned}
\end{align}
where $\hat{\mathcal{P}}_t$ is the imputed price matrix, and the Frobenius norm $\|\cdot\|_F$ in the numerator is calculated only over the set of masked entries.

\begin{table}
	\TABLE
	{Imputation error for different methods. \label{tab:Imputation error}}
	{
	\setlength\tabcolsep{2pt}{\begin{tabular*}{\textwidth}{@{\extracolsep{\fill}}ccccccccccccc}
		\toprule
		Mask  &       & \multicolumn{3}{c}{1 minute} &       & \multicolumn{3}{c}{5 minute} &       & \multicolumn{3}{c}{10 minute} \\
		\cmidrule{3-5}\cmidrule{7-9}\cmidrule{11-13}    probability &       & PI    & LI    & NN    &       & PI    & LI    & NN    &       & PI    & LI    & NN \\
		\midrule
		\multicolumn{13}{c}{Absolute error} \\
		\midrule
		0.1   &       & 0.1088 & 0.0757 & 0.0681 &       & 0.2070 & 0.1426 & 0.1225 &       & 0.2853 & 0.1947 & 0.1646 \\
		0.2   &       & 0.1151 & 0.0785 & 0.0711 &       & 0.2195 & 0.1488 & 0.1289 &       & 0.3027 & 0.2033 & 0.1737 \\
		0.3   &       & 0.1220 & 0.0819 & 0.0747 &       & 0.2338 & 0.1558 & 0.1363 &       & 0.3213 & 0.2128 & 0.1839 \\
		0.4   &       & 0.1311 & 0.0864 & 0.0794 &       & 0.2514 & 0.1648 & 0.1459 &       & 0.3455 & 0.2254 & 0.1973 \\
		0.5   &       & 0.1427 & 0.0925 & 0.0857 &       & 0.2739 & 0.1764 & 0.1582 &       & 0.3768 & 0.2425 & 0.2153 \\
		0.6   &       & 0.1586 & 0.1006 & 0.0942 &       & 0.3036 & 0.1922 & 0.1749 &       & 0.4178 & 0.2650 &  0.2390 \\
		0.7   &       & 0.1814 & 0.1128 & 0.1067 &       & 0.3485 & 0.2170 & 0.2008 &       & 0.4781 & 0.3004 & 0.2759 \\
		\midrule
		\multicolumn{13}{c}{Relative error} \\
		\midrule
		0.1   &       & 0.0270 & 0.0188 & 0.0169 &       & 0.0509 & 0.0351 & 0.0300 &       & 0.0699 & 0.0478 & 0.0402 \\
		0.2   &       & 0.0286 & 0.0195 & 0.0176 &       & 0.0540 & 0.0366 & 0.0316 &       & 0.0742 & 0.0499 & 0.0425 \\
		0.3   &       & 0.0303 & 0.0204 & 0.0185 &       & 0.0575 & 0.0383 & 0.0334 &       & 0.0788 & 0.0522 & 0.0449 \\
		0.4   &       & 0.0326 & 0.0215 & 0.0197 &       & 0.0618 & 0.0405 & 0.0357 &       & 0.0847 & 0.0553 & 0.0482 \\
		0.5   &       & 0.0354 & 0.0230 & 0.0212 &       & 0.0673 & 0.0434 & 0.0388 &       & 0.0924 & 0.0595 & 0.0526 \\
		0.6   &       & 0.0394 & 0.0250 & 0.0233 &       & 0.0746 & 0.0473 & 0.0429 &       & 0.1024 & 0.0650 & 0.0584 \\
		0.7   &       & 0.0450 & 0.0280 & 0.0264 &       & 0.0857 & 0.0534 & 0.0492 &       & 0.1172 & 0.0737 & 0.0674 \\
		\bottomrule
	\end{tabular*}}
}
{See \eqref{eq:compute error} for error calculations in this table. Key to abbreviations: ``LI'' stands for linear interpolate; ``PI'' stands for previous-tick interpolate; ``NN'' stands for our nuclear norm.}
\end{table}%

Table \ref{tab:Imputation error} presents several key findings. First and foremost, our proposed method consistently and significantly outperforms the conventional PI method, yielding imputation errors that are approximately half the size of those from the PI approach. This superiority is robust across all tested sampling frequencies and masking probabilities. We therefore conclude that relying on stale prices, which inherently ignores both cross-sectional dependence and dynamic time-series patterns, leads to substantial and avoidable inaccuracies in price imputation. In contrast, our global optimization framework, by recovering an underlying low-rank structure, provides a much more accurate approximation of the true, unobserved high-frequency prices.

Second, our method also demonstrates a clear advantage over the LI method. A plausible explanation for this result is that the gains from capturing cross-sectional dependence, a key feature of our model, outweigh the benefits of simple temporal interpolation. Furthermore, the performance gap between our method and the benchmarks widens as the masking probability increases. This highlights the particular strength of our approach in scenarios with high data sparsity, where traditional methods like PI become increasingly unreliable.

Finally, we observe that the imputation performance of all methods tends to degrade at lower frequencies (e.g., 10-minute). This presents a general challenge for high-frequency data imputation, as longer time intervals between observations can obscure the underlying price dynamics. Despite this challenge, the superiority of our method remains striking. For instance, even under a severe 70\% masking probability, our method achieves a lower error rate than the PI method does under a minimal 10\% masking probability. This compelling result underscores the remarkable efficiency and robustness of our proposed imputation framework.

\subsection{Applications to Portfolio Selection}\label{subsec:Applications to Portfolio Selection}
In this subsection, we evaluate the economic significance of the covariance matrix estimates by examining their performance in a practical portfolio allocation context. We consider the following constrained minimum-variance portfolio problem:
\begin{equation}\label{eq:portfolio problem}
	\min_{\bm{w}} \bm{w}'\bm{\Sigma}\bm{w}, ~~s.t.~~\bm{w}'\bm{1}=1, \|\bm{w}\|_1\leq c,
\end{equation}
where $\|\bm{w}\|_1\leq c$ impose an risk-exposure constraint, with $\|\cdot\|_1$ denoting the $L_1$ norm. A value of $c=1$ corresponds to a long-only portfolio (no short sales), whereas $c>1$ allows for short selling. 

Our empirical analysis uses intraday data only, excluding overnight returns to avoid complications from dividend issuances and stock splits. To ensure robustness to price jumps, we further apply a five-standard-deviation truncation rule. The truncation threshold is calibrated using the bipower variation estimator of \cite{barndorff2004power}, adjusted for diurnal volatility patterns following the methodology of \cite{li2017jump} and \cite{bollerslev2024optimal}.

Following the literature \cite{fan2012vast}, \cite{ait2017using}, and \cite{cui2024regularized}, we adopt a monthly rebalancing strategy. At the end of each month, we construct the optimal portfolio weights by solving problem \eqref{eq:portfolio problem} using an estimate of the integrated covariance matrix. This covariance matrix is estimated using the high-frequency data from the past month, first imputed by one of the competing methods and then processed using the estimator of \cite{ait2017using}. We then evaluate the out-of-sample performance of these portfolios over the next month under various exposure constraints $c$.

The results are presented in Table \ref{tab:Portfolio}, where we report three key performance metrics: out-of-sample annualized average return (AR), annualized standard deviation (SD), and the Sharpe ratio (SR).

\begin{table}
	\TABLE
{The out-of-sample performance of monthly-rebalanced optimal portfolios. 	\label{tab:Portfolio}}
{
	{\begin{tabular*}{\textwidth}{@{\extracolsep{\fill}}cccccccccccccc}
		\toprule
		&       & \multicolumn{3}{c}{Constraint: 2} & \multicolumn{3}{c}{Constraint: 3} & \multicolumn{3}{c}{Constraint: 4} & \multicolumn{3}{c}{Constraint: 5} \\
		\cmidrule(lr){3-5} \cmidrule(lr){6-8} \cmidrule(lr){9-11} \cmidrule(lr){12-14}
		G. &      & {PI}  & {LI}  & {NN}  & {PI}  & {LI}  & {NN}  & {PI}  & {LI}  & {NN}  & {PI}  & {LI}  & {NN} \\
		\midrule
		1 & AR    & 11.796 & 12.925 & 13.777 & 12.782 & 12.852 & 13.836 & 12.782 & 12.852 & 13.938 & 12.782 & 12.852 & 13.938 \\
		& SD    & 23.179 & 24.265 & 24.439 & 23.345 & 24.576 & 25.191 & 23.345 & 24.576 & 25.281 & 23.345 & 24.576 & 25.281 \\
		& SR    & 0.508  & 0.532  & \bfseries 0.564 & 0.547  & 0.522  & \bfseries 0.549 & 0.547  & 0.522  & \bfseries 0.551 & 0.547  & 0.522  & \bfseries 0.551 \\
		\addlinespace
		2 & AR    & 6.743  & 6.999  & 6.993  & 6.155  & 6.321  & 6.861  & 6.115  & 6.064  & 6.537  & 6.115  & 6.064  & 6.537 \\
		& SD    & 20.954 & 21.072 & 21.355 & 20.819 & 20.923 & 21.220 & 20.811 & 20.927 & 21.262 & 20.811 & 20.927 & 21.262 \\
		& SR    & 0.321  & \bfseries 0.332 & 0.327 & 0.295  & 0.302  & \bfseries 0.323 & 0.293  & 0.289  & \bfseries 0.307 & 0.293  & 0.289  & \bfseries 0.307 \\
		\addlinespace
		3 & AR    & 11.373 & 13.375 & 13.858 & 11.617 & 13.809 & 14.575 & 11.598 & 13.974 & 14.660 & 11.598 & 13.974 & 14.660 \\
		& SD    & 24.311 & 24.467 & 24.513 & 24.624 & 24.962 & 24.978 & 24.681 & 25.055 & 25.094 & 24.681 & 25.055 & 25.094 \\
		& SR    & 0.467  & 0.546  & \bfseries 0.565 & 0.471  & 0.553  & \bfseries 0.584 & 0.469  & 0.557  & \bfseries 0.584 & 0.469  & 0.557  & \bfseries 0.584 \\
		\addlinespace
		4 & AR    & 5.655  & 5.431  & 5.844  & 6.097  & 6.011  & 6.690  & 6.351  & 6.204  & 6.936  & 6.350  & 6.204  & 6.935 \\
		& SD    & 23.790 & 23.950 & 23.844 & 23.604 & 23.703 & 23.567 & 23.566 & 23.669 & 23.538 & 23.566 & 23.669 & 23.538 \\
		& SR    & 0.237  & 0.226  & \bfseries 0.245 & 0.258  & 0.253  & \bfseries 0.283 & 0.269  & 0.262  & \bfseries 0.294 & 0.269  & 0.262  & \bfseries 0.294 \\
		\addlinespace
		5 & AR    & 4.002  & 4.183  & 4.377  & 4.028  & 4.226  & 4.312  & 4.023  & 4.212  & 4.316  & 4.023  & 4.212  & 4.316 \\
		& SD    & 21.204 & 21.268 & 21.334 & 21.469 & 21.539 & 21.580 & 21.500 & 21.574 & 21.627 & 21.500 & 21.574 & 21.627 \\
		& SR    & 0.188  & 0.196  & \bfseries 0.205 & 0.187  & 0.196  & \bfseries 0.199 & 0.187  & 0.195  & \bfseries 0.199 & 0.187  & 0.195  & \bfseries 0.199 \\
		\bottomrule
	\end{tabular*}}
}
	{This table reports the out-of-sample performance of portfolios constructed using different covariance matrix estimators, evaluated monthly from 2016 to 2022. We report the annualized average return (AR), annualized standard deviation (SD), and the Sharpe ratio (SR). ``G.'' refers to the stock groups defined before, and $c$ denotes the gross exposure constraint from the optimization problem \eqref{eq:portfolio problem}. Values in bold indicate the highest Sharpe ratio for each given case.}
\end{table}%

Table \ref{tab:Portfolio} reports the out-of-sample portfolio performance from 2016 to 2022 across various gross exposure constraints ($c$) and for five equally-sized stock groups. These groups are formed by sorting stocks based on their degree of data missingness, from highest (Group 1) to lowest (Group 5), allowing us to assess performance under different data quality scenarios.

A primary finding is that portfolios constructed using our imputation method achieve the highest Sharpe ratios in nearly all scenarios. This result underscores the economic value of accurately estimating the covariance matrix by accounting for asynchrony. While the Sharpe ratios tend to decline as leverage increases (i.e., as $c$ grows), they stabilize for $c>4$. It is important to note that while more advanced covariance estimators, such as that of \cite{cui2024regularized}, could potentially further enhance performance, our focus here is on isolating the impact of the imputation method itself, for which we use a standard estimation approach.

A striking pattern emerges from the inter-group comparison: portfolios of stocks with more missing data (e.g., Groups 1-2) consistently generate higher average returns and Sharpe ratios than those with less missing data (e.g., Groups 4-5). The Sharpe ratio of Group 1, for instance, is more than double that of Group 5. This observation aligns with the well-documented liquidity premium in asset pricing. Infrequent trading, which directly causes data missingness in our high-frequency setting, is a hallmark of illiquidity. Seminal works by \cite{amihud1986asset} and \cite{pastor2003liquidity} have established that illiquid stocks tend to offer higher expected returns to compensate investors for higher trading costs and exposure to systematic liquidity risk.

However, the link is not one-to-one. Price staleness and traditional liquidity measures are highly correlated but distinct concepts \citep{bandi2020zeros, bandi2024systematic}. \cite{bandi2024systematic} posit that staleness can be decomposed into a systematic component, driven by factors like capital ``shadow costs'', and an idiosyncratic component, related to asset-specific spreads. The performance differentials we observe across groups may therefore reflect a complex interplay of these factors, leading to conclusions that may differ from those based on low-frequency data alone.

Finally, it is worth noting that relying on price staleness for out-of-sample portfolio allocations may reduce out-of-sample performance, which is consistent with the findings of \cite{kong2024staleness}. Moreover, here we also find that when missing values are severe (staleness probability is high), the gap between portfolio allocation using estimates of effective prices (NN method) and using stale prices (PI method) may be larger.

\subsection{Spot Beta}\label{subsec:Spot Beta}
In this subsection, we investigate the impact of price staleness on the estimation of spot betas for individual stocks relative to an exchange-traded fund (ETF). We estimate the spot beta using the fixed-$k$ local regression method proposed by \cite{bollerslev2024optimal}, to which we refer the reader for detailed computational procedures. To the best of our knowledge, the effect of price staleness (as induced by the PI method) on high-frequency beta estimation has not been systematically studied, and this analysis aims to fill that gap. This is a critical gap, as recent literature demonstrates the profound impact of high-frequency beta estimation on asset pricing tests. For example, \cite{hollstein2020conditional} find that the empirical failures of the Conditional CAPM can be largely resolved by moving from daily to high-frequency betas. While their study successfully underscores the value of high-frequency data, it relies on lag-based adjustments to mitigate non-synchronicity. Our analysis, therefore, also serves to show how our synchronization framework can provide cleaner and more reliable beta estimates, which are essential inputs for the type of asset pricing tests conducted by \cite{hollstein2020conditional}, thereby preventing potentially misleading inferences caused by data artifacts like price staleness.

Our market proxy is the SPDR S\&P 500 ETF Trust (SPY). While the S\&P 500 Index (SPX) is the theoretical benchmark, it is not directly tradable. The SPY, designed to track the SPX, is one of the most liquid and widely traded ETFs globally, making it an ideal instrument for this analysis.

To highlight the effects of asynchrony, we focus our analysis on a period of extreme market turbulence: the ten trading days from March 5 to March 18, 2020. This period witnessed the onset of the COVID-19 financial crisis, characterized by an oil price war (March 9, ``Black Monday I'' crash), the declaration of a global pandemic (March 11), and emergency actions by the Federal Reserve (March 15), leading to extreme price volatility. In our regression framework, the SPY return series serves as the regressor, and the individual stock return series is the dependent variable.\footnote{The period's turmoil ignited on March 9 when an oil price war erupted, sparking recession fears and the ``Black Monday I'' crash. The crisis deepened on March 11 as the WHO declared a global pandemic and the U.S. announced a European travel ban, devastating key industries. In a dramatic response, the Federal Reserve executed an emergency rate cut to zero on March 15, but this was perceived as a panic move, failing to reassure investors and triggering another severe market plunge.}

We specifically compare two types of stocks: those with a high number of missing observations during this period and those with very few. For instance, Mettler-Toledo International (MTD, Industrials) and Verizon Communications (VZ, Communication Services) represent the extremes of high and low data missingness, respectively, within our sample for this period. Our subsequent analysis will focus on these two representative stocks.

In our empirical tests, we primarily focus on testing the null hypothesis of zero beta ($H_0:\beta=0$). This focus is motivated by the strong idiosyncratic nature of individual stock returns, which often makes it statistically challenging to detect a consistently significant, non-zero beta. Following \cite{bollerslev2024optimal}, our main analysis uses a 15-minute estimation window ($k=15$). Robustness checks using $k=5$ and $k=10$ are provided in the Supplementary Appendix.

\begin{figure}
	\FIGURE
	{\includegraphics[scale=0.55]{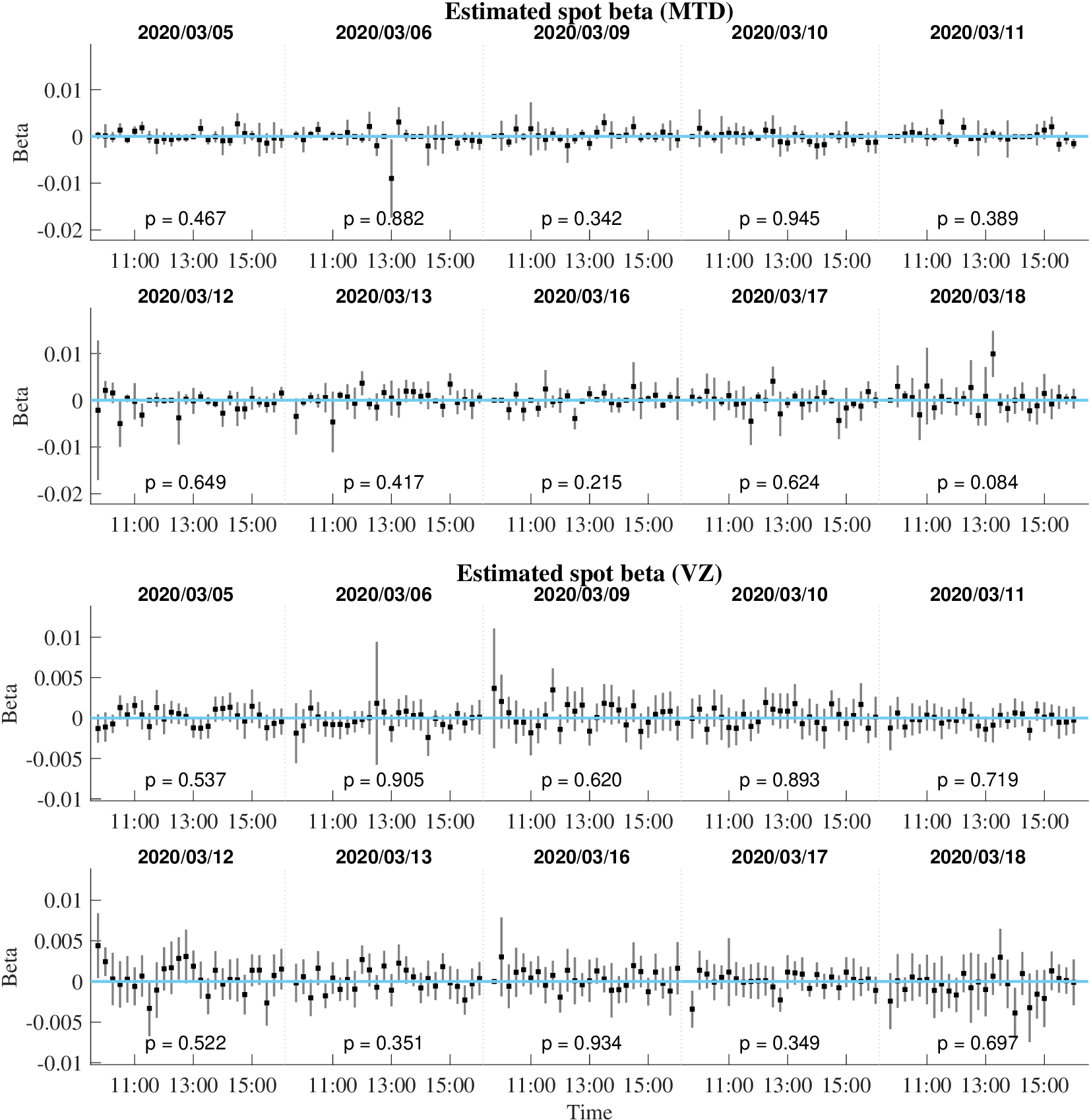}}
	{Spot beta for the presence of price staleness. \label{fig:Beta PI}}
	{This figure plots the estimated spot betas of Mettler-Toledo International (MTD) and Verizon Communications (VZ) against the SPY. The betas are estimated using 1-minute price data over 15-minute rolling windows, along with their corresponding 90\% confidence intervals. The analysis covers the two-week period of high market volatility from March 5, 2020, to March 18, 2020. The $p$-value reported in each panel corresponds to a test of the functional null hypothesis that the entire spot beta process for a given day is equal to zero ($H_0:\beta_t=0$ for all $t$).}
\end{figure}

The upper panel of Figure \ref{fig:Beta PI} reveals a striking pattern for the MTD stock: the spot beta estimates are frequently exactly zero, with confidence intervals degenerating to zero width. This is not a reflection of economic reality but rather a methodological artifact induced by the PI imputation. As established in the classic literature on non-synchronous trading \citep{scholes1977estimating}, when a stock's price does not update, the PI method records a zero return. This mechanically sets the stock's local covariance with the market (i.e. SPY) to zero, forcing the corresponding beta estimate to be zero as well. This issue is, unsurprisingly, most severe for the less liquid stock, MTD.

The consequences extend beyond distorted point estimates; this artifact critically undermines the validity of statistical inference. For instance, on March 9, 2020 (``Black Monday I''), the PI-based data yields an exceptionally high $p$-value of 0.342 for the functional test of the null hypothesis that the beta process is identically zero. This misleading result leads to a failure to reject the null of zero systematic risk during a major market crash.

\begin{figure}
	\FIGURE
	{\includegraphics[scale=0.55]{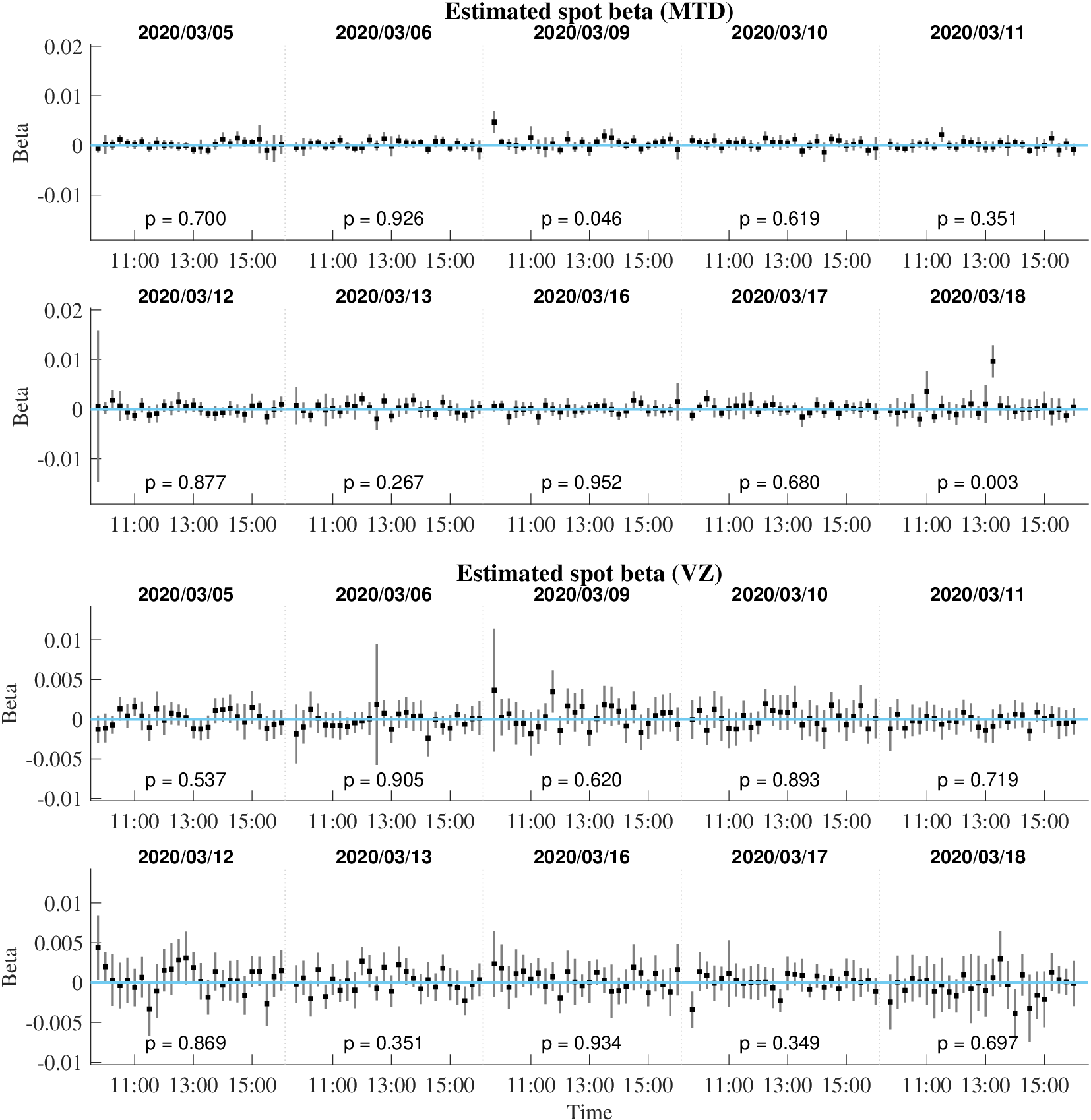}}
	{Spot beta for the absence of price staleness. \label{fig:Beta NN}}
	{This figure plots the estimated spot betas of Mettler-Toledo International (MTD) and Verizon Communications (VZ) against the SPY. The betas are estimated using 1-minute price data over 15-minute rolling windows, along with their corresponding 90\% confidence intervals. The analysis covers the two-week period of high market volatility from March 5, 2020, to March 18, 2020. The $p$-value reported in each panel corresponds to a test of the functional null hypothesis that the entire spot beta process for a given day is equal to zero ($H_0:\beta_t=0$ for all $t$).}
\end{figure}

In stark contrast, when using the effective prices recovered by our method (see Figure \ref{fig:Beta NN}), the $p$-value for the same test on the same day is merely 0.046, leading to a clear rejection of the null hypothesis. This discrepancy highlights a fundamental flaw: the zero-inflated return series generated by the PI method violates the ``local Gaussianity'' assumption that underpins the inferential framework of \cite{bollerslev2024optimal}. Consequently, any statistical tests based on such contaminated data become unreliable.

In stark contrast to the erratic zero-beta estimates produced by the PI method, the beta paths derived from our approach (Figure \ref{fig:Beta NN}) are notably smoother and more continuous. This demonstrates that once the microstructure noise from non-synchronous trading is properly filtered, a stock's systematic risk exposure is revealed to be a dynamically evolving intraday process, not a series of binary jumps. This finding aligns with the modern consensus in high-frequency finance that asset price processes are well approximated by continuous semi-martingales \citep{ait2014high}.

The resulting statistical inference is also substantially more credible. For MTD on March 18, 2020, for example, the uniform test yields a $p$-value of 0.003, allowing for a decisive rejection of the zero-beta null hypothesis and accurately capturing the stock's significant market risk on that day. Even on days with less statistical power, the confidence intervals remain informative, showing the beta fluctuating around a non-zero mean. This underscores how a robust imputation method, when combined with the optimal inference framework of \cite{bollerslev2024optimal}, enables reliable inference even from short estimation windows.

The period of March 2020 included several market-wide ``circuit breaker'' trading halts (on March 9, 12, 16, and 18), and Figure \ref{fig:Beta NN} provides a clear window into how risk characteristics evolved under such extreme stress. For instance, VZ, a defensive telecommunications stock, generally exhibits a beta near zero, consistent with its sector profile. However, during the most turbulent sessions, the volatility of its beta and the width of its confidence bands visibly expand, reflecting a heightened spillover of systematic risk even to traditionally ``safe'' assets. This provides compelling visual evidence for the intraday variation of systematic risk, a phenomenon documented by \cite{andersen2021recalcitrant}. Furthermore, MTD's beta path on March 18 reveals a distinct intraday pattern, trending upwards in the afternoon, potentially reflecting the market's continuous repricing of the stock's risk as new information was digested. Such fine-grained dynamics are completely obscured in daily or lower-frequency data.

In summary, the proper handling of non-synchronous trading is a critical prerequisite for drawing valid economic conclusions from high-frequency spot regressions. The conventional previous-tick method not only produces severely biased zero-beta estimates but also invalidates statistical inference. By employing a robust imputation technique, we can uncover the true, dynamic, and economically meaningful evolution of a stock's beta, particularly during periods of market turmoil. This capacity for precise, real-time risk measurement has significant practical implications for risk management, algorithmic trading, and event study analysis, allowing market participants to assess and respond to changes in asset risk with a high degree of precision when it matters most.

\section{Conclusion}\label{sec:Conclusion}
Asynchronous trading is a fundamental challenge in high-frequency finance that biases risk estimates and impairs asset allocation. We address this by recasting data synchronization as a constrained matrix completion problem. Our framework recovers the complete matrix of synchronous price increments by minimizing its nuclear norm—capturing the underlying low-rank factor structure—subject to linear constraints derived from observed, asynchronous prices.

Theoretically, we prove the existence and uniqueness of our estimator, establish its convergence rate, and show that it efficiently pools information across both liquid and illiquid assets, overcoming a key limitation of traditional methods. Empirically, using extensive simulations and a large panel of S\&P 500 stocks, we demonstrate that our approach substantially outperforms established benchmarks. It corrects systematic biases in risk estimates and, most critically, generates portfolios with economically and statistically significant higher out-of-sample Sharpe ratios. Our research provides a powerful and practical tool for uncovering the true dynamics of asset prices, opening promising avenues for future work in areas such as microstructure modeling and mixed-frequency analysis. By providing a theoretically sound and empirically validated solution to a long-standing problem, this paper enables more precise risk measurement and offers a clearer lens into the dynamic nature of modern financial markets.

\ACKNOWLEDGMENT{Authors are listed in alphabetical order.}

\theendnotes

\bibliographystyle{informs2014} 
\bibliography{ref.bib} 



\ECSwitch


\ECHead{Technical Proofs and Additional Results}

This appendix contains technical proofs and additional results for the paper.
\begin{itemize}
    \item Appendix \ref{sec:Proofs} provides the proofs of theorems in the main text.
    \item Appendix \ref{sec:Additional Simulation} provides the additional simulation results.
    \item Appendix \ref{sec:Additional Empirical Analysis} provides additional empirical results.
\end{itemize}

\section{Proofs}\label{sec:Proofs}
\proof{Proof of Theorem \ref{th1}} The proof proceeds by establishing the restricted isometry property (RIP) for the linear operator $\mathcal{A}$. We will prove the two bounds, \eqref{bound-th1} and \eqref{bound-th1-1}, separately.

\noindent\textbf{Proof of the Bound in \eqref{bound-th1}}

Recall the model setting that
\begin{equation}
	dX_t=\mu_tdt+\sigma_tdW_t+\sigma^*_tdW_t^*=:dX^{\mu}+dX^c_t+dX^*_t,
\end{equation}
which implies a corresponding decomposition for the increment matrix, where $\sigma_t=\sigma^0\Sigma_t$. Let $V=\left(\int^{t_j}_{t_{j-1}}\Sigma_tdW_t\right)_{r\times n}$, $\Pi^{\mu}=\left(X^{\mu}_{it_j}-X^{\mu}_{it_{j-1}}\right)_{N\times r}$, $\Pi=\left(X^c_{it_j}-X^c_{it_{j-1}}\right)_{N\times n}$ and $\Pi^{*,\mu}=\left(X^*_{it_j}-X^*_{it_{j-1}}\right)_{N\times n}+\Pi^{\mu}:=\Pi^{*}+\Pi^{\mu}$. Now, we first prove that the solution for $\Pi$ in (\ref{nuclear norm2}) is equal to $\Pi_0$ with probability approaching one (conditional on $\{R_j, D_{ik}; i\leq N, j\leq n, k\leq n\}$). Rearranging the second equation in (\ref{nuclear norm2}), we have $\mathcal{A}(\Pi)=b-\mathcal{A}(\Pi^{*,\mu}):=b^*$ and the solution is to find a matrix of rank no larger than $r$ so that this constraint is satisfied and $\|\Pi\|_*$ is minimized. To this end, we first prove that $\Pi$ satisfies the nearly isometry condition for small enough $\delta$,
\begin{eqnarray*}
	&&P_{R,D}\left((1-\delta)\|\Pi\|_F \leq \|\mathcal{A}(\Pi)\| \leq (1+\delta)\|\Pi\|_F  \ \mbox{for all} \ \sigma^0\in\Sigma^0\right)\nonumber\\
	&\geq & 1-C\exp\left\{-c^2r^2/\left(\sum^n_{j=1}R_j^2\right)^{1/4}\right\}-C\exp\{-c^2r^2\delta^2N^{\alpha}/\overline{L}_1\},
\end{eqnarray*}
which amounts to the following condition for another small $\delta$ ($\delta$ is a generic small constant that may vary at different appearance),
\begin{eqnarray}\label{isometry}
	&&P_{R,D}\left(|\frac{\|\mathcal{A}(\Pi)\|^2-\|\Pi\|_F^2}{\|\Pi\|_F^2}|\leq \delta \ \mbox{for all} \ \sigma^0\in\Sigma^0\right)\nonumber\\
	&\geq & 1-C\exp\left\{-c^2r^2/\left[\left(\sum^n_{j=1}R_j^2\right)^{1/4}\right]\right\}-C\exp\{-c^2r^2\delta^2N^{\alpha}/\overline{L}_1\}.
\end{eqnarray}

Since
$\|\Pi\|_F^2=N^{\alpha} \|V\|_F^2$,
(\ref{isometry}) reduces to the following relative concentration condition for $V$.
\begin{eqnarray}\label{isometry1}
	&& P_{R,D}\left(|\frac{\|\mathcal{A}(\sigma^0V)\|^2-\|\sigma^0V\|_F^2}{\|V\|_F^2}| > N^{\alpha}\delta \ \mbox{for some} \ \sigma^0\in\Sigma^0\right)\nonumber\\
	&\leq & C\exp\left\{-c^2r^2\delta^2N^{\alpha}/\overline{L}_1\right\}+C\exp\left\{-c^2r^2/\left(\sum^n_{j=1}R_j^2\right)^{1/4}\right\}.
\end{eqnarray}

\noindent\textbf{Step 1.1: Bounding the Denominator $\|V\|_F^2$.}

In the sequel, we aim to prove (\ref{isometry1}). We first consider the limit of $\|V\|_F^2$ in the denominator which is irrelevant to $\sigma^0$. Notice that $V$ consists of $r$ rows of martingale differences with shrinking intervals, by standard stochastic calculus, for $k=1,..., r$,
\begin{eqnarray*}
	&& P_{R,D}\left(\|V_{k\cdot}\|^2-\int^T_0\Sigma_{tk\cdot}\Sigma_{t k\cdot}^{\prime}dt > x\sqrt{\sum^n_{j=1}R_j^2} \right)\\
	&\leq & E\exp{\left(-\theta x\sqrt{\sum^n_{j=1}R_j^2}+\theta \sum^n_{j=1}\left[(V_{kj}-V_{k(j-1)})^2-\int^{t_j}_{t_{j-1}}\Sigma_{t, k\cdot}\Sigma_{t, k\cdot}^{\prime}dt\right]\right)}\\
	&\leq & \exp{\left(-\theta x\sqrt{\sum^n_{j=1}R_j^2}\right)}E\left\{\prod^{n-1}_{j=1}e^{\theta\left[\left(V_{kj}-V_{k(j-1)}\right)^2-\int^{t_j}_{t_{j-1}}\Sigma_{t, k\cdot}\Sigma_{t, k\cdot}^{\prime}dt\right]}\right.\\
	&&\times \left.E_{\mathcal{F}_{t_{n-1}}}e^{\theta \left[\left(V_{kn}-V_{k(n-1)}\right)^2-\int^{t_n}_{t_{n-1}}\Sigma_{t, k\cdot}\Sigma_{t, k\cdot}^{\prime}dt\right]}\right\}.
\end{eqnarray*}

Now we compute the conditional expectation inside the unconditional expectation. By the change of time, there exists a standard Brownian motion $B$ so that
$$
(V_{kn}-V_{k(n-1)})^2-\int^{t_n}_{t_{n-1}}\Sigma_{t, k\cdot}\Sigma_{t, k\cdot}^{\prime}dt=B_{<V_k>_{t_{n-1}}^{t_n}}^2-<V_k>_{t_{n-1}}^{t_n},
$$
where $<V_k>_a^b$ is the quadratic variation of $V_k$ in the interval $(a, b]$. Notice that $e^{\theta[(V_{kj}-V_{k(j-1)})^2-\int^{t_j}_{t_{j-1}}\Sigma_{t, k\cdot}\Sigma_{t, k\cdot}^{\prime}dt]}$ is the end point of a submartingale for $\theta>0$, by Assumption \ref{ass2}, $<V_k>_{t_{j-1}}^{t_j}\leq CR_j$, and hence
\begin{eqnarray}\label{change of time}
	&&E_{\mathcal{F}_{t_{n-1}}}e^{\theta \left[(V_{kn}-V_{k(n-1)})^2-\int^{t_n}_{t_{n-1}}\Sigma_{t, k\cdot}\Sigma_{t, k\cdot}^{\prime}dt\right]}\nonumber\\
	&\leq & Ee^{\theta \left[B_{CR_n}^2-CR_n\right]}=Ee^{\theta CR_n(Z^2-1)}\leq e^{2\theta^2C^2R_n^2},
\end{eqnarray}
by Lemma 1 of \cite{fan2012vast} for $|\theta C R_n|\leq 1/4$, where $Z$ is a standard normal random variable. Iteratively using (\ref{change of time}),
\begin{eqnarray*}
	&& P_{R,D}\left(\sum^n_{j=1}(V_{kj}-V_{k(j-1)})^2-\int^T_{0}\Sigma_{t, k\cdot}\Sigma_{t, k\cdot}^{\prime}dt > x\sqrt{\sum^n_{j=1}R_j^2} \right)\nonumber\\
	&\leq & \exp{\left(-\theta x\sqrt{\sum^n_{j=1}R_j^2}+2\theta^2 C^2 \sum^n_{j=1}R_j^2\right)},
\end{eqnarray*}
which is minimized when $\theta=x/(4C^2\sqrt{\sum^n_{j=1}R_j^2})$ and the minimum is $e^{(-x^2/(8C^2))}$. By the range of $\theta$, the range of $x$ is $0<x\leq C\sqrt{\sum^n_{j=1}R_j^2}/R_j$ for all $j=1,...,n$. Therefore,
\begin{eqnarray*}\label{concentrate1-1}
	&&P_{R,D}\left(\|V\|_F^2-\sum^r_{k=1}\int^T_0\Sigma_{t, k\cdot}\Sigma_{t, k\cdot}^{\prime}dt> x\sqrt{\sum^n_{j=1}R_j^2}
	\right) \nonumber\\
	&\leq & \sum^r_{k=1}P\left(\sum^n_{j=1}(V_{kj}-V_{k(j-1)})^2-\int^T_0\Sigma_{t,k\cdot}\Sigma_{t,k\cdot}^{\prime}dt>x\sqrt{\sum^n_{j=1}R_j^2}/r\right)\nonumber\\
	&\leq & r \max_{k\leq r} P\left(\sum^{n}_{j=1}(V_{kj}-V_{k(j-1)})^2-\int^T_0\Sigma_{t,k\cdot}\Sigma_{t,k\cdot}^{\prime}dt>x\sqrt{\sum^n_{j=1}R_j^2}/r\right).
\end{eqnarray*}
Similarly, we can obtain the other side
\begin{equation*}
	P_{R,D}\left(\|V\|_F^2-\sum^r_{k=1}\int^T_0\Sigma_{t,k\cdot}\Sigma_{t,k\cdot}^{\prime}dt<-x\sqrt{\sum^n_{j=1}R_j^2}
	\right)\leq e^{-x^2/(8C^2)+\log{r}}.
\end{equation*}
Summarizing the above two equations, we have
\begin{equation}\label{VC}
	P_{R,D}\left(|\|V\|_F^2-\sum^r_{k=1}\int^T_0\Sigma_{t,k\cdot}\Sigma_{t,k\cdot}^{\prime}dt|> x\sqrt{\sum^n_{j=1}R_j^2}
	\right)\leq 2e^{-x^2/(8C^2)+\log{r}}.
\end{equation}
We take $x=cr/\{2(\sum^n_{j=1}R_j^2)^{1/4}\}$ for $c$ small enough which satisfies the range condition for $x$ because of the assumption that $\frac{(\sum^n_{j=1}R_j^2)^{3/4}}{\max_jR_j}\geq \frac{cr}{2C}$.
By the condition that $\int^T_0\Sigma_{t,k\cdot}\Sigma_{t,k\cdot}^{\prime}dt\geq c$ and choose $c$ small,
\begin{equation}\label{denominator}
	P_{R,D}\left( \|V\|_F^2 \leq \frac{cr}{2} \right) \leq 2\exp\left\{-c^2r^2/\left\{32C^2\left(\sum^n_{j=1}R_j^2\right)^{1/2}\right\}\right\}.
\end{equation}

\noindent\textbf{Step 1.2: Bounding the Numerator $\|\mathcal{A}(\sigma^0V)\|_F^2-\|\sigma^0V\|_F^2$.}

Next, we consider the numerator $\|\mathcal{A}(\sigma^0V)\|_F^2-\|\sigma^0V\|_F^2$. To express the difference by a sum of crossing products, we recall
$$
d_{ik}:=\sharp\{j; \tau_{i(l-1)}\leq t_j\leq t_{k-1}\leq \tau_{il} \ \mbox{for some} \ 1\leq l\leq n_i\}
$$
to be the number of potential increments before $\Delta^n_{k}V$ in the observed interval $(\tau_{i(l-1)}, \tau_{il}]$ which contains $t_{k-1}$. We make a convention that $d_{ik}:=0$ if the set is empty and the resulting sum $\sum^{0}_{l=1}\Delta^n_{k-l}V=0$. We rewrite
$$
\|\mathcal{A}(\sigma^0V)\|_F^2-\|\sigma^0V\|_F^2=\sum^n_{k=1}\sum^N_{i=1}(\sigma^{0}(i,\cdot)\Delta^n_kV)(\sigma^0(i,\cdot)\sum^{d_{ik}}_{l=1}\Delta^n_{k-l}V).
$$
Recall that $\overline{L}_1=\sum^n_{k=1}R_k\sum^N_{i=1}\|\sigma^0(i,\cdot)\|^2(\sum^{D_{ik}}_{l=1}R_{k-l})\log{(\sum^{D_{ik}}_{l=1}R_{k-l})}$.
\begin{eqnarray}\label{A}
	&&P_{R,D}\left(\|\mathcal{A}(\sigma^0V)\|_F^2-\|\sigma^0V\|_F^2>N^{\alpha} x\right) \nonumber\\
	&\leq & \exp\left\{-\theta N^{\alpha}x\right\} E\prod^{n-1}_{k=1}\exp\left\{\theta \sum^N_{i=1}(\sigma^0(i,\cdot)\Delta^n_kV)(\sigma^0(i,\cdot)\sum^{d_{ik}}_{l=1}\Delta^n_{k-l}V)\right\} \nonumber\\
	&& \times E_{\mathcal{F}_{n-1}}\exp\left\{\theta\sum^N_{i=1}(\sigma^0(i,\cdot)\Delta^n_nV)(\sigma^0(i,\cdot)\sum^{d_{in}}_{l=1}\Delta^n_{n-l}V)\right\}\nonumber\\
	&\leq &\exp\left\{-\theta N^{\alpha} x + Cr\theta^2 \sum^n_{k=1}R_k \|\sum^N_{i=1}\sigma^0(i,\cdot)\sum^{D_{ik}}_{l=1}\Delta^n_{k-l}V\sigma^0(i, \cdot)\|_F^2 \right\} \nonumber\\
	&\leq & \exp\left\{-\theta N^{\alpha} x + Cr\theta^2 \sum^n_{k=1}R_k \sum^r_{m=1}\sum^N_{i=1}(\sigma^0(i,m))^2\sum^N_{i=1}(\sigma^0(i,\cdot)\sum^{D_{ik}}_{l=1}\Delta^n_{k-l}V)^2\right\}\nonumber\\
	&\leq & \exp\left\{-\theta N^{\alpha} x + Cr^2\theta^2 N^{\alpha}\overline{L}_1 \right\},
\end{eqnarray}
for any $\theta>0$, where in the last step we have made use of the fact that $|\sum^{d_{ik}}_{l=1}\Delta^n_{k-l}V|\leq C\left(\sum^{D_{ik}}_{l=1}R_{k-l}\right)^{1/2}\log{\left(\sum^{D_{ik}}_{l=1}R_{k-l}\right)^{1/2}}$ by Assumption \ref{ass2} and the law of iterated logarithm for diffusion paths, and the fact that given $\mathcal{F}_{k-1}$,
$$
\sum^N_{i=1}(\sigma^0(i,\cdot)\Delta^n_kV)(\sigma^0(i,\cdot)\sum^{d_{ik}}_{l=1}\Delta^n_{k-l}V)
$$
is an end point of a continuous martingale and hence can be represented by $B_{\mathcal{T}_{k,nN}}$ for some stopping time $\mathcal{T}_{k,nN}$ satisfying
$$
\mathcal{T}_{k,nN}\leq CR_kN^{\alpha}\sum^N_{i=1}\|\sigma^0(i,\cdot)\|^2\left(\sum^{D_{ik}}_{l=1}R_{k-l}\right)\log{\left(\sum^{D_{in}}_{l=1}R_{k-l}\right)}.
$$
A simple use of the optional stopping theorem for submartingales leads to the last inequality.

\noindent\textbf{Step 1.3: Combining the Bounds.}

Take $\theta=x/(2Cr^2\overline{L}_1)$, the upper bound of (\ref{A}) is minimized with the minimum being $\exp\{-x^2N^{\alpha}/(4Cr^2\overline{L}_1)\}$ for any $x>0$. 
Combining the bounds for the numerator and the denominator \eqref{denominator} via a union bound, we have:
\begin{eqnarray}\label{denominator+numerator}
	&& P_{R,D}\left(\frac{\|\mathcal{A}(\Pi)\|_F^2-\|\Pi\|_F^2}{\|\Pi\|_F^2}>\delta\right)\nonumber\\
	&\leq & P_{R,D}\left(\|V\|_F^2\leq cr/2\right)+P\left(|\|\mathcal{A}(\Pi)\|_F^2-\|\Pi\|_F^2|>N^{\alpha}\delta cr/2\right)\nonumber\\
	&\leq & 2\exp\left\{-c^2r^2/\left(\sum^n_{j=1}R_j^2\right)^{1/2}\right\}+C\exp\left\{-c^2\delta^2N^{\alpha}/\overline{L}_1\right\},
\end{eqnarray}
where $c$ is taken small and the first probability bound is irrelevant to $\sigma^0$. (\ref{denominator+numerator}) proves the theorem for fixed $\sigma^0$.

\noindent\textbf{Proof of the Bound in \eqref{bound-th1-1}}

The proof follows a similar strategy, but the analysis is more complex due to the presence of the idiosyncratic component $\Pi^*$. We use the decomposition $\Delta=\Pi+\Pi^*$.

\noindent\textbf{Step 2.1: Bounding the Denominator $\|\Delta\|_F^2$.}

To prove (\ref{bound-th1-1}), we present a decomposition, $\|\Delta\|_F^2=N^{\alpha}\|V\|_F^2+\|\Pi^{*,\mu}\|_F^2+2tr(\Pi^{\prime}\Pi^{*,\mu})$. The result for $\|V\|_F^2$ is already given as above. To derive the concentration result for $\|\Pi^*\|_F^2$, we decompose
$$
\Pi^*=\Pi^*_1+\Pi^*_2:=\left(\int^{t_j}_{t_{j-1}}\sigma_{t_{j-1}}^*dW_t^*\right)_{N\times n}+\left(\int^{t_j}_{t_{j-1}}(\sigma^*_t-\sigma_{t_{j-1}}^*)dW_t^*\right)_{N\times n}.
$$
Notice that $\|\Pi^*_2\|_F^2=\sum^N_{i=1}\sum^n_{j=1}\left(\int^{t_j}_{t_{j-1}}(\sigma^*_{it}-\sigma^*_{it_{j-1}})dW_{it}^*\right)^2$ and $\int^{t_j}_{t_{j-1}}(\sigma^*_{it}-\sigma^*_{it_{j-1}})^2dt\leq CR_j^{2-\epsilon}$. For $|\theta C R_j^{2-\epsilon}|<1/4$,
\begin{eqnarray}\label{concentration2-1}
	&&P_{R,D}\left(|\|\Pi^*_2\|_F^2-\sum^N_{i=1}\sum_{j=1}^n\int^{t_j}_{t_{j-1}}\left(\sigma^*_{it}-\sigma^*_{it_{j-1}}\right)^2dt|> Ny\sqrt{\sum^n_{j=1}R_j^{4-2\epsilon}}\right)\nonumber\\
	&\leq & N\max_i P\left(|\sum^n_{j=1}\left\{\left(\int^{t_j}_{t_{j-1}}\left(\sigma_{it}^*-\sigma_{it_{j-1}}^*\right)dW_t^*\right)^2-\int^{t_{j}}_{t_{j-1}}\left(\sigma_{it}^*-\sigma_{it_{j-1}}^*\right)^2dt\right\}|\right.\nonumber\\
	&&\left.>y\sqrt{\sum^n_{j=1}R_j^{4-2\epsilon}}\right)  \nonumber\\
	&\leq & 2N \exp\left\{-\theta y\sqrt{\sum^n_{j=1}R_j^{4-2\epsilon}}+2\theta^2C^2\sum^n_{j=1}R_j^{4-2\epsilon}\right\},
\end{eqnarray}
which is minimized at $\theta=y/(4C^2\sqrt{\sum^n_{j=1}R_j^{4-2\epsilon}})$ with $2 e^{\{-y^2/(8C^2)+\log(N)\}}$ being the minimum for $0<y<C\sqrt{\sum^n_{j=1}R_j^{4-2\epsilon}}/R_j^{2-\epsilon}$ for all $j=1,...,n$.

For $\Pi^*_1$, we have
$$
\|\Pi^*_1\|_F^2-\sum^N_{i=1}\sum^n_{j=1}\sigma_{it_{j-1}}^{*2}(t_j-t_{j-1})=\sum^n_{j=1}(\sigma_{t_{j-1}}^*)^2(t_j-t_{j-1})\sum^N_{i=1}[(Z_i)^2-1],
$$
where $Z_i$'s are independent standard normal random variables. 
By Lemma 2.27 of \cite{wainwright2019high}, for independent standard Gaussian random variables $Z_1,...,Z_N$ and $\frac{C\theta^2R_j^2}{2}<1/4$,
\begin{eqnarray*}
	&& E_{\mathcal{F}_{t_{j-1}}}\exp\left\{\theta(\sigma_{t_{j-1}}^*)^2(t_j-t_{j-1})\sum^N_{i=1}[(Z_i)^2-1]\right\}\nonumber\\
	&\leq & E_{\mathcal{F}_{t_{j-1}}} \exp\left\{\frac{\theta^2(\sigma_{t_{j-1}}^*)^4\pi^2(t_j-t_{j-1})^2}{2}\|\rho^*(Z_1,...,Z_N)^{\prime}\|_2^2\right\} \nonumber\\
	&\leq & E_{\mathcal{F}_{t_{j-1}}} \exp\left\{\frac{C\theta^2R_j^2}{2}\sum^N_{i=1}Z_i^2\right\}  \leq \exp\left\{\frac{CNR_j^2\theta^2}{2}\left(1+C\theta^2R_j^2\right)\right\} \nonumber\\
	&\leq & \exp\left\{\frac{3CNR_j^2\theta^2}{4}\right\}.
\end{eqnarray*}
Again, following the steps in proving the result for $\|V\|_F^2$, we have
\begin{eqnarray}\label{concentration2}
	&&P_{R,D}\left(|\|\Pi^*_1\|_F^2-\sum^N_{i=1}\sum^n_{j=1}\sigma_{it_{j-1}}^{*2}(t_j-t_{j-1})|> z \sqrt{N\sum^n_{j=1}R_j^2}
	\right)\nonumber\\
	&\leq & 2\exp\left\{-\theta z\sqrt{N\sum^n_{j=1}R_j^2}+\frac{3C\theta^2}{4}N\sum^n_{j=1}R_j^2\right\},
\end{eqnarray}
which is minimized at $\theta=2z/(3C\sqrt{N\sum^n_{j=1}R_j^2})$ with the minimum $e^{-z^2/3C}$ for $0<z\leq 3\sqrt{C}\sqrt{N\sum^n_{j=1}R_j^2}/(2\sqrt{2}R_j)$ for all $j$.

Now we summarize the results to give the concentration result for $\Delta$. Let $IV=N^{\alpha}\int^T_0(\sum^r_{k=1}\Sigma_{t, k\cdot}\Sigma_{t, k\cdot}^{\prime})dt+\sum^N_{i=1}\int^T_0(\sigma_{it}^*)^2dt=:N^{\alpha}IV_1+IV_2$. For any $0<u<1$,
\begin{eqnarray*}
	&&P_{R,D}\left(|\|\Delta\|_F^2-IV|>Nu \ \mbox{for some} \ \sigma^0\in \mathcal{B}(N,r)\right)\nonumber\\
	&\leq & P_{R,D}\left(N^{\alpha}|\|V\|_F^2-IV_1|>Nu/3\right)\nonumber\\
	&&+P_{R,D}\left(\|\Pi^{\mu}\|_F^2>Nu/3\right)+P_{R,D}\left(|\|\Pi^*\|-IV_2|>Nu/3\right).
\end{eqnarray*}
For the first term, (\ref{VC}) shows that
$$
P_{R,D}\left(N^{\alpha}|\|V\|_F^2-IV_1|>Nu/3\right)\leq 2\exp\left\{-N^{2-2\alpha}u^2/\left(72C^2\sum^n_{j=1}R_j^2\right)\right\}
$$
By the boundedness of the drift coefficient, for $n$ large enough,
$$
P_{R,D}\left(\|\Pi^{\mu}\|_F^2>Nu/3\right)=0.
$$
Taking $\theta=1/(4CR_j^{2-\epsilon})$ and $y=u/(3\sqrt{\sum^n_{j=1}R_j^{4-2\epsilon}})$ in (\ref{concentration2-1}), and $\theta=\frac{\epsilon^*}{\sqrt{2C}R_j}$ and $z=\sqrt{N}u/(3\sqrt{\sum^n_{j=1}R_j^2})$ for $\epsilon^*$ small enough in (\ref{concentration2}), we have for small but fixed $u$,
\begin{eqnarray*}
	&&P_{R,D}\left(|\|\Pi^*\|-IV_2|>Nu/3\right)\\
	&\leq & 2N\exp\left\{-\frac{u}{12C\max_{j}R_j^{2-\epsilon}}+\frac{\sum^n_{j=1}R_j^{4-2\epsilon}}{8(\max_jR_j^{2-\epsilon})^2}\right\}\\
	&& + 2\exp\left\{-\frac{Nu\epsilon^*}{3\sqrt{2C}\max_jR_j}+\frac{3N(\epsilon^*)^2\sum^n_{j=1}R_j^2}{8(\max_jR_j)^2}\right\}\\
	&\leq & 2N\exp\left\{-\frac{u}{12C\max_{j}R_j^{2-\epsilon}}\right\}+ 2\exp\left\{-\frac{Nu\epsilon^*}{3\sqrt{2C}\max_jR_j}\right\}.
\end{eqnarray*}
Putting pieces together, for $N$ and $n$ large enough,
\begin{eqnarray*}\label{IV1}
	&& P_{R,D}\left(|\|\Delta\|_F^2-IV|>Nu \ \mbox{for some} \ \sigma^0\in \mathcal{B}(N,r)\right)\\
	&\leq &  2\exp\left\{-N^{2-2\alpha}u^2/\left(72C^2\sum^n_{j=1}R_j^2\right)\right\}+2N\exp\left\{-\frac{u}{12C\max_{j}R_j^{2-\epsilon}}\right\}\\
	&&+ 2\exp\left\{-\frac{Nu\epsilon^*}{3\sqrt{2C}\max_jR_j}\right\}.
\end{eqnarray*}
Due to Assumption \ref{ass2}, we further deduce that
\begin{eqnarray*}
	&& P_{R,D}\left(\|\Delta\|_F^2\leq Nc/2 \ \mbox{for some} \ \sigma^0\in \mathcal{B}(N,r)\right)\nonumber\\
	&\leq &  2\exp\left\{-N^{2-2\alpha}c^2/\left(128C^2\sum^n_{j=1}R_j^2\right)\right\}+2N\exp\left\{-\frac{c}{24C\max_{j}R_j^{2-\epsilon}}\right\}\nonumber\\
	&&+ 2\exp\left\{-\frac{Nc\epsilon^*}{6\sqrt{2C}\max_jR_j}\right\}.
\end{eqnarray*}

\noindent\textbf{Step 2.2: Bounding the Numerator $\|\mathcal{A}(\Delta)\|_F^2-\|\Delta\|_F^2$.}

So far we have given a bound for the denominator of $\{\|\mathcal{A}(\Delta)\|_F^2-\|\Delta\|_F^2\}/\|\Delta\|_F^2$. Next, we investigate into the numerator, i.e., the difference $\|\mathcal{A}(\Delta)\|_F^2-\|\Delta\|_F^2$. We begin with the difference between $\|\mathcal{A}(\Pi^*)\|_F^2$ and $\|\Pi^*\|_F^2$. This difference can also be expressed as
$$
\|\mathcal{A}(\Pi^*)\|_F^2-\|\Pi^*\|_F^2=\sum^n_{k=1}\sum^N_{i=1}\Pi^*(i, k)\sum^{d_{ik}}_{l=1}\Pi^*(i, k-l).
$$
Recall the notation $\overline{L}_2$. Similar to the derivation in (\ref{A}), we have
\begin{eqnarray}\label{A*}
	&&P_{R,D}\left(|\|\mathcal{A}(\Pi^*)\|_F^2-\|\Pi^*\|_F^2|>y\sqrt{\overline{L}_2}\right)\nonumber\\
	&\leq & \exp\left\{-\theta y \sqrt{\overline{L}_2}\right\}E\prod^{n-1}_{k=1}\exp\left\{\theta\sum^N_{i=1}\Pi^*(i, k)\sum^{d_{ik}}_{l=1}\Pi^*(i, k-l)\right\} \nonumber\\
	&& \times E_{\mathcal{F}_{n-1}}\exp\left\{\theta \sum^N_{i=1}\Pi^*(i, n)\sum^{d_{in}}_{l=1}\Pi^*(i, n-l)\right\}\nonumber\\
	&\leq & \exp\left\{-\theta y \sqrt{\overline{L}_2}+C\theta^2\overline{L}_2\right\},
\end{eqnarray}
for any $\theta>0$ and where in the last step, again, we have made use of the optional stopping theorem for submartingales and the L\'{e}vy representation theorem for continuous martingales. Take $\theta=y/(2C\sqrt{\overline{L}_2})$, the upper bound of (\ref{A*}) is minimized at $e^{-y^2/(4C)}$ for any $y>0$.

Lastly, we consider the difference between $\|\mathcal{A}(\Pi^{\mu})\|_F^2$ and $\|\Pi^{\mu}\|_F^2$. By the local boundedness of $\mu_t$ in Assumption \ref{ass2}, we soon have
$$
|\|\mathcal{A}(\Pi^{\mu})\|_F^2-\|\Pi^{\mu}\|_F^2|\leq C\sum^N_{i=1}\sum^n_{k=1}R_k\sum^{D_{ik}}_{l=1}R_{k-l}=o(N),
$$
by Assumption \ref{ass1}. This shows that
\begin{eqnarray*}
	P_{R,D}\left(|\|\mathcal{A}(\Pi^{\mu})\|_F^2-\|\Pi^{\mu}\|_F^2|>N\epsilon/3\right)=0,
\end{eqnarray*}
for large enough $N$ and $n$. Taking
$$
x=\frac{\epsilon N^{1-\alpha}}{3}, \ y=\frac{\epsilon N}{3\sqrt{\overline{L}_2}},
$$
we have
\begin{eqnarray*}
	&&P_{R,D}\left(|\|\mathcal{A}(\Delta)\|_F^2-\|\Delta\|_F^2|>N\epsilon\right)\nonumber\\
	&\leq & 2\exp\left\{-\frac{N^{2-\alpha}\epsilon^2}{36Cr^2\overline{L}_1}\right\}+ 2 \exp\left\{-\frac{N^2\epsilon^2}{36C\overline{L}_2}\right\}.
\end{eqnarray*}
\noindent\textbf{Step 2.3: Combining the Bounds.}

Finally, combining the probability bounds for the numerator and denominator of $\frac{\|\mathcal{A}(\Delta)\|_F^2-\|\Delta\|_F^2}{\|\Delta\|_F^2}$ leads to the desired result in \eqref{bound-th1-1}:
\begin{eqnarray}\label{combine1}
	&&P_{R,D}\left(|\frac{\|\mathcal{A}(\Delta)\|_F^2-\|\Delta\|_F^2}{\|\Delta\|_F^2}|>\delta \right)\nonumber\\
	&\leq & P_{R,D}\left(|\|\mathcal{A}(\Delta)\|_F^2-\|\Delta\|_F^2|>\frac{c N\delta}{2}\right) + P_{R,D}\left(\|\Delta\|_F^2 \leq \frac{Nc}{2}\right)\nonumber\\
	&\leq & 2\exp\left\{-\frac{N^{2-\alpha}c^2\delta^2}{144Cr^2\overline{L}_1}\right\}+ 2 \exp\left\{-\frac{N^2c^2\delta^2}{144C\overline{L}_2}\right\}\nonumber\\
	&& + 2\exp\left\{-c^2N^{2-2\alpha}/\left(128C^2\sum^n_{j=1}R_j^2\right)\right\}+2N\exp\left\{-\frac{c}{24C\max_{j}R_j^{2-\epsilon}}\right\}\nonumber\\
	&&+ 2\exp\left\{-\frac{Nc\epsilon^*}{6\sqrt{2C}\max_jR_j}\right\}.
\end{eqnarray}

\Halmos
\endproof

We next prove Theorem \ref{th2} that says the upper bounds in (\ref{denominator+numerator}) and (\ref{combine1}) hold uniformly in $\sigma^0\in \mathcal{B}(N, r)$.

\proof{Proof of Theorem \ref{th2}}
First, we recall that $U$ is an arbitrary subspace of $N\times n$ matrices with dimension $r$. Without loss of generality, we assume that $\|\Pi\|_F^2\leq 1$ and let $M$ be the maximum of $\mathcal{A}(\Pi)$ for $\Pi\in U$. Theorem \ref{th1} demonstrates that (\ref{bound-th1}) holds for each $Q$, and thus $(1-\delta/2)\|Q\|_F\leq \|\mathcal{A}(Q)\|\leq (1+\delta/2)\|Q\|_F$ for large enough $N$ and $n$ (replace $\delta$ in (\ref{bound-th1}) by $\delta/2$). Then by the triangular inequality,
$$
\|\mathcal{A}(\Pi)\|\leq \|\mathcal{A}(Q)\|+\|\mathcal{A}(\Pi-Q)\|\leq 1+\delta/2+M\delta/4\leq 1+\delta,
$$
where we noticed that $M\leq 1+\delta$ due to $M\leq 1+\delta/2+M\delta/4$, and
$$
\|\mathcal{A}(\Pi)\|\geq \|\mathcal{A}(Q)\|-\|\mathcal{A}(\Pi-Q)\|\geq 1-\delta/2-(1+\delta)\delta/4\geq 1-\delta.
$$
This proves that for any $\delta\in (0, 1)$, for $N$ and $n$ large enough,
\begin{eqnarray}\label{U}
	&&P_{R,D}\left(|\frac{\|\mathcal{A}(\Pi)\|_F^2-\|\Pi\|_F^2}{\|\Pi\|_F^2}|>\delta/2 \ \mbox{for some} \ \sigma^0\in U \right)\nonumber\\
	&\leq &
	2\exp\left\{-c^2r^2/\left(\sum^n_{j=1}R_j^2\right)^{1/2}\right\}+C\left(\frac{24}{\delta}\right)^r\exp\{-c^2\delta^2N^{\alpha}/\overline{L}_1\}.
\end{eqnarray}
Notice that the term $(\frac{24}{\delta})^r$ is only multiplied to the second term in the upper bound of Theorem \ref{th1} because the first term in the bound is irrelevant to $\sigma^0$.
By Lemma 4.4 and equation (4.17) of \cite{recht2010guaranteed},
\begin{eqnarray}\label{Sigmai}
	\sup_{\sigma^0\in \{U; \rho(U, U_i)\leq \epsilon/2 \}}|\frac{\|\mathcal{A}(\Pi)\|_F-\|\Pi\|_F}{\|\Pi\|_F}|\leq \delta^{\prime}:=\delta/2+(\|\mathcal{A}\|+1)\epsilon.
\end{eqnarray}
To make $\delta^{\prime}<\delta$, we should have $\|\mathcal{A}\|=\lambda_{\max}^{1/2}(AA^{\prime})\leq \delta/(2\epsilon)-1$ where $AA^{\prime}$ is a diagonal matrix. In the subsequent, we choose $\epsilon=(\delta/4)(\sqrt{Nn/p}+1)^{-1}$. By the definition of $\mathcal{A}$ and the Assumption on $\|\mathcal{A}\|$, we have
\begin{eqnarray}\label{normA}
	P_{R,D}\left(\|\mathcal{A}\|>\frac{\delta}{2\epsilon}-1\right)\leq P_{R,D}\left(\lambda_{\max}^{1/2}(AA^{\prime})>2\sqrt{Nn/n^*+1}\right)\leq C\exp\{-\gamma Nn\}.
\end{eqnarray}
Combining (\ref{Sigmai}) and (\ref{normA}), we have
\begin{eqnarray*}
	P_{R,D}\left(\sup_{\sigma^0\in \{U; \rho(U, U_i)\leq \epsilon/2 \}}|\frac{\|\mathcal{A}(\Pi)\|_F-\|\Pi\|_F}{\|\Pi\|_F}\| > \delta\right) \leq C\exp\{-\gamma Nn\}.
\end{eqnarray*}
By (\ref{U}) and the covering number of $\mathcal{B}(N, r)$,
\begin{eqnarray*}
	&& P_{R,D}\left(|\frac{\|\mathcal{A}(\Pi)\|_F^2-\|\Pi\|_F^2}{\|\Pi\|_F^2}|>\delta/2, \ \mbox{for some} \ \sigma^0\in U \right)\nonumber\\
	&\leq &
	2\left(\frac{2C_0}{\epsilon}\right)^{r(N-r)}\left(\frac{24}{\delta}\right)^r\exp\left\{-c^2\delta^2N^{\alpha}/\overline{L}_1\right\}+2\exp\left\{-c^2r^2/\left(\sum^n_{j=1}R_j^2\right)^{1/2}\right\}\nonumber\\
	&\leq & C\exp\{-c^2\delta^2N^{\alpha}/\overline{L}_1\}+2\exp\left\{-c^2r^2/\sum^n_{j=1}R_j^2\right\},
\end{eqnarray*}
for $c$ small enough and $C$ large enough which depend on $\delta$ and may change across lines, where in the last step, we have made use of the condition that
$$
r(N-r)\log{\left(\sqrt{\frac{Nn}{n^*}}+1\right)}=o\left\{\frac{N^{\alpha}}{\overline{L}_1}\right\}.
$$
This proves (\ref{equation-th2}).

Next, we prove the uniqueness. Suppose that there is another matrix $\Pi_1$ of rank at most $r$ so that $\mathcal{A}(\Pi_1)=b$ or $\mathcal{A}(\Pi_1)=b-\mathcal{A}(\Pi^*)$ and $\Pi_0\not=\Pi_1$. Then $\Pi_1-\Pi_0$ is a nonzero matrix of rank at most $2r$ and $\mathcal{A}(\Pi_1-\Pi_0)=0$. However, $0=\|\mathcal{A}(\Pi_1-\Pi_0)\|\geq (1-\delta)\|\Pi_1-\Pi_0\|_F>0$, where $\delta$ implicitly depends on $2r$ here. This contradiction shows the uniqueness of $\Pi$.

Similar to the proof of (\ref{equation-th2-1}), due to
$$
r(N-r)\log{\sqrt{Nn/n^*}+1}=o(N^{2-\alpha}/\overline{L}_1)
$$
we have
\begin{eqnarray*}
	&&P_{R,D}\left(|\frac{\|\mathcal{A}(\Delta)\|_F^2-\|\Delta\|_F^2}{\|\Delta\|_F^2}|>\delta \ \mbox{for some} \ \sigma^0\in\mathcal{B}(N, r) \right)\nonumber\\
	&\leq & P_{R,D}\left(|\|\mathcal{A}(\Delta)\|_F^2-\|\Delta\|_F^2|>\frac{c N\delta}{2} \ \mbox{for some} \ \sigma^0\in\mathcal{B}(N, r) \right) + P_{R,D}\left(\|\Delta\|_F^2\leq \frac{Nc}{2}\right)\nonumber\\
	&\leq & 2\exp\left\{-\frac{N^{2-\alpha}c^2\delta^2}{144Cr^2\overline{L}_1}\right\}+ 2 \exp\left\{-\frac{N^2c^2\delta^2}{144C\overline{L}_2}\right\}\nonumber\\
	&& + 2\exp\left\{-c^2N^{2-2\alpha}/\left(128C^2\sum^n_{j=1}R_j^2\right)\right\}+2N\exp\left\{-\frac{c}{24C\max_{j}R_j^{2-\epsilon}}\right\}\nonumber\\
	&&+ 2\exp\left\{-\frac{Nc\epsilon^*}{6\sqrt{2C}\max_jR_j}\right\}.
\end{eqnarray*}
Let $\Delta_1$ be another solution of (\ref{nuclear norm2}), similarly, we have
$$
0=\|\mathcal{A}(\Delta_1)-\mathcal{A}(\Delta_0)\|^2\geq (1-\delta)\|\Delta_1-\Delta_0\|_F^2>0.
$$
This proves the uniqueness of $\Delta$ solving (\ref{nuclear norm2}).

\Halmos
\endproof

\proof{Proof of Theorem \ref{th3}} Let $(\widehat{\Pi}_*, \widehat{\Delta}_*)$ be a linear combination of $(\widehat{\Pi}, \widehat{\Delta})$ and $(\widehat{\Pi}_0, \widehat{\Delta}_0)$, i.e., $(\widehat{\Pi}_*, \widehat{\Delta}_*)=\frac{\delta}{\beta}(\widehat{\Pi}, \widehat{\Delta})+(1-\frac{\delta}{\beta})(\widehat{\Pi}_0, \widehat{\Delta}_0)$ for $\beta>\delta$. Then we have $\|(\widehat{\Pi}_*, \widehat{\Delta}_*)-(\widehat{\Pi}_0, \widehat{\Delta}_0)\|_F=\delta$ and $\|(\widehat{\Pi}, \widehat{\Delta})-(\widehat{\Pi}_0, \widehat{\Delta}_0)\|_F=\beta$. By the convexity of the function $G_{n}(\Pi,\Delta)$, we have
$$
\frac{\delta}{\beta}G_{n}(\widehat{\Pi}, \widehat{\Delta})+(1-\frac{\delta}{\beta})G_{n}(\widehat{\Pi}_0,\widehat{\Delta}_0)\geq G_{n}(\widehat{\Pi}_*,\widehat{\Delta}_*).
$$
This together with (\ref{closeness}) yields
\begin{eqnarray}\label{convexity}
	\frac{\delta}{\beta}(G_{n}(\widehat{\Pi}, \widehat{\Delta})-G_{n}(\widehat{\Pi}_0, \widehat{\Delta}_0))&\geq & G_{n}(\widehat{\Pi}_*,\widehat{\Delta}_*)-G_{n}(\widehat{\Pi}_0,\widehat{\Delta}_0) \nonumber\\
	&\geq & G_{n0}(\widehat{\Pi}_*,\widehat{\Delta}_*)-G_{n0}(\widehat{\Pi}_0,\widehat{\Delta}_0)-2Na_{Nn}\nonumber\\
	&\geq & c(1+\lambda)(\|\widehat{\Pi}_*-\widehat{\Pi}_0\|_F^2+\|\widehat{\Delta}_*-\widehat{\Delta}_0\|)-2Na_{Nn}\nonumber\\
	&=& c(1+\lambda)\delta^2-2Na_{Nn},
\end{eqnarray}
for some $c>0$ with probability approaching one, where in the last step we have made use of Theorem 4.2 in \cite{zhang2016no}. The inequality (\ref{convexity}) demonstrates that
\begin{eqnarray*}
	0\geq \frac{\delta}{\beta}\left[G_{n}(\widehat{\Pi},\widehat{\Delta})-G_{n}(\widehat{\Pi}_0,\widehat{\Delta}_0)\right]\geq c(1+\lambda)\delta^2-2Na_{Nn}.
\end{eqnarray*}
This shows that $G_{n}(\Pi, \Delta)$ can not be minimized outside a $\delta$ neighborhood of $(\widehat{\Pi}_0,\widehat{\Delta}_0)$ when $\delta>\sqrt{\frac{cNa_{Nn}}{1+\lambda}}$, and hence $\|\widehat{\Pi}-\widehat{\Pi}_0\|_F=O_p\left(\sqrt{\frac{Na_{Nn}}{1+\lambda}}\right)$ and $\|\widehat{\Delta}-\widehat{\Delta}_0\|_F=O_p\left(\sqrt{\frac{Na_{Nn}}{1+\lambda}}\right)$.

\Halmos
\endproof

\proof{Proof of Theorem \ref{th4}} The proof relies on a standard error decomposition. By adding and subtracting the oracle estimators $\widehat{\Delta}_0$ and $\widehat{\Pi}_0$, we can write:
$$
\widehat{\Delta}-\Delta_0=\widehat{\Delta}-\widehat{\Delta}_0+\widehat{\Delta}_0-\Delta_0, \quad \widehat{\Pi}-\Pi_0=\widehat{\Pi}-\widehat{\Pi}_0+\widehat{\Pi}_0-\Pi_0.
$$
Applying the triangle inequality to the norms of these expressions, the results of Theorem \ref{th4} follow directly from the bounds provided in Theorem \ref{th3} and the definitions of the oracle estimators $\widehat{\Delta}_0$ and $\widehat{\Pi}_0$.

\Halmos
\endproof

\proof{Proof of Theorem \ref{th5}} The proof strategy is to decompose the total error into two components: the error of the de-biased estimator relative to the initial estimator, and the error of the initial estimator itself, which is bounded by Theorem \ref{th3}. We begin with the following decompositions:
\begin{eqnarray}
\label{decom1}	\widetilde{\Delta}-\Delta_0&=&\widehat{\Delta}-\widehat{\Delta}_0+\left[(\widetilde{\Delta}-\widehat{\Delta})+(\widehat{\Delta}_0-\Delta_0)\right],\\
\label{decom2}	\widetilde{\Pi}-\Pi_0&=&\widehat{\Pi}-\widehat{\Pi}_0+\left[(\widetilde{\Pi}-\widehat{\Pi})+(\widehat{\Pi}_0-\Pi_0)\right].
\end{eqnarray}
We will analyze the second term on the right-hand side for both cases. Let $U_n$ and $V_n$ be the matrices of left and right singular vectors of $\widehat{\Delta}$, respectively. By the $SIN(\theta)$ theorem and Theorem \ref{th3},
\begin{equation}\label{sintheta}
	\|U_n-U_*\|/\sqrt{N}+\|V_n-V_*\|/\sqrt{N}=O_p\left(\sqrt{\frac{a_{Nn}}{1+\lambda}}\right).
\end{equation}
Let $D_b$ be the difference between the matrices of the singular values of $\widetilde{\Delta}$ and $\widehat{\Delta}$, then we have by (\ref{sintheta})
\begin{eqnarray*}
	&&\|\widetilde{\Delta}-\widehat{\Delta}+\widehat{\Delta}_0-\Delta_0\|=\|U_nD_bV_n^{\prime}-U_*D_bV_*^{\prime}\|=O_P\left(\|D_b\|\sqrt{\frac{Na_{Nn}}{1+\lambda}}\right).
\end{eqnarray*}
This together with (\ref{decom1}) and Theorem \ref{th3} proves the results for $\|\widetilde{\Delta}-\Delta_0\|$. For the closeness in the Frobenius norm, the steps are the same but with the operator norm replaced by the Frobenius norm.

The proofs for $\widetilde{\Pi}$ are similar to that for $\widetilde{\Delta}$ except for noticing that the difference between the matrices of the singular values of $\widetilde{\Pi}$ and $\widehat{\Pi}$ has $J$ nonzero singular values all equal to $\lambda(1+1/\mu)$, while the difference between the matrices of the singular values of $\widehat{\Pi}_0$ and $\Pi_0$ has the nonzero singular values being $(\underbrace{-\lambda(1+1/\mu),...,-\lambda(1+1/\mu)}_J, -\lambda_{J+1}^*,...,-\lambda_{r}^*)$ when $J<r$, being $(\underbrace{-\lambda(1+1/\mu),...,-\lambda(1+1/\mu)}_r, \lambda_{r+1}^*-\lambda(1+1/\mu),...,\lambda_{J}^*-\lambda(1+1/\mu))$ when $J>r$, and being $(\underbrace{-\lambda(1+1/\mu),...,-\lambda(1+1/\mu)}_r)$ when $J=r$. Then by the Wyle theorem and the $SIN(\theta)$ theorem,
\begin{eqnarray*}\label{sintheta1}
	&& \|\widetilde{\Pi}-\widehat{\Pi}+\widehat{\Pi}_0-\Pi_0\|/\sqrt{N}\\
	&=& O_p\left\{\left[\lambda\left(1+\frac{1}{\mu}\right)+1\right]\sqrt{\frac{a_{Nn}}{1+\lambda}}+\frac{\lambda^*_{J+1}I(J<r)+\lambda^*_{r+1}I(J>r)+0I(J=r)}{\sqrt{N}}\right\},
\end{eqnarray*}
and replacing the operator norm by the Frobenius norm,
\begin{eqnarray*}\label{sintheta2}
	&& \|\widetilde{\Pi}-\widehat{\Pi}+\widehat{\Pi}_0-\Pi_0\|_F/\sqrt{N}\\
	&=& O_p\left\{\left[\lambda\left(1+\frac{1}{\mu}\right)+1\right]\sqrt{\frac{a_{Nn}}{1+\lambda}}+\frac{\sum\limits^r_{l=J+1}\lambda^*_lI(J<r)+\sum\limits^J_{l=r+1}\lambda^*_{r+1}I(J>r)+0I(J=r)}{\sqrt{N}}\right\}.
\end{eqnarray*}
This together with (\ref{decom2}) and Theorem \ref{th3} proves the results for $\widetilde{\Pi}-\Pi_0$.

\Halmos
\endproof

\section{Additional Simulation}\label{sec:Additional Simulation}
\begin{figure}
\FIGURE
	{\includegraphics[scale=0.42]{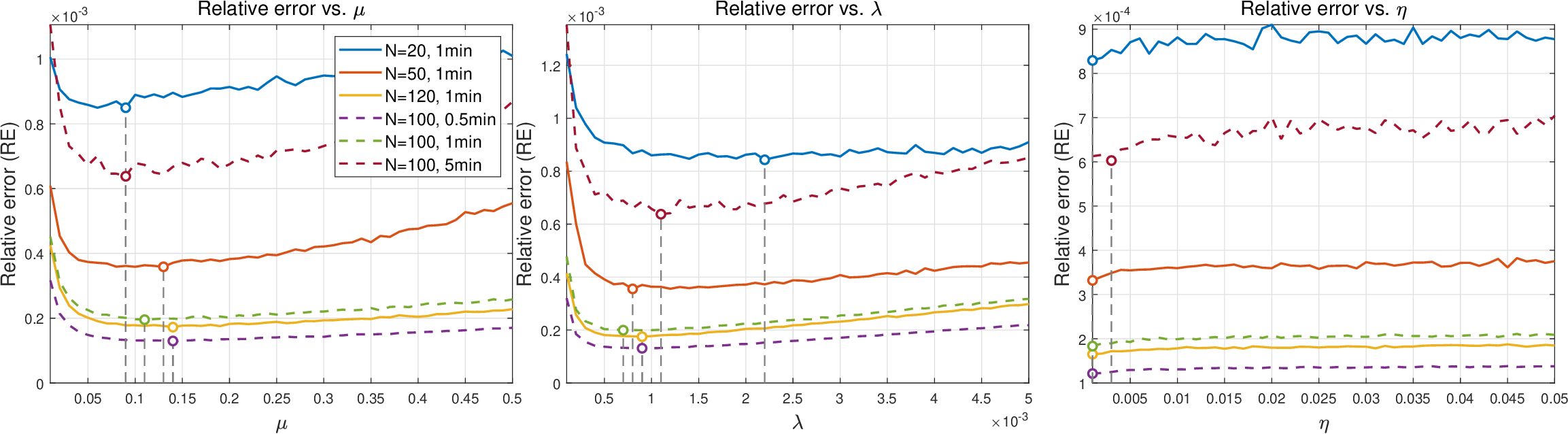}}
	{Relative error as a function of tuning parameters $\mu$, $\lambda$, and $\eta$. \label{fig:TuningPara_RE}}
	{The relative error is calculated according to \eqref{eq:compute error0}. The vertical dotted line in each panel indicates the parameter value that minimizes the error.}
\end{figure}

In this subsection, we show the results of another tuning parameter selection in the main text, based on the relative error, as in Figure \ref{fig:TuningPara_RE}.

The analysis based on relative errors yields conclusions that are largely consistent with those derived from absolute errors. This consistency across different error metrics further underscores the robustness of our findings and the stability of the method with respect to its tuning parameters.

\section{Additional Empirical Analysis}\label{sec:Additional Empirical Analysis}
\subsection{Factors}\label{subsec:Factors}
\begin{figure}
	\FIGURE
	{\includegraphics[scale=0.55]{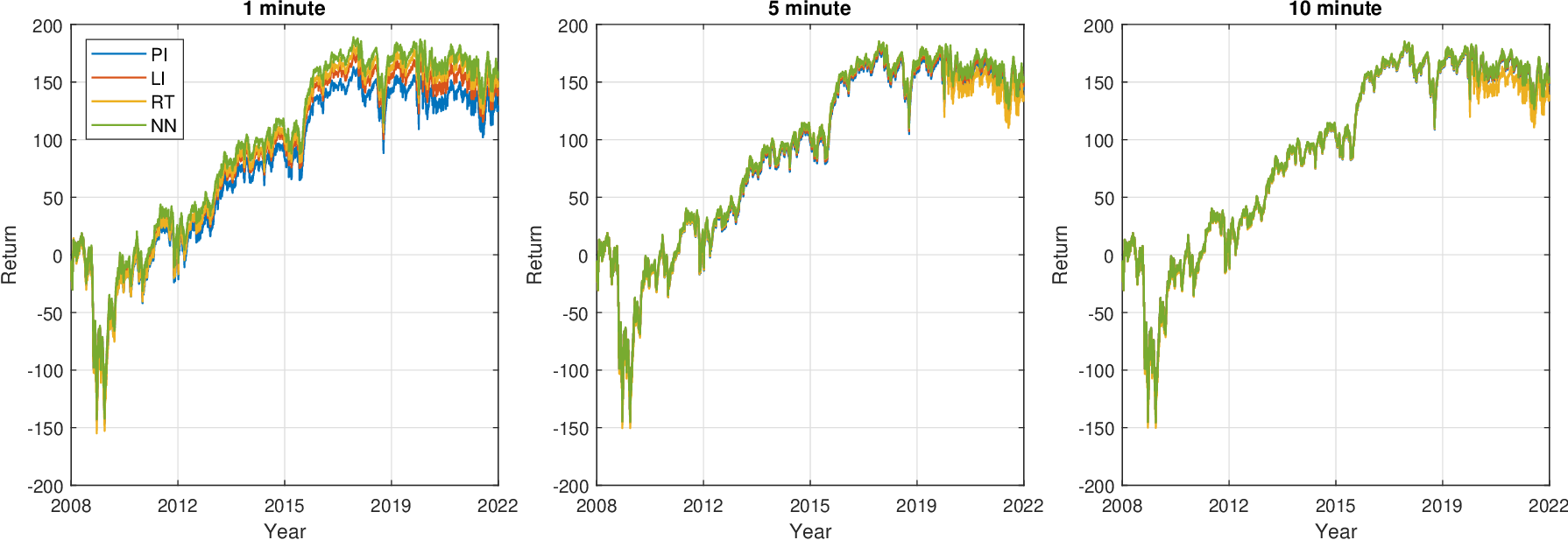}}
	{Normalized cumulative factor returns for intraday returns. \label{fig:CumFactor}}
	{}
\end{figure}

In this subsection, we further evaluate the economic implications of different imputation methods by constructing and analyzing a high-frequency return factor. Specifically, we extract the first principal component (the ``market factor'') from the imputed intraday return matrices and compute its cumulative performance over time, excluding overnight returns. Figure \ref{fig:CumFactor} plots the cumulative return of this factor, derived from data imputed by four different methods (NN, PI, LI, RT), across three sampling frequencies. As expected, the first latent factor captures a significant portion of the total variation under all imputation methods.

The results for the 1-minute frequency reveal substantial performance discrepancies among the methods. The factor constructed from our NN-imputed data yields the highest cumulative return over the 15-year sample period. In contrast, the factor derived from the PI method performs the worst. The performance gap is economically significant: the cumulative return of the top-performing factor (NN) exceeds that of the worst-performing factor (PI) by more than 30\%.

At lower frequencies (5-minute and 10-minute), the performance gap between the NN, PI, and LI methods narrows considerably. This is consistent with our earlier findings that the distorting effects of price staleness are less severe in lower-frequency data. However, the factor derived from the RT method continues to exhibit markedly inferior performance. This is attributable to the inherent drawback of the RT scheme, which discards a vast amount of valid price information by subsampling to a sparse, synchronized time grid, thereby compromising the quality of the extracted factor.

\subsection{Additional Beta Results}\label{subsec:Additional Beta Results}
\begin{figure}
	\FIGURE
	{\includegraphics[scale=0.55]{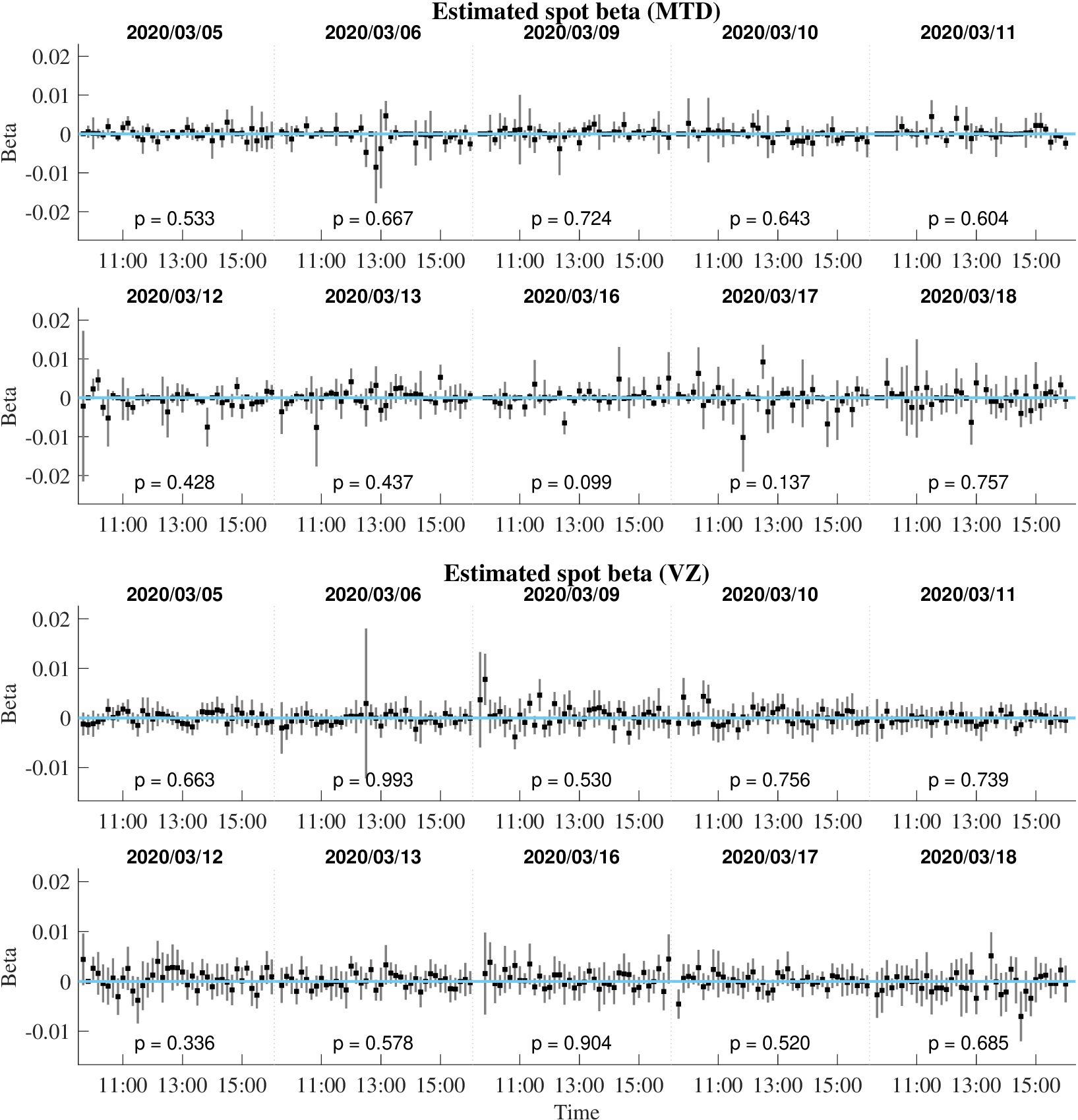}}
	{Spot beta for the presence of price staleness. \label{fig:Beta PI 5min}}
	{This figure plots the estimated spot betas of Mettler-Toledo International (MTD) and Verizon Communications (VZ) against the SPY. The betas are estimated using 1-minute price data over 5-minute rolling windows, along with their corresponding 90\% confidence intervals. The analysis covers the two-week period of high market volatility from March 5, 2020, to March 18, 2020. The $p$-value reported in each panel corresponds to a test of the functional null hypothesis that the entire spot beta process for a given day is equal to zero ($H_0:\beta_t=0$ for all $t$).}
\end{figure}

\begin{figure}
	\FIGURE
	{\includegraphics[scale=0.55]{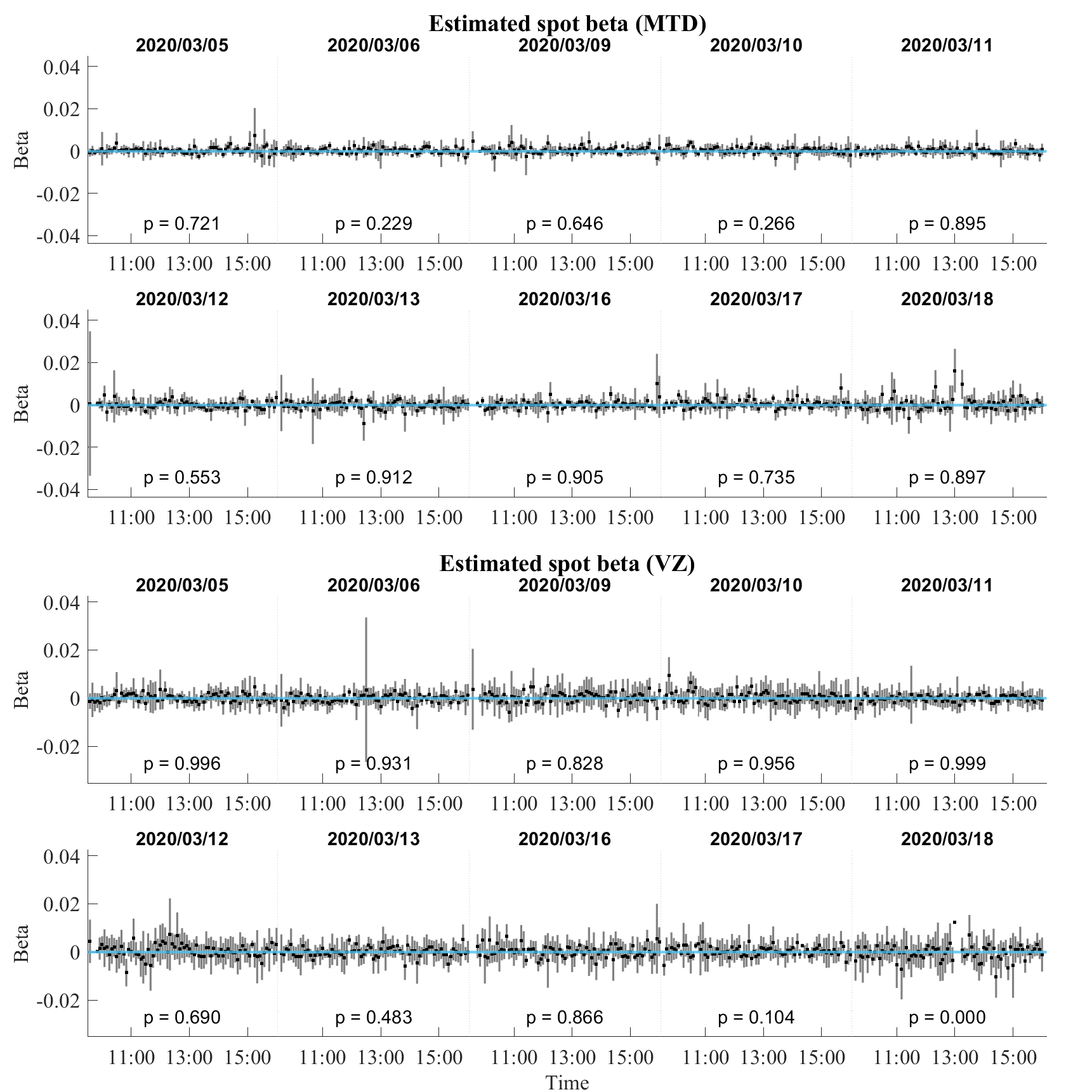}}
	{Spot beta for the absence of price staleness. \label{fig:Beta NN 5min}}
	{This figure plots the estimated spot betas of Mettler-Toledo International (MTD) and Verizon Communications (VZ) against the SPY. The betas are estimated using 1-minute price data over 5-minute rolling windows, along with their corresponding 90\% confidence intervals. The analysis covers the two-week period of high market volatility from March 5, 2020, to March 18, 2020. The $p$-value reported in each panel corresponds to a test of the functional null hypothesis that the entire spot beta process for a given day is equal to zero ($H_0:\beta_t=0$ for all $t$).}
\end{figure}

\begin{figure}
	\FIGURE
	{\includegraphics[scale=0.55]{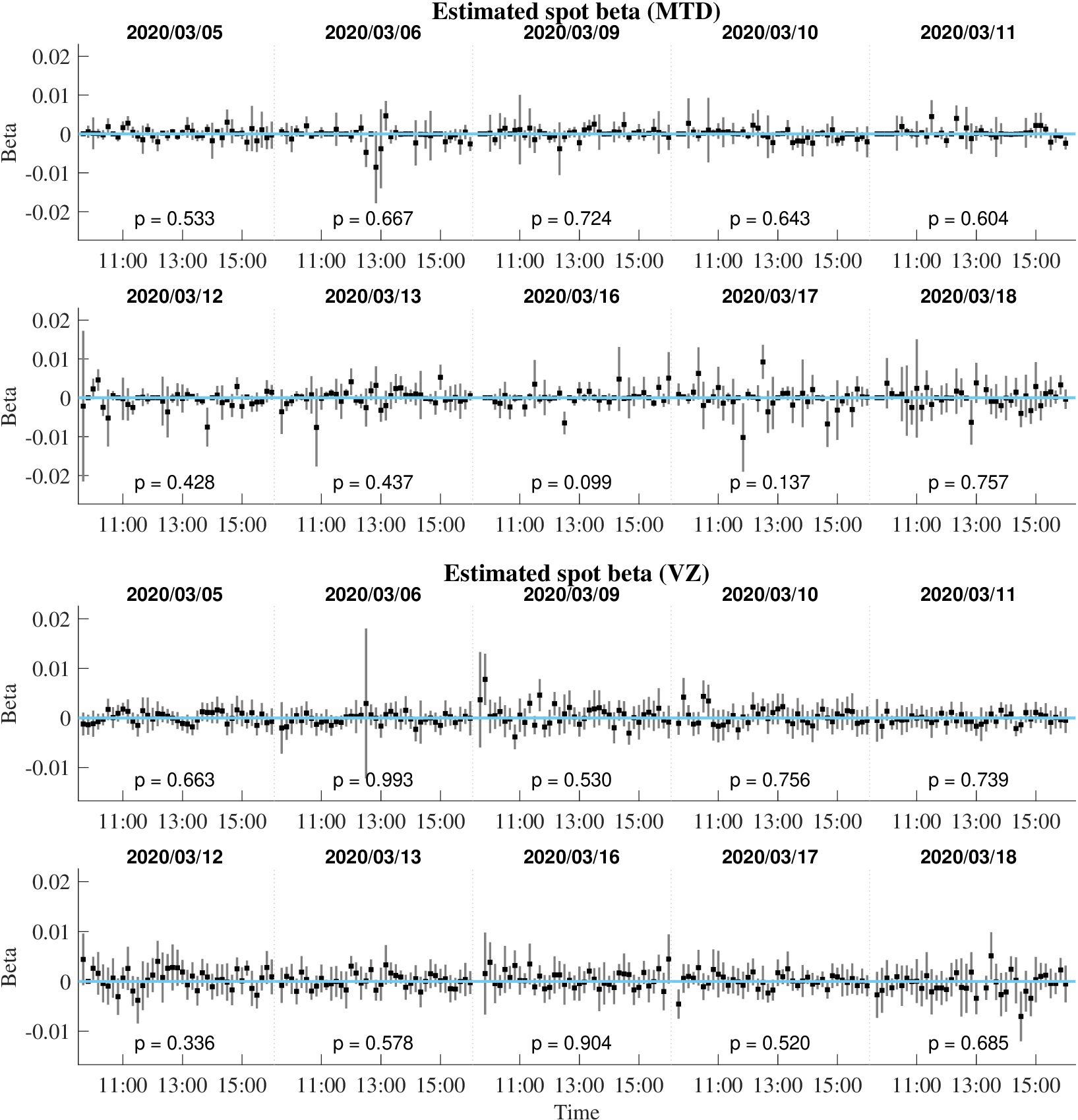}}
	{Spot beta for the presence of price staleness. \label{fig:Beta PI 10min}} 
	{This figure plots the estimated spot betas of Mettler-Toledo International (MTD) and Verizon Communications (VZ) against the SPY. The betas are estimated using 1-minute price data over 10-minute rolling windows, along with their corresponding 90\% confidence intervals. The analysis covers the two-week period of high market volatility from March 5, 2020, to March 18, 2020. The $p$-value reported in each panel corresponds to a test of the functional null hypothesis that the entire spot beta process for a given day is equal to zero ($H_0:\beta_t=0$ for all $t$).}
\end{figure}

\begin{figure}
	\FIGURE
	{\includegraphics[scale=0.55]{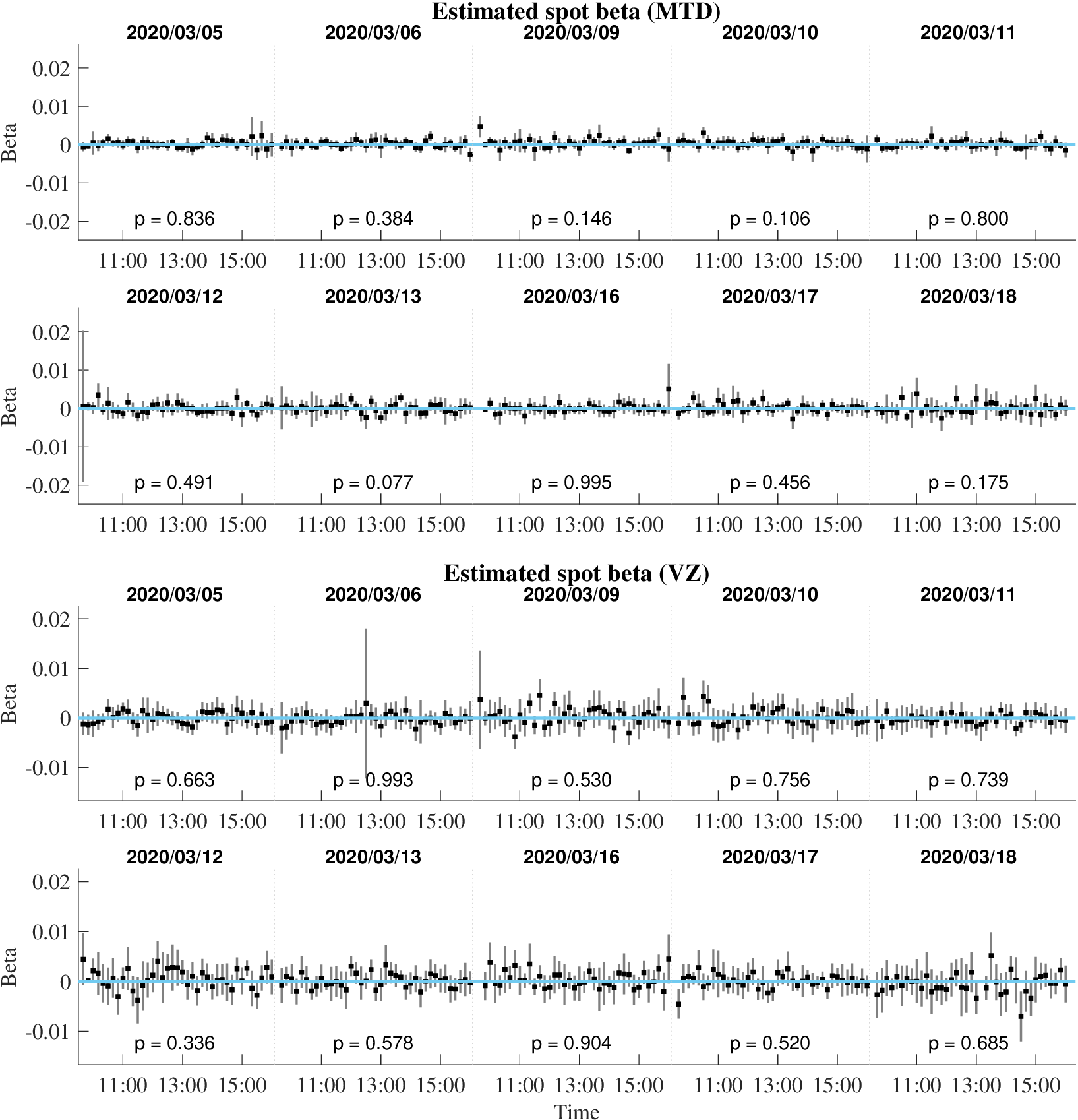}}
	{Spot beta for the absence of price staleness. \label{fig:Beta NN 10min}} 
	{This figure plots the estimated spot betas of Mettler-Toledo International (MTD) and Verizon Communications (VZ) against the SPY. The betas are estimated using 1-minute price data over 10-minute rolling windows, along with their corresponding 90\% confidence intervals. The analysis covers the two-week period of high market volatility from March 5, 2020, to March 18, 2020. The $p$-value reported in each panel corresponds to a test of the functional null hypothesis that the entire spot beta process for a given day is equal to zero ($H_0:\beta_t=0$ for all $t$).}
\end{figure}

Building on our initial analysis using a 15-minute estimation window, we now conduct a series of robustness checks by re-estimating the spot betas for Mettler-Toledo International (MTD) and Verizon Communications (VZ) using 5-minute (Figures \ref{fig:Beta PI 5min} and \ref{fig:Beta NN 5min}) and 10-minute (Figures \ref{fig:Beta PI 10min} and \ref{fig:Beta NN 10min}) windows. This multi-window approach allows us to directly examine the practical implications of the bias-variance trade-off discussed in \cite{bollerslev2024optimal} and to verify the stability of our core conclusions. The findings strongly reaffirm our initial results and provide deeper insights into the properties of the estimation methods.

Our initial analysis revealed that the ``previous-tick'' method, which generates stale prices, produces artificial zero-beta estimates. The robustness checks demonstrate that this is not an idiosyncratic issue tied to a 15-minute window but a fundamental flaw of the method. Comparing the ``price staleness'' plots across windows (Figures \ref{fig:Beta PI 5min}, \ref{fig:Beta NN 5min}, and the original 15-minute plot) reveals that the problem of zero betas for the less-liquid MTD becomes more severe as the estimation window shrinks. In the 5-minute window (Figure \ref{fig:Beta PI 5min}), there are visibly more instances where the beta estimate collapses to zero than in the 10- or 15-minute windows. This is perfectly logical: the probability of a stock not trading is higher over a 5-minute interval than a 15-minute one. This confirms that the zero-beta phenomenon is a direct, mechanical consequence of price staleness, as first explored in early literature by \cite{scholes1977estimating}. Across all window sizes, the statistical inference from the price-staleness method remains unreliable. 

As expected from theory, the spot beta estimates become smoother as the window size $k$ increases. The 5-minute beta paths (Figure \ref{fig:Beta NN 5min}) are visibly ``noisier'' and have wider confidence intervals, reflecting higher variance from using fewer observations. The 10-minute paths (Figure \ref{fig:Beta PI 10min}) are smoother, and the original 15-minute paths are the smoothest of all, reflecting lower estimation variance. This is the classic bias-variance trade-off in action. Despite the differences in variance, the underlying economic narrative remains remarkably consistent across all windows for the our method.

The consistency of our findings across different temporal resolutions greatly strengthens our conclusions about intraday risk dynamics, a topic of growing interest (e.g., \citealt{andersen2021recalcitrant}). The general intraday shapes of the beta paths, especially on the most volatile days, are preserved across the 5, 10, and 15-minute horizons. This suggests that the primary intraday risk fluctuations for these stocks occur at a frequency that is well-captured even by a 15-minute window. A shorter window like 5 minutes provides a more granular view but confirms the same broader patterns, increasing our confidence that these are genuine features of the market and not artifacts of a specific window choice.

This analysis provides practical guidance. For assets where theory or observation suggests very high-frequency changes in risk exposure, a smaller $k$ (like 5 minutes) might be preferable, despite higher variance. For assets with smoother risk profiles or for analyses focused on broader intraday trends, a larger $k$ (like 15 minutes) can provide a clearer signal by reducing noise. The fact that our conclusions hold across this range indicates the robustness of both the underlying economic phenomena and our methodology itself.

The comprehensive robustness analysis confirms and strengthens our initial findings. We have shown that the flaws of the ``price staleness'' method are systematic, while the robust estimation procedure provides stable and economically meaningful results that are not sensitive to the specific choice of the estimation window. This demonstrates the power and reliability of the framework proposed by \cite{bollerslev2024optimal}, providing a credible tool for uncovering the rich, dynamic nature of systematic risk at high frequencies, even under the most extreme market conditions.

\end{document}